\documentclass[11pt,letterpaper]{article}
\usepackage{jheppub}

\usepackage{graphicx}

\usepackage{amsmath}
\usepackage{subfigure}
\usepackage{url}

\usepackage{multirow}
\usepackage{threeparttable}
\usepackage{paralist}

\newcommand{\vsh}{\vspace{-.5cm}}

\newcommand{\GeV}{\text{GeV}}

\DeclareRobustCommand{\Sec}[1]{Sec.~\ref{#1}}

\DeclareRobustCommand{\Tab}[1]{Table~\ref{#1}}

\DeclareRobustCommand{\Fig}[1]{Fig.~\ref{#1}}
\DeclareRobustCommand{\Figs}[2]{Figs.~\ref{#1} and \ref{#2}}
\DeclareRobustCommand{\Eq}[1]{Eq.~(\ref{#1})}

\DeclareRobustCommand{\Ref}[1]{Ref.~\cite{#1}}
\DeclareRobustCommand{\Refs}[1]{Refs.~\cite{#1}}

\newcommand{\be}{\begin{equation}}
\newcommand{\ee}{\end{equation}}

\title{Maximizing Boosted Top Identification by Minimizing N-subjettiness}

\author{Jesse Thaler}
\author{and Ken Van Tilburg}

\emailAdd{jthaler@jthaler.net}
\emailAdd{kvt@mit.edu}

\affiliation{Center for Theoretical Physics, Massachusetts Institute of Technology, Cambridge, MA 02139, USA}


\abstract{$N$-subjettiness is a jet shape designed to identify boosted hadronic objects such as top quarks.  Given $N$ subjet axes within a jet, $N$-subjettiness sums the angular distances of jet constituents to their nearest subjet axis.  Here, we generalize and improve on $N$-subjettiness by minimizing over all possible subjet directions, using a new variant of the $k$-means clustering algorithm.  On boosted top benchmark samples from the BOOST2010 workshop, we demonstrate that a simple cut on the 3-subjettiness to 2-subjettiness ratio yields 20\% (50\%) tagging efficiency for a 0.23\% (4.1\%) fake rate, making $N$-subjettiness a highly effective boosted top tagger.  $N$-subjettiness can be modified by adjusting an angular weighting exponent, and we find that the jet broadening measure is preferred for boosted top searches.  We also explore multivariate techniques, and show that additional improvements are possible using a modified Fisher discriminant.  Finally, we briefly mention how our minimization procedure can be extended to the entire event, allowing the event shape $N$-jettiness to act as a fixed $N$ cone jet algorithm.}

\begin{document}

\hfill MIT-CTP 4287

\maketitle

\section{Introduction}
\label{sec:introduction}

With over one inverse femtobarn delivered by the Large Hadron Collider (LHC), the ATLAS and CMS experiments are truly exploring the high energy frontier of particle physics.  Jets are an important probe for physics beyond the standard model, and both experiments have demonstrated a high level of sophistication in their study of jets.  Using modern infrared- and collinear-safe jets algorithms \cite{Cacciari:2008gp,Salam:2009jx}, the LHC experiments are searching for new physics in monojet production \cite{Aad:2011xw,Chatrchyan:2011nd}, high-mass dijet resonances \cite{Aad:2011aj,Collaboration:2011ns}, as well as multijet final states \cite{daCosta:2011qk,Collaboration:2011ida}, and these searches have an impressive reach for new phenomena.

In addition, both experiments have started to use boosted hadronic objects as a probe of new physics in data \cite{CMS-PAS-JME-10-013,CMS-PAS-EXO-11-006,ATLAS-CONF-2011-073} (see also \Ref{Abazov:2011vh,Aaltonen:2011pg} for Tevatron measurements).  When hadronically decaying resonances---such as top quarks, Higgs bosons, or $W/Z$ bosons---are produced with a large enough Lorentz boost factor, they form a ``fat jet'' where the decay products are highly collimated.  Jet mass is the most basic observable for distinguishing a boosted object from an ordinary quark- or gluon-initiated jet, but there has also been an explosion of interest in using jet substructure techniques to further distinguish, say, ``top jets'' from ``QCD jets''.   The experimental and theoretical progress in jet substructure has been summarized in a report following the BOOST2010 workshop \cite{Abdesselam:2010pt}, where the various tagging methods were roughly grouped as follows:  algorithmic procedures to directly identify subjets within a fat jet \cite{Seymour:1993mx,Butterworth:2002tt,Brooijmans:1077731,Thaler:2008ju,Kaplan:2008ie,Plehn:2009rk,Plehn:2010st}; jet shape techniques to measure the energy flow in a jet \cite{Almeida:2008yp,Gallicchio:2010sw,Hook:2011cq,Jankowiak:2011qa}; and grooming methods to improve jet mass resolution by reducing jet contamination \cite{Butterworth:2008iy,Ellis:2009su,Ellis:2009me,Krohn:2009th,Soper:2010xk}.  There has also been work on template and matrix element methods \cite{Almeida:2010pa,Soper:2011cr}.

Recently, we introduced a new method to tag boosted hadronic objects using a jet shape called $N$-subjettiness \cite{Thaler:2010tr}.  Denoted by $\tau_N$ and adapted from the event shape $N$-jettiness \cite{Stewart:2010tn}, $N$-subjettiness measures the degree to which radiation within a jet is aligned along $N$ candidate subjet axes.  As a jet shape, $N$-subjettiness is interesting in its own right, since it is a calculable property of jets that generalizes the notion of jet angularities \cite{Berger:2003iw,Almeida:2008yp,Ellis:2010rwa}.  As a boosted object tagger, $N$-subjettiness exhibits a number of advantages, combining the flexibility of jet shape techniques with the tagging performance of algorithmic procedures.  As a proof of concept, we found in \Ref{Thaler:2010tr} that a simple one-dimensional cut on the ratio $\tau_3/\tau_2$ is particularly effective for identifying boosted hadronic tops.  An alternative version of $N$-subjettiness defined in the jet rest frame was introduced by Kim in \Ref{Kim:2010uj} and applied to boosted Higgs searches.  Recently, $N$-subjettiness has been applied to boosted ditau resonances \cite{Englert:2011iz} and technipions \cite{Bai:2011mr}.

\begin{figure}[tp]
  \begin{center}
    \subfigure[]{\label{fig:master_pt500}\includegraphics[trim = 0mm 0mm 0mm 0mm, clip, width = 0.48\textwidth]{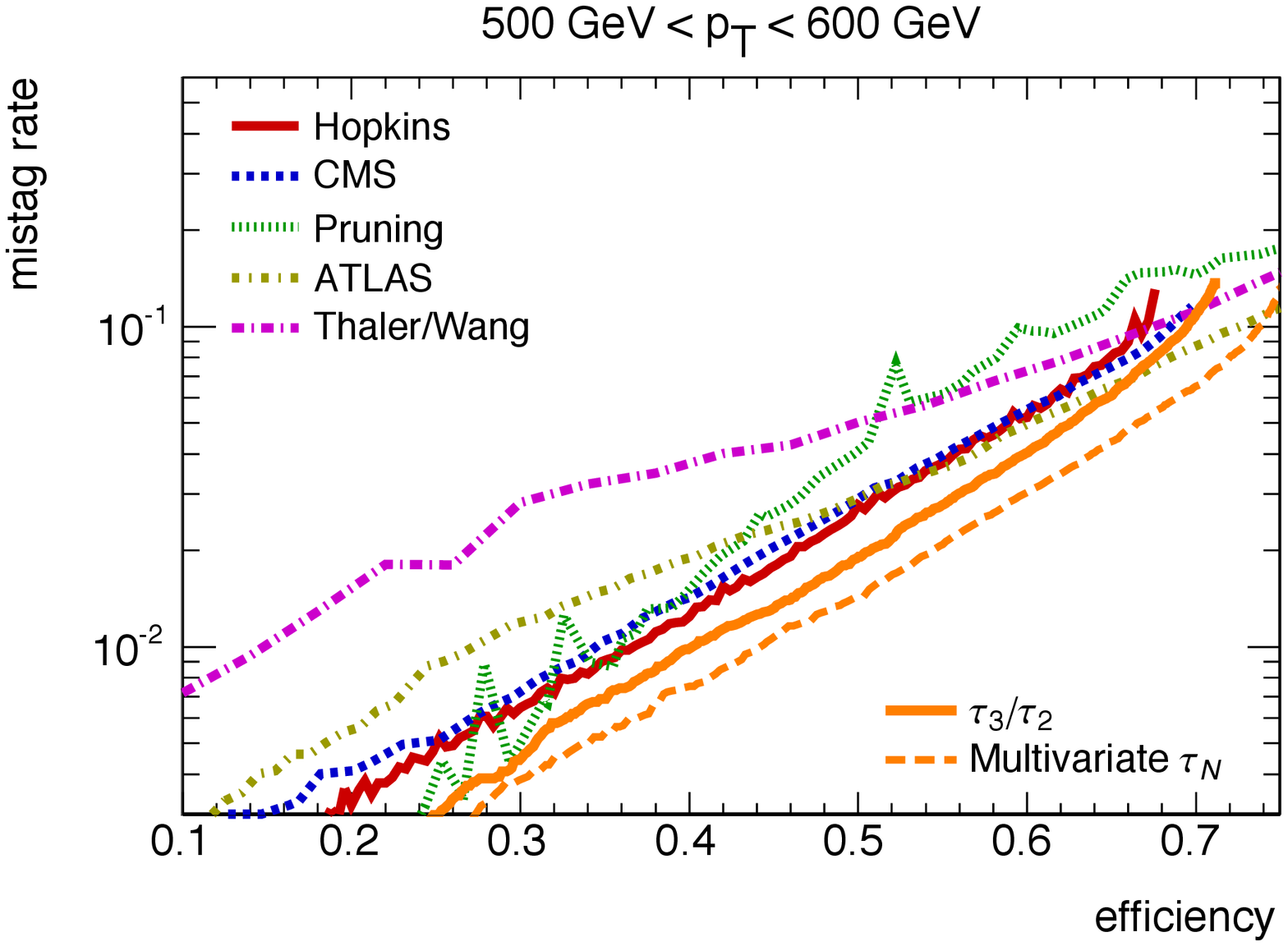}}\;\;\;
    \subfigure[]{\label{fig:master_ptall}\includegraphics[trim = 0mm 0mm 0mm 0mm, clip, width = 0.48\textwidth]{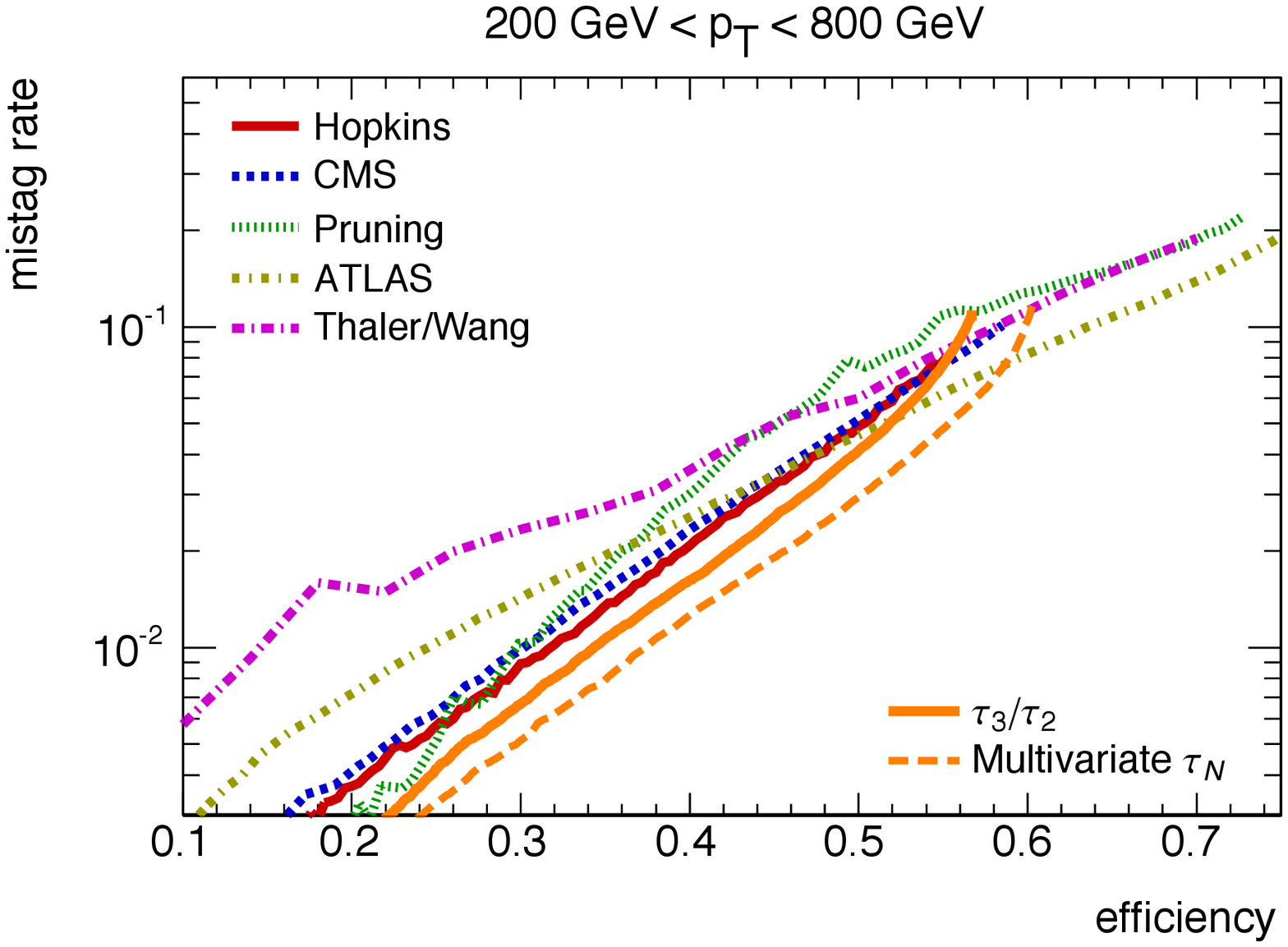}}
 \end{center}
 \vsh
  \caption{Comparison of $N$-subjettiness to other boosted top taggers using benchmark samples from the BOOST2010 report \cite{Abdesselam:2010pt}.  These  efficiency/mistag curves are taken from \Ref{Abdesselam:2010pt} and then overlayed with our results from \Fig{fig:TopSigEff} (for a one-dimensional $\tau_3/\tau_2$ cut) and  \Fig{fig:FisherEfficiency} (for a multivariate $\tau_N$ method).  Details about these curves are given in \Sec{sec:topTaggingPerformance}, and we will use a different range for the vertical axis in subsequent figures to highlight the small mistag rate region.  Except for the very high efficiency region, $N$-subjettiness outperforms previous top tagging methods.}
  \label{fig:executiveSummary}
\end{figure}

In this paper, we will show how the tagging performance of $N$-subjettiness can be improved through minimization, focusing on the case of boosted tops.  As originally defined in \Ref{Thaler:2010tr}, $N$-subjettiness required an external algorithm to determine the $N$ candidate subjet axes within a jet, as it relied on axes from the exclusive $k_T$ clustering algorithm \cite{Catani:1993hr,Ellis:1993tq} to calculate $\tau_N$.  Here, we will show how to find the subjet axes which \emph{minimize} $\tau_N$, using a variant of the so-called $k$-means clustering algorithm \cite{Lloyd82leastsquares}.  This is analogous to how the event shape thrust \cite{Farhi:1977sg} is defined, since thrust can be measured with respect to any axis, but what we call ``thrust'' is determined by the axis that minimizes thrust.

Using the minimum $\tau_N$, we will demonstrate the excellent tagging performance of $N$-subjettiness using the boosted top benchmark samples prepared for the BOOST2010 report \cite{Abdesselam:2010pt}.  Analogous to jet angularities, $N$-subjettiness can incorporate different angular weighting exponents, and we will find that the best tagging performance is achieved for the ``jet broadening'' measure \cite{Catani:1992jc}.  Different minimization procedures are needed for different angular weighting exponents, and we will see that  $k$-means clustering minimizes the thrust measure, while a new algorithm is introduced for more general angular measures such as the jet broadening measure.\footnote{After we developed our algorithm, we learned of  \Ref{Ding:2006:RPR:1143844.1143880}  on $R1$-$k$-means clustering, which implements a similar procedure for the jet broadening measure alone.}  The tagging performance of $N$-subjettiness is summarized in \Fig{fig:executiveSummary}, which demonstrates the excellent performance of both a one-dimensional cut on $\tau_3/\tau_2$ as well as a modified Fisher discriminant based on $N$-subjettiness and jet mass information.   While we focus on boosted 3-prong tops in this paper, we expect the same minimization technique to improve $\tau_2/\tau_1$ for boosted 2-prong identification as well (i.e. $W$/$Z$ or Higgs bosons).

Finally, turning to the event as a whole, we will show that the same $\tau_N$ minimization procedure can be applied to the event shape $N$-jettiness \cite{Stewart:2010tn}, allowing $N$-jettiness to act like a fixed $N$ cone jet algorithm.  We will briefly comment on how such a procedure might be useful for boosted object searches.

The remainder of this paper is organized as follows.  In \Sec{sec:n-subjettinessVariations}, we review the definition of $N$-subjettiness and describe two generalizations.  We then introduce the procedure to minimize $N$-subjettiness in \Sec{sec:minimizationProcedure}.\footnote{The minimization algorithm is available at \url{http://www.jthaler.net/jets/} as a plugin to \texttt{FastJet} \cite{FastJet,Cacciari:2005hq}.}  We study the top tagging performance of $N$-subjettiness in \Sec{sec:topTaggingPerformance}, using the BOOST2010 benchmark samples.  We briefly describe how our minimization procedure can be extended to convert $N$-jettiness into a jet algorithm in \Sec{sec:jetAlgorithm}, and conclude in \Sec{sec:conclusion}.  

\section{Generalizing N-subjettiness}
\label{sec:n-subjettinessVariations}

Boosted hadronic tops have a radiation pattern that is distinctly different from gluon- or quark-initiated jets, owing to the 3-prong nature of the top decay.  $N$-subjettiness exploits this difference in expected energy flow by ``counting'' the number of hard lobes of energy within a jet.  Here, we will generalize the original definition of $N$-subjettiness from \Ref{Thaler:2010tr} in two ways, first by including an angular weighting exponent and second by minimizing $N$-subjettiness over all possible candidate subjet axes.

Consider a fat jet reconstructed using some jet algorithm.  $N$-subjettiness is defined with respect to $N$ candidate subjet axes, that is, $N$ light-like directions $\hat{n}_J$ within a jet that are chosen to align with the dominant radiation directions.  We will use a tilde to indicate $N$-subjettiness measured with respect to generic subjet axes:
\be
 \tilde{\tau}^{(\beta)}_N = \frac{1}{d_0} \sum\limits_i p_{T,i} \min \left\lbrace (\Delta R_{1,i})^{\beta}, (\Delta R_{2,i})^{\beta}, \dots, (\Delta R_{N,i})^{\beta} \right\rbrace.
 \label{eq:tilde_tau_N}
\ee
Here, $i$ runs over the constituent particles in a given jet, $p_{T,i}$ are their transverse momenta, and $ \Delta R_{J,i} = \sqrt{(\Delta y_{J,i})^2+(\Delta \phi_{J,i})^2}$ is the distance in the rapidity-azimuth plane between a candidate subjet $J$ and a constituent particle $i$.  Compared to \Ref{Thaler:2010tr}, we have included an angular weighting exponent $\beta$, and we will often drop the $^{(\beta)}$ superscript for notational simplicity.  The normalization factor $d_0$ is taken as
\be
\label{eq:normalizationfactor}
d_0 = \sum\limits_i p_{T,i} (R_0)^{\beta},
\ee
where $R_0$ is the characteristic jet radius used in the original jet clustering algorithm.

The choice of subjet axes is crucial for defining $N$-subjettiness, since \Eq{eq:tilde_tau_N} partitions the jet constituents into $N$ so-called Voronoi regions centered on the subjet axes.  In \Ref{Thaler:2010tr}, the exclusive $k_T$ algorithm \cite{Catani:1993hr,Ellis:1993tq} was used to find the directions $\hat{n}_J$.  Here, we will focus on the axes which minimize $\tilde{\tau}_N$, removing the tilde:
\begin{equation}
 \tau^{(\beta)}_N = \min_{\hat{n}_1, \hat{n}_2, \ldots, \hat{n}_N} \tilde{\tau}^{(\beta)}_N.
 \label{eq:tau_N}
\end{equation}
In particular, $\tilde{\tau}_N$ is a function of the $N$ light-like subjet axes $\hat{n}_J$, and $\tau_N$ is the value of this function at its (global) minimum.  This minimization over candidate subjet directions is not a trivial step and may at first seems computationally daunting, but in \Sec{sec:minimizationAlgorithm} we present an efficient algorithm to perform this task.  Once the minimum  is found, then the normalization factor in \Eq{eq:normalizationfactor} ensures that $0 \le \tau_N \le 1$.

The angular weighting exponent $\beta$ is analogous to the parameter $a$ in angularities \cite{Berger:2003iw}, with the correspondence $a \equiv 2- \beta$.  Collinear safety requires $\beta \geq 0$.  In \Ref{Thaler:2010tr}, we found that $\beta = 1$ (corresponding to the jet broadening measure \cite{Catani:1992jc}) was particularly effective for boosted object identification, and this finding will be confirmed in \Sec{sec:topTaggingPerformance}.  Interestingly, the choice $\beta = 1$ is also preferred for discriminating light-quark jets from gluon jets \cite{Gallicchio:2011xq}.  As we will see in \Sec{sec:minimizationAlgorithm}, $\beta = 2$ (corresponding to the thrust measure \cite{Farhi:1977sg}) is a special value from a minimization point of view.  In addition, when we discuss jet finding in \Sec{sec:jetAlgorithm}, $\beta = 2$ will correspond most closely to iterative cone algorithms.

\begin{figure}[!t]
  \begin{center}
    \subfigure[]{\label{fig:eventDisplayTop1}\includegraphics[trim = 0mm 0mm 0mm 0mm, clip, width = 0.30\textwidth]{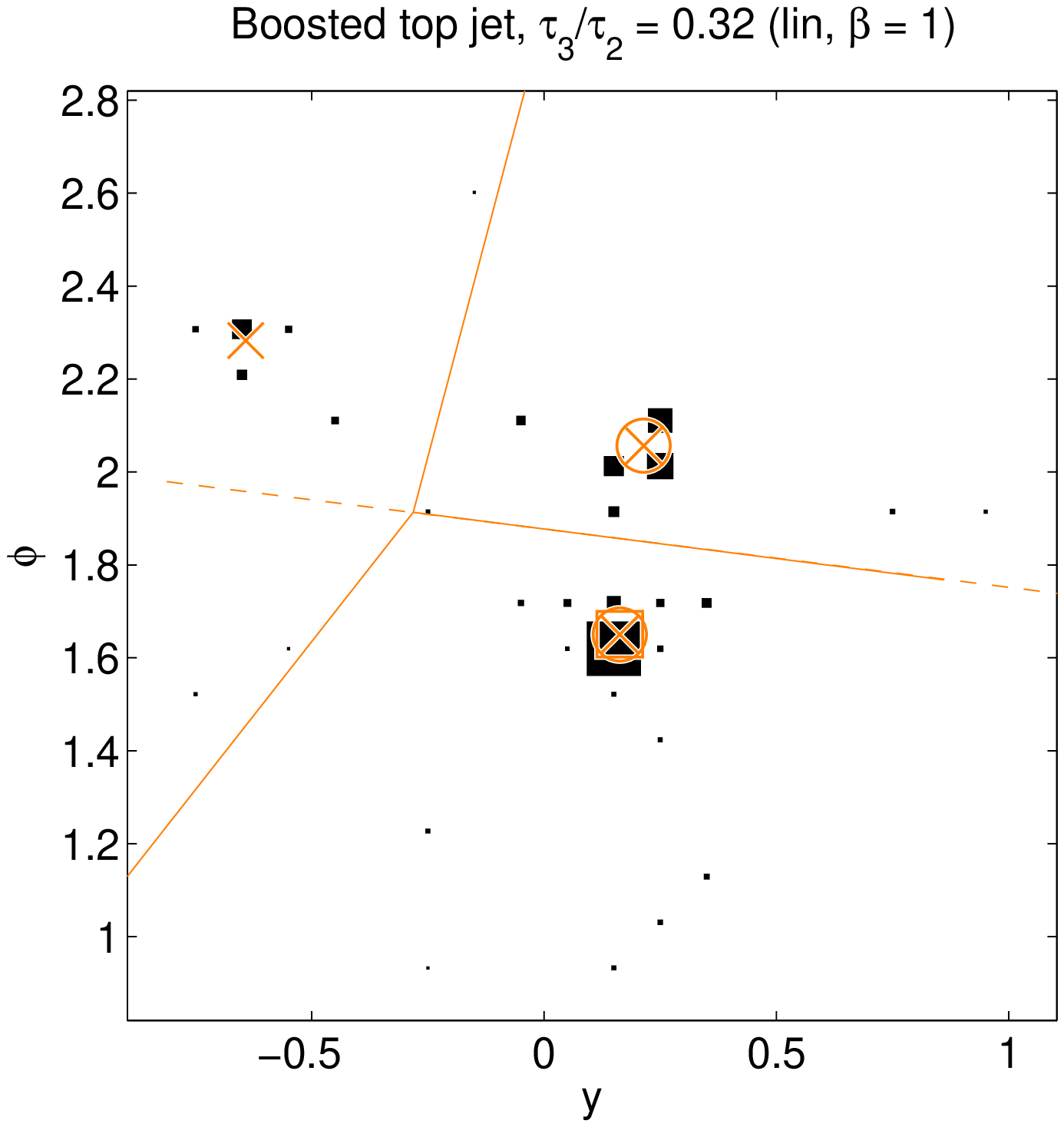}} \;\;\;
    \subfigure[]{\label{fig:eventDisplayTop2}\includegraphics[trim = 0mm 0mm 0mm 0mm, clip, width = 0.30\textwidth]{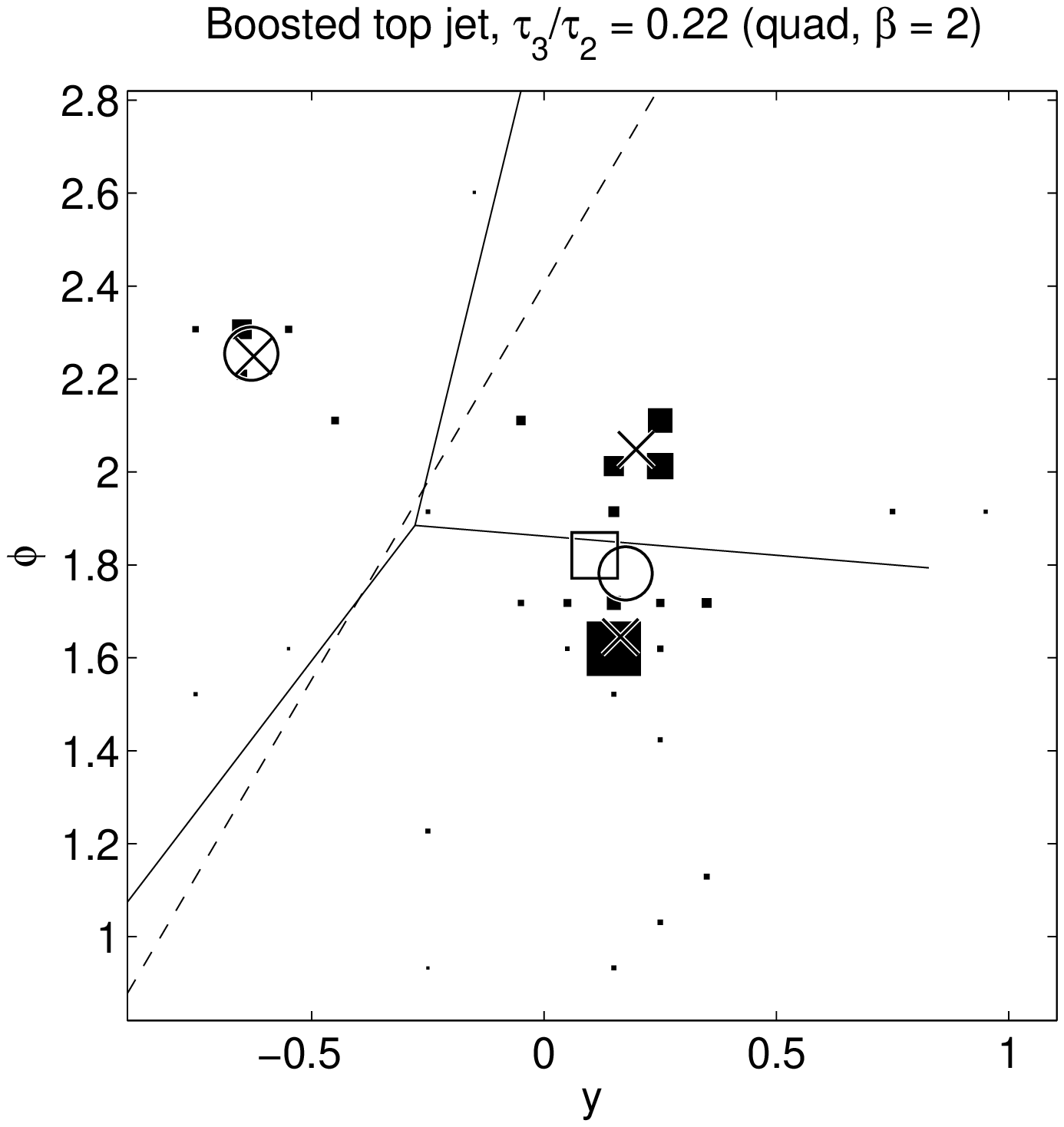}} \;\;\;
    \subfigure[]{\label{fig:eventDisplayTop3}\includegraphics[trim = 0mm 0mm 0mm 0mm, clip, width = 0.30\textwidth]{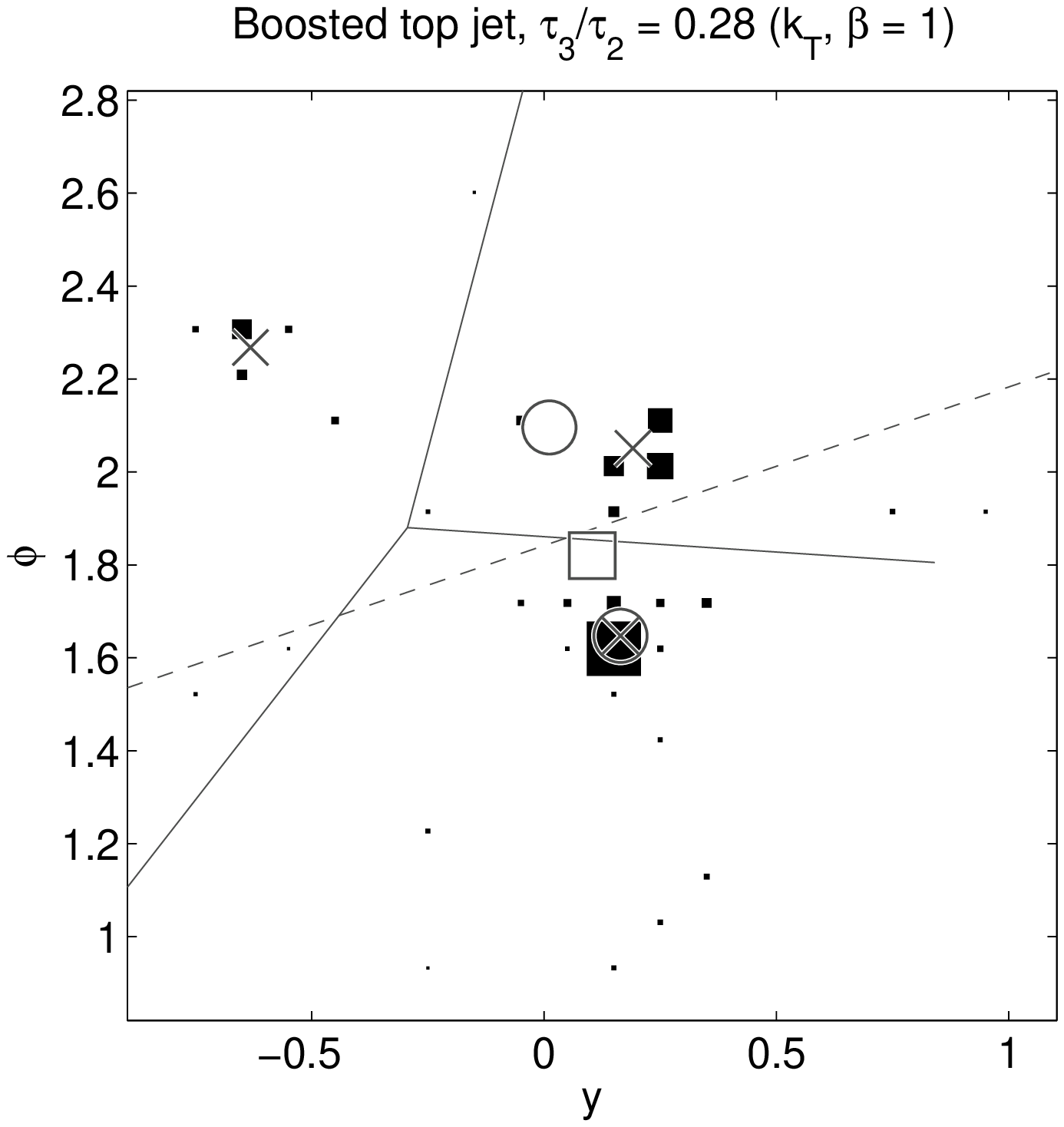}}\\
    \subfigure[]{\label{fig:eventDisplayQCD1}\includegraphics[trim = 0mm 0mm 0mm 0mm, clip, width = 0.30\textwidth]{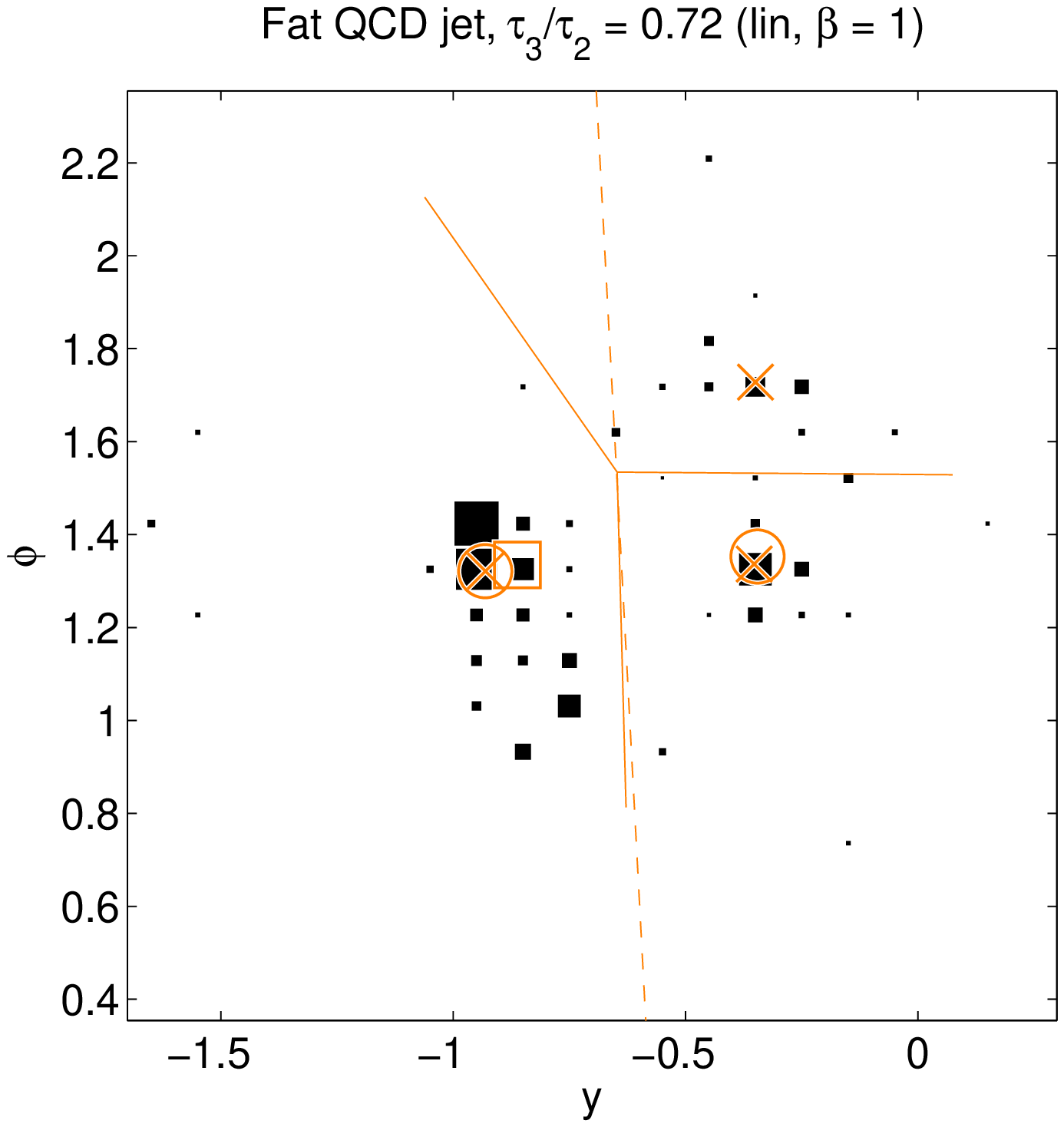}} \;\;\;
    \subfigure[]{\label{fig:eventDisplayQCD2}\includegraphics[trim = 0mm 0mm 0mm 0mm, clip, width = 0.30\textwidth]{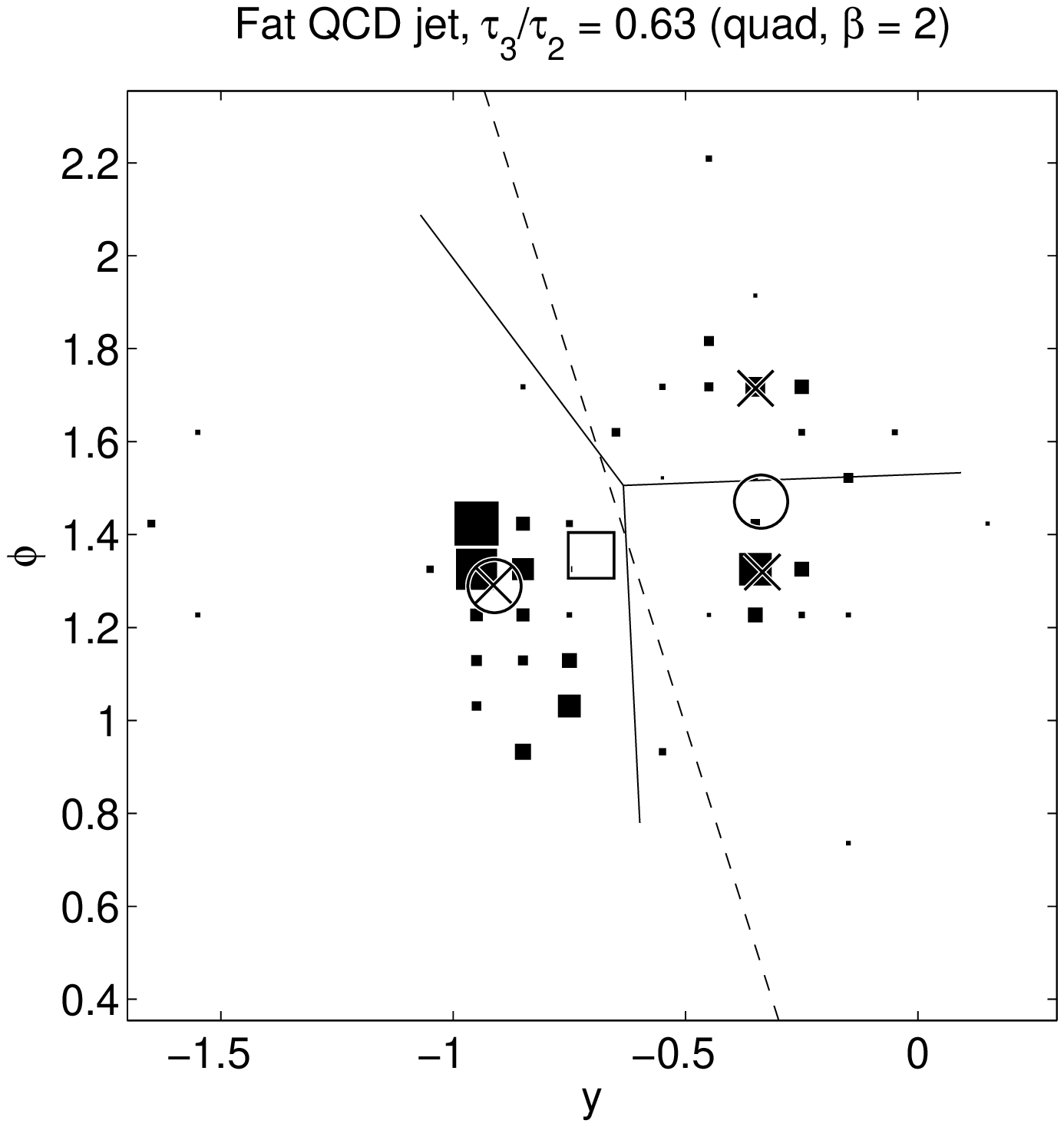}} \;\;\;
    \subfigure[]{\label{fig:eventDisplayQCD3}\includegraphics[trim = 0mm 0mm 0mm 0mm, clip, width = 0.30\textwidth]{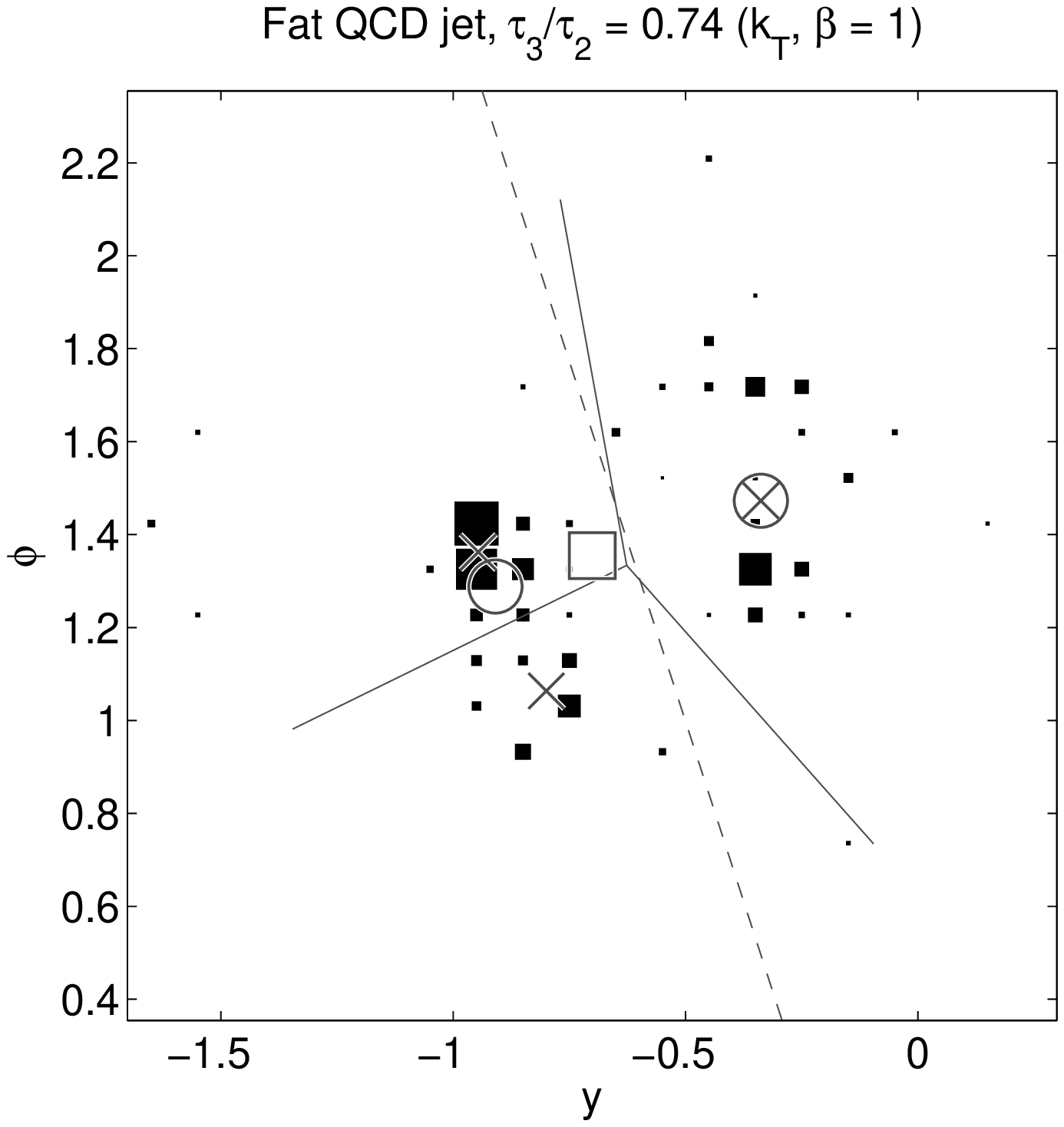}}
 \end{center}
 \vsh
  \caption{Top row: Event displays for a typical top jet with invariant mass near $m_{\rm top}$.  In (a), the orange square, circles, and crosses indicate the axes that minimize $\tilde{\tau}_1$, $\tilde{\tau}_2$, and $\tilde{\tau}_3$, respectively, for $\beta = 1$ (``linear'' minimization).  The dashed orange line indicates the edge of the two Voronoi regions for the axes minimizing $\tilde{\tau}^{(1)}_2$, and the solid orange lines indicate the Voronoi edges for the axes minimizing $\tilde{\tau}^{(1)}_3$.  In (b), we show the same top jet with equivalent information for $\beta = 2$ and the ``quadratic minimization'' in black, and in (c) for $\beta = 1$ and the axes found by the exclusive $k_T$ algorithm in gray.  In this and subsequent event displays, the particles are clustered into virtual calorimeter cells of size 0.1 by 0.1, and the marker area for each cell is proportional its scalar transverse momentum.  Bottom row: similar diagrams for a fat QCD jet with mass near $m_{\rm top}$.}
  \label{fig:eventDisplays}
\end{figure}

In \Fig{fig:eventDisplays}, we demonstrate how $N$-subjettiness works on a boosted top jet compared to a QCD jet with mass near $m_{\rm top}$.  Shown are the subjet axes and Voronoi regions determined by the minimum $\tau_N$ with $\beta = 1$ and $\beta = 2$, as well as $\tilde{\tau}_N$ using subjets from the exclusive $k_T$ algorithm.  Note that the partitioning depends crucially on the choice of subjet axes.  Also, unlike recursive clustering procedures like the $k_T$  \cite{Catani:1993hr,Ellis:1993tq} or Cambridge-Aachen \cite{Dokshitzer:1997in,Wobisch:1998wt} methods, the regions determined by minimizing $\tilde{\tau}_N$ are not directly correlated with the regions determined by minimizing $\tilde{\tau}_{N-1}$.\footnote{In what follows, we will only compare the minimization axes to the axes found by the exclusive $k_T$ algorithm.   Even though the Cambridge-Aachen algorithm also has an exclusive version which returns a fixed number of subjets, the nature of its clustering procedure allows far-away soft radiation to be clustered into the jet last, yielding anomalously large values for $\tilde{\tau}_N$.}  The axes that minimize $\beta =1$ tend to point in the direction of actual jet radiation (like a ``median''), while the axes that minimize $\beta =2$ tend to point in direction determined by the  average subjet energy (like a ``mean'').

It is straightforward to see why $\tau_N$ quantifies how $N$-subjetty a particular jet is, or in other words, to what degree it can be regarded as a jet composed of $N$ subjets.  Jets with $\tau_N \simeq 0$ have all their radiation aligned with the subjet directions $\hat{n}_J$ and therefore have $N$ (or fewer) subjets.  Jets with $\tau_N \gg 0$ have a large fraction of their energy distributed away from the  subjet directions $\hat{n}_J$ and therefore have at least $N+1$ subjets.  Therefore, jets that are very ``$N$-subjetty" should have a relatively large difference in their $\tau_N$ and $\tau_{N-1}$ values.  In practice, the purely geometrical, dimensionless ratio $\tau_N/\tau_{N-1}$ is the best (simple) discriminant for $N$-prong hadronic decays, a point we will further elaborate on in \Sec{sec:results}.

\section{Minimization Procedure}
\label{sec:minimizationProcedure}

A key ingredient in the definition of $N$-subjettiness is an appropriate choice of candidate subjet directions $\hat{n}_J$.  Ideally, one would determine $\tau_N$ by minimizing over all possible candidate subjet directions, analogously to how the event shape thrust is defined \cite{Farhi:1977sg}.  In that case, $\tau_N$ is a strictly decreasing function of $N$ with $0 < \tau_{N}/\tau_{N-1} < 1$, since adding an additional subjet axis can always decrease the Voronoi distances.  

In \Ref{Thaler:2010tr}, it was (erroneously) believed that a search for the global minimum of $\tilde{\tau}_N$ would be too computationally intensive, which is why candidate subjet directions were determined using the exclusive $k_T$ algorithm.  While this was found to work reasonably well for boosted object tagging, it introduced residual algorithmic dependence and a certain sense of arbitrariness in the jet shape.  Here, we present a fast minimization procedure to determine the candidate subjet directions which minimize $\tilde{\tau}_N$, using a generalization of $k$-means clustering.  

\subsection{Minimization Algorithm}
\label{sec:minimizationAlgorithm}

Minimizing the function $\tilde{\tau}^{(\beta)}_N$ in \Eq{eq:tilde_tau_N} is similar to the classic computer science problem of finding $k$ clusters in a data set.\footnote{The ``$k$'' is standard notation in the computer science literature, while ``$N$'' is standard notation for jet counting in particle physics.   We will use ``$k$'' when referring to the $k$-means algorithm, and ``$N$'' to the jet shape $N$-subjettiness, but of course $k=N$ throughout.}   For $\beta = 2$, this is the $k$-means clustering problem, which is to find the $k$ cluster centers (or ``means'') that minimize the in-cluster variance (i.e.\ the weighted sum of the distances \emph{squared} between data points and their nearest cluster center).
One solution to this problem is Lloyd's algorithm \cite{Lloyd82leastsquares}, which terminates in polynomial time and produces $k$ means which form a (local) minimum of the cluster variance.  Combined with ``sufficiently'' many reseedings of the initial $k$ cluster centers, Lloyd's algorithm can find the global minimum of the cluster variance.  Below we generalize Lloyd's algorithm beyond $\beta = 2$, to an algorithm capable of minimizing $\tilde{\tau}_N$ for $1 \le \beta < 3$.

Let us motivate an adaptation of Lloyd's algorithm, which aims to minimize $N$-subjettiness also for $\beta \not=2$.  For simplicity, suppose for a moment that we want to minimize 1-subjettiness for a particular cluster $C$ by adjusting a single subjet axis $(y_0,\phi_0)$:  
\be
\label{eq:1subjettiness}
\tilde{\tau}^{(\beta)}_1(y_0,\phi_0) = \frac{1}{d_0} \sum_{i \in C} p_{T,i} \left[(y_0-y_i)^2 + (\phi_0 - \phi_i)^2\right]^{\beta/2}.
\ee
Taking first-order partial derivatives of $\tilde{\tau}_1$ and setting them to zero gives:
\begin{align}
 \frac{\partial \tilde{\tau}^{(\beta)}_1}{\partial y_0} & = \frac{1}{d_0} \frac{\beta}{2} \sum_{i \in C} p_{T,i} (y_i - y_0)\left[(y_i - y_0)^2 + (\phi_i - \phi_0)^2\right]^{\frac{\beta-2}{2}} = 0, \nonumber \\
 \frac{\partial \tilde{\tau}^{(\beta)}_1}{\partial \phi_0} &= \frac{1}{d_0}\frac{\beta}{2} \sum_{i \in C} p_{T,i} (\phi_i - \phi_0)\left[(y_i - y_0)^2 + (\phi_i - \phi_0)^2\right]^{\frac{\beta-2}{2}} = 0 \label{eq:minaxesequations}.
\end{align}
Any pair $(y_0,\phi_0)$ which solves these two equations for a given distribution of particles corresponds to a local minimum of $\tilde{\tau}_1$.  For $\beta = 2$, the equations are easily solved by finding the weighted centroid of the cluster
\be
\beta = 2:  \qquad (y_0,\phi_0) = \left(\frac{\sum_i p_{T,i} y_i}{\sum_j p_{T,j}}, \frac{\sum_i p_{T,i} \phi_i}{\sum_j p_{T,j}}\right),
\label{eq:minfinderbeta2}
\ee
and this observation forms the basis of Lloyd's algorithm.  Interestingly, for recursive clustering algorithms, the (sub)jet axis is also aligned with the weighted centroid,\footnote{Strictly speaking, this is only true for recursive clustering algorithms that use the $p_T$ scheme \cite{Butterworth:2002xg} for defining the jet axis.  Modern recursive clustering algorithms use the $E$ scheme, which maintains information about the mass of a jet axis, so the rapidity distance is modified compared to using a light-like axis in the $p_T$ scheme.} so there is a relatively small difference between jet axes found with $k_T$-like algorithms and jet axes found by minimizing 1-(sub)jettiness with $\beta = 2$.  (See also \Sec{sec:previousliterature} for a discussion of iterative cone algorithms.)

For general $\beta$, \Eq{eq:minaxesequations} does not have a closed form solution.  However, there is a fast iterative algorithm to find a local minimum $(y_0,\phi_0)$ to arbitrary precision.  Suppose we already have a ``guess'' or initial seeding of the candidate subjet direction; call it $(y^{(0)}_0,\phi^{(0)}_0$).  We can then define a recursive procedure $(y^{(n)}_0,\phi^{(n)}_0) \rightarrow (y_0^{(n+1)},\phi_0^{(n+1)})$ as
\be
\label{eq:updateStep}
1 \le \beta < 3:  \qquad  y^{(n+1)}_0 = \frac{\sum\limits_{i \in C} p_{T,i} y_i \left[\left(y_i - y^{(n)}_0\right)^2 + \left(\phi_i - \phi^{(n)}_0\right)^2\right]^{\frac{\beta-2}{2}} }{\sum\limits_{j \in C} p_{T,j} \left[\left(y_j - y^{(n)}_0\right)^2 + \left(\phi_j - \phi^{(n)}_0\right)^2\right]^{\frac{\beta-2}{2}} },
\ee
and similarly for $\phi^{(n+1)}_0$.  It is straightforward to see that if $(y_0^{(n+1)},\phi_0^{(n+1)}) = (y_0^{(n)},\phi_0^{(n)})$, then we have found a local minimum.  Furthermore, we argue in \Sec{sec:uniqueness} that any cluster of particles has only \emph{one} local minimum of $\tilde{\tau}_1$ for $\beta \ge 1$, which is thus the global minimum.\footnote{For $\beta = 1$ and certain fine-tuned particle configurations, it is possible to have a degenerate line of local minima.  However, as $\tilde{\tau}^{(1)}_1$ is constant on this line, convergence of the recursive algorithm means that $\tau^{(1)}_1$ itself (the value of the global minimum) is still found.}  

\begin{figure}[!p]
  \begin{center}
\includegraphics[trim = 0mm 0mm 0mm 0mm, clip,width = 0.8\textwidth]{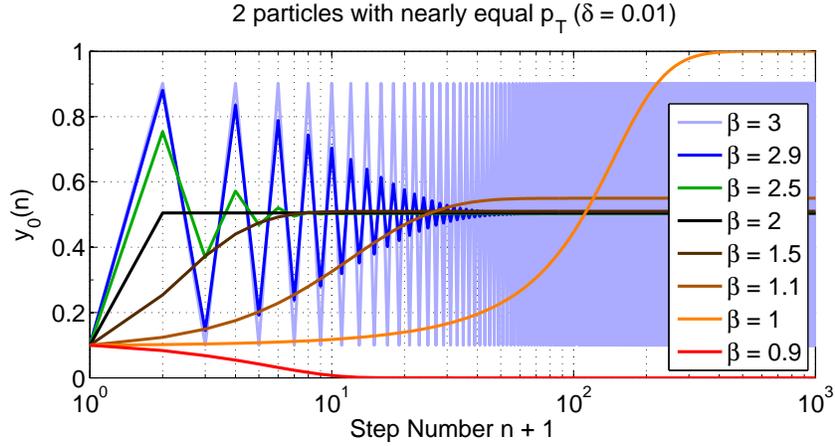}
  \end{center}
  \vsh
  \caption{Convergence of the minimization algorithm for notable values of $\beta$ on a one-dimensional two-particle configuration.  One particle is located at $y = 0$ with $p_T = e^{- \delta}$, the other at $y=1$ with $p_T = e^{+ \delta}$, and the global minimum of  $\tilde{\tau}_1^{(\beta)}$ is located at $y_0 = \frac{1}{2}(1 + \tanh \frac{\delta}{\beta - 1})$.  The algorithm is initialized at $y_0^{(0)} = 0.1$, which is closer to the softer particle.  Convergence to the global minimum of $\tilde{\tau}_1^{(\beta)}$ is reached for $1 \le \beta < 3$.  The algorithm can converge to a non-global minimum for $\beta <  1$ if the initial axis is chosen too close to the softer particle (here shown by $\beta = 0.9$), and the algorithm diverges for $\beta \ge 3$ (here shown for the critical case $\beta = 3$).  For $\beta = 2$, the algorithm finds the global minimum in one step, as expected from Lloyd's algorithm.}  
  \label{fig:convergencePath}
\end{figure}

\begin{figure}[!p]
  \begin{center}
    \subfigure[]{\label{fig:eventDisplayPath1}\includegraphics[trim = 0mm 0mm 0mm 0mm, clip, width = 0.30\textwidth]{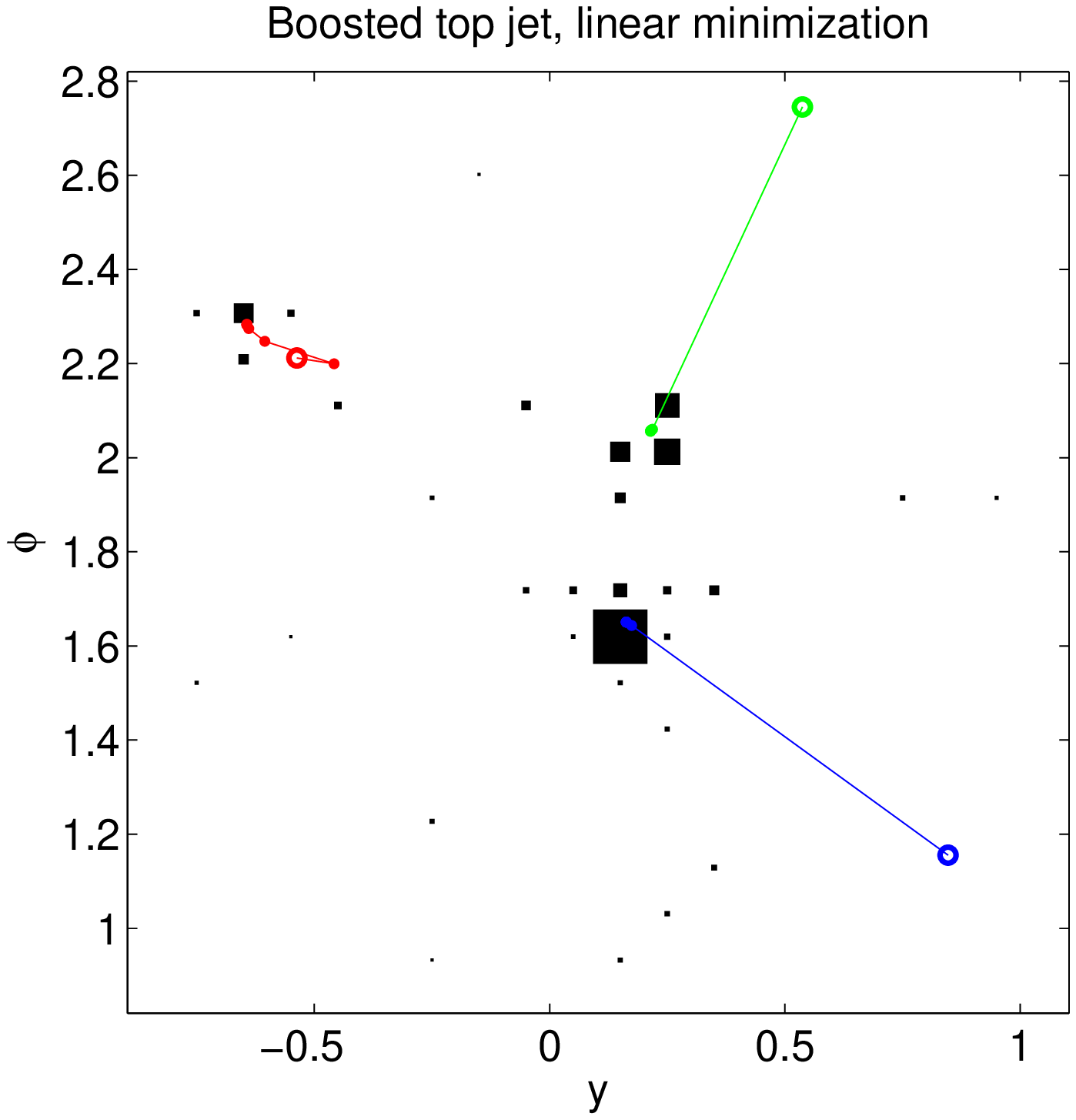}}\;\;\;
    \subfigure[]{\label{fig:eventDisplayPath2}\includegraphics[trim = 0mm 0mm 0mm 0mm, clip, width = 0.30\textwidth]{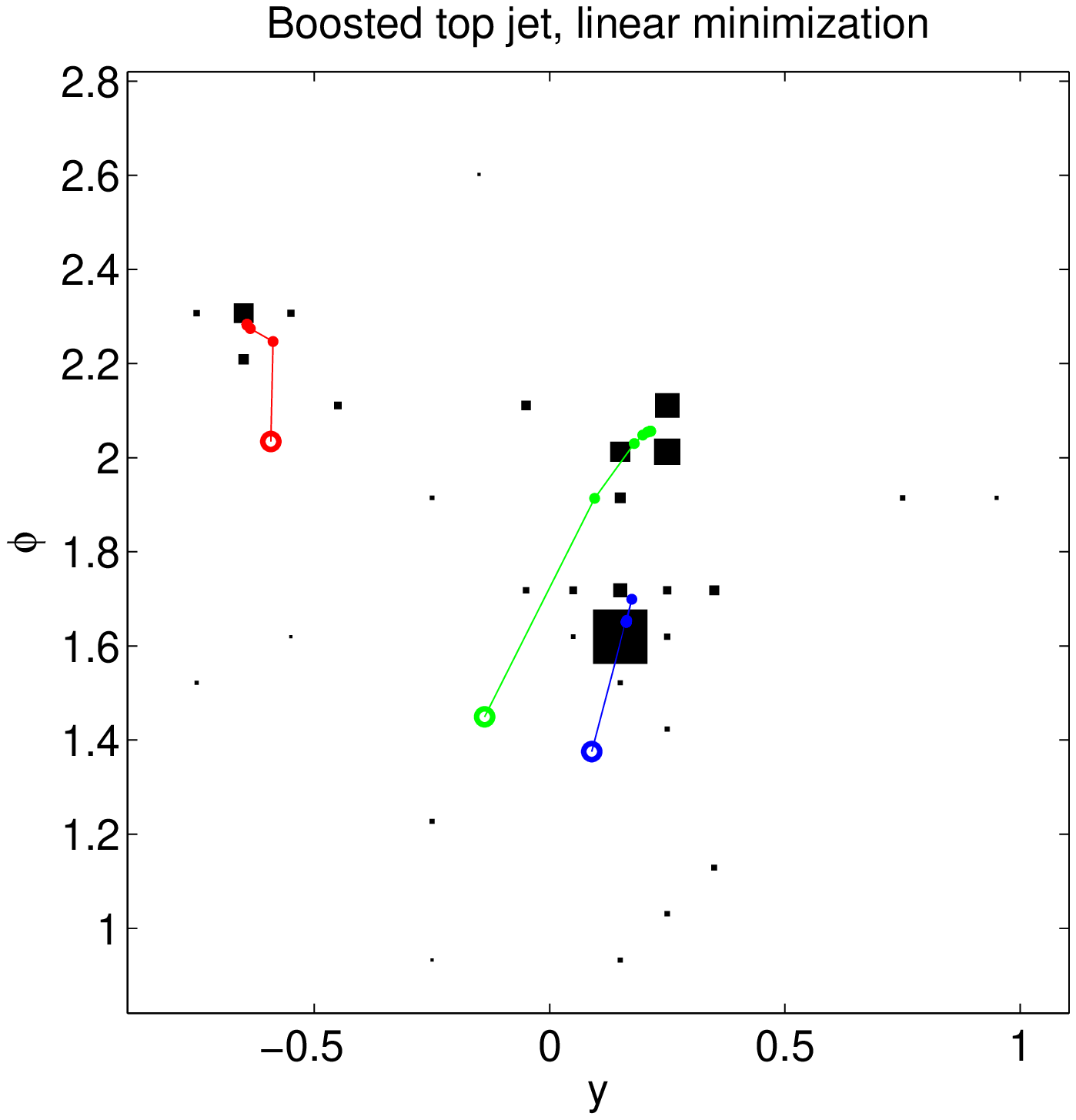}}\;\;\;
    \subfigure[]{\label{fig:eventDisplayPath3}\includegraphics[trim = 0mm 0mm 0mm 0mm, clip, width = 0.30\textwidth]{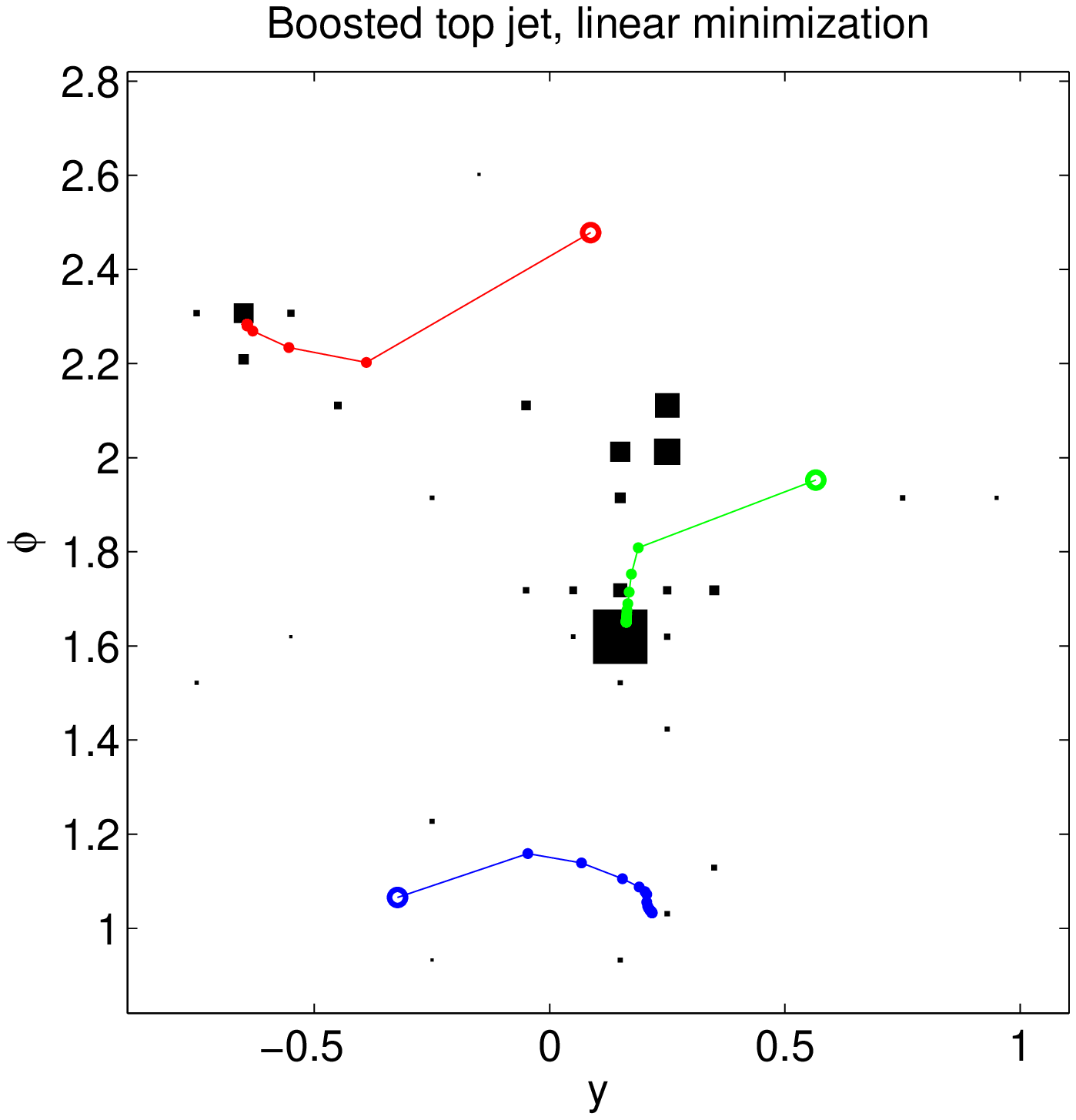}}
    \end{center}
    \vsh
  \caption{Convergence path of the minimization algorithm for $N=3$ and $\beta = 1$.  Shown is the same top jet as in \Fig{fig:eventDisplays}.  Panels (a), (b) and (c) show three different initial seedings for our modified $k$-means clustering procedure.  The open circle is the seed position, the dots are the updated positions, and a line connecting them is drawn to guide the eye.  The first two seeds find the correct global minimum in a small number of steps, while the third seed gets trapped at a local minimum.}
  \label{fig:eventDisplayPath}
\end{figure}

The sequence $(y_0^{(n)},\phi_0^{(n)})$ does not generally yield an exact solution to \Eq{eq:minaxesequations} in finite time, but for $1 \le \beta < 3$ it will quickly asymptote to the desired value.  This is demonstrated for a two-particle configuration in \Fig{fig:convergencePath}, where convergence is shown for $1 \le \beta < 3$.  For $\beta \ge 3$, the recursive procedure in \Eq{eq:updateStep} tends to diverge, though this behavior can be remedied if a dampening factor is included in the recursion.\footnote{That is, instead of using $y^{(n+1)}_0$ directly, one uses a modified $\tilde{y}^{(n+1)}_0 = d \, y^{(n)}_0 + (1-d) \, y^{(n+1)}_0$, with a dampening factor $0 \le d < 1$ (the undamped case is $d = 0$).  With $1/2 < d < 1$, the minimization algorithm does converge for all $\beta \ge 1$.}  For $\beta < 1$, $\tau^{(\beta)}_1$ has many local minima, and finding the global minimum becomes computationally impractical.  In the range $1 \le \beta < 3$, we can efficiently find the global minimum of $\tilde{\tau}_1$ to arbitrary precision.

The generalization from 1-subjettiness to $N$-subjettiness is then straightforward.  From \Eq{eq:tilde_tau_N}, we see that $\tilde{\tau}_N$ partitions the jet constituents into $N$ Voronoi subclusters and sums the $\tilde{\tau}_1$ values for each of the subclusters.  Since the procedure in \Eq{eq:updateStep} asymptotes to the (global) minimum of $\tilde{\tau}_1$ for any given subcluster, we can iteratively determine a (local) minimum of $\tilde{\tau}_N$ by repeatedly applying \Eq{eq:updateStep} for each subcluster and then recalculating the Voronoi regions.  As is the case for Lloyd's algorithm, this procedure is not guaranteed to find the global minimum for $\tilde{\tau}_N$ (except for $N=1$ and $1 \le \beta < 3$), though as we will discuss in \Sec{sec:infraredSafetyAndConvergence}, we have found that with sufficiently many starting seeds, the global minimum is obtained within the desired precision.

Our algorithm for minimizing $N$-subjettiness works essentially the same as Lloyd's algorithm with a modified assignment step and stopping criterion:
\begin{itemize}
 \item \textbf{Initialization Step}: Pick initial seed axes $\hat{n}^{(0)}_J \equiv (y^{(0)}_{0,J},\phi^{(0)}_{0,J})$ for ${J \in \lbrace 1,\dots,N \rbrace}$, and set the iteration number $n=0$.  We will discuss the choice of seed axes in more detail in \Sec{sec:infraredSafetyAndConvergence}.
 \item \textbf{Assignment Step}: Divide the jet into clusters $C_J$ by assigning the particles to the closest subjet direction.  In other words, $i \in C_J$ if and only if $\Delta R(p_i, \hat{n}^{(n)}_J) < \Delta R (p_i, \hat{n}^{(n)}_M)$ for all $M \not=J$.
 \item \textbf{Update Step}: Update each cluster axis according to \Eq{eq:updateStep}, yielding a new set of $\hat{n}^{(n+1)}_J$ axes.
 \item \textbf{Iteration}: Repeat the Assignment and Update Steps until the average directional change of the subjets
 	\begin{equation} 
	\overline{\Delta}^{(n+1)} \equiv \frac{1}{N} \sum_{J = 1}^N \Delta R (\hat{n}^{(n)}_J, \hat{n}^{(n+1)}_J)
	\end{equation}
is smaller than the desired precision threshold ($\overline{\Delta} < 10^{-4}$ in this paper).
\end{itemize}
In the computer science literature, a similar algorithm called $R1$-$k$-means was proposed in \Ref{Ding:2006:RPR:1143844.1143880} for $\beta = 1$.  The $R1$-$k$-means algorithm should not be confused with the $k$-medians algorithm, as $k$-medians is not rotationally symmetric. 

\subsection{Infrared Safety and Seed Choices}
\label{sec:infraredSafetyAndConvergence}

Given candidate subjet axes (not necessarily the minimum axes), $\tilde{\tau}_N$ is an infrared- and collinear-safe observable. Since \Eq{eq:tilde_tau_N} is linear in each of the constituent particle's transverse momenta, the addition of infinitesimally soft particles does not change $N$-subjettiness (infrared safety).  This linear $p_T$ dependence combined with smooth angular dependence ($\beta \ge 0$) ensures that the same $\tilde{\tau}_N$ value is obtained for collinear splittings (collinear safety).

Crucially, the candidate subjet axes used in $N$-subjettiness must be determined via a method that is also infrared- and collinear-safe.  Certainly, the subjet axes which determine the global minimum of $\tilde{\tau}_N$ are infrared safe.  But even if the algorithm in \Sec{sec:minimizationAlgorithm} can only find a local minimum, the minimization procedure is still infrared safe as long as the method to determine the seed axes is infrared safe.\footnote{In particular, the addition of infinitesimally soft particles does not affect the minimization procedure as they cannot create extra local minima.  Also, since the number of cluster regions is fixed at $N$, soft radiation certainly cannot change the number of subjets.}

Of course, this still leaves an ambiguity as to the exact method for choosing the initial seed subjet axes.
Note that randomly choosing initial subjet axes is a non-deterministic procedure and therefore gives an ambiguous definition of $N$-subjettiness, since there is a chance the algorithm will converge to a non-global minimum.  A deterministic (and infrared- and collinear-safe) option would be to use the output of a recursive subjet clustering algorithm (such as exclusive $k_T$) as the seed axes, and only do one pass at minimization, though there is still no guarantee of converging to the global minimum.  However, given the speed of the minimization algorithm in \Sec{sec:minimizationAlgorithm} and the fact that the definition of \Eq{eq:tilde_tau_N} is such that a jet typically has relatively few local minima, one can almost always find the global minimum of $\tilde{\tau}_N$ by brute force reinitialization with random seed axes.  

Throughout the paper, we use random initialization (repeated 100 times) and keep the axes that yield the smallest $\tau_N$.  More precisely, we first recluster the jet with the exclusive $k_T$ algorithm into exactly $N$ candidate subjets and add random noise, uniformly distributed in a $0.8 \times 0.8$ square, to the rapidity-azimuth coordinates of these axes.  We then use these shifted coordinates as 100 different sets of seed axes, and the outcome of the minimization algorithm which yields the lowest $\tilde{\tau}_N$ is identified as the global minimum.  In \Fig{fig:eventDisplayPath}, we show three typical minimization paths for $\beta = 1$ and different initial seeds.  Even when the seed axes are quite far from the (local) minimum, only a small number of iterations are typically needed to achieve $\overline{\Delta} < 10^{-1}$, and the majority of initializations converge to the global minimum.\footnote{We note that our minimization algorithm becomes less sensitive to the initial seeding for higher values of $\beta$.  In that respect, $\beta = 1$, which we use extensively throughout this paper, is not optimal though still manageable for our purposes, as only 100 seeds are sufficient to obtain $\tilde{\tau}_N$ values within $\ll 1\%$ of the values found after thousands of seeds.  In other applications, one may find it useful to use $\beta \simeq 1.1$, which in many respects (including tagging performance) has similar behavior to $\beta = 1$ but decreased seed sensitivity and faster convergence (see \Fig{fig:convergencePath}).  That said, the positions of the minimum axes can be very different for $\beta \simeq 1.1$ compared to $\beta = 1$.}

\begin{figure}[tp]
  \begin{center}
    \subfigure[]{\label{fig:t1CompBeta1}\includegraphics[trim = 0mm 0mm 0mm 0mm, clip, width = 0.31\textwidth]{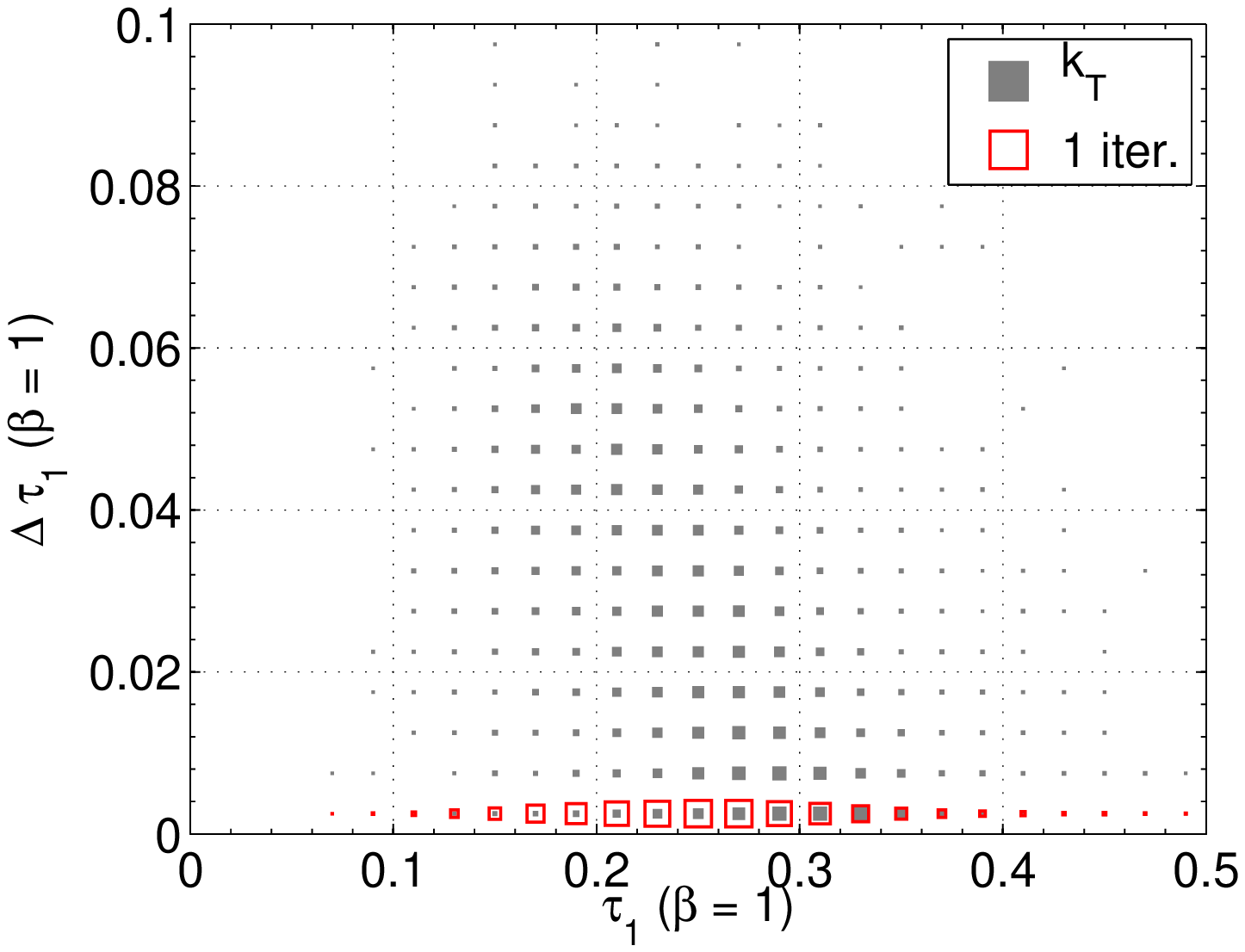}}\;\;
    \subfigure[]{\label{fig:t2CompBeta1}\includegraphics[trim = 0mm 0mm 0mm 0mm, clip, width = 0.31\textwidth]{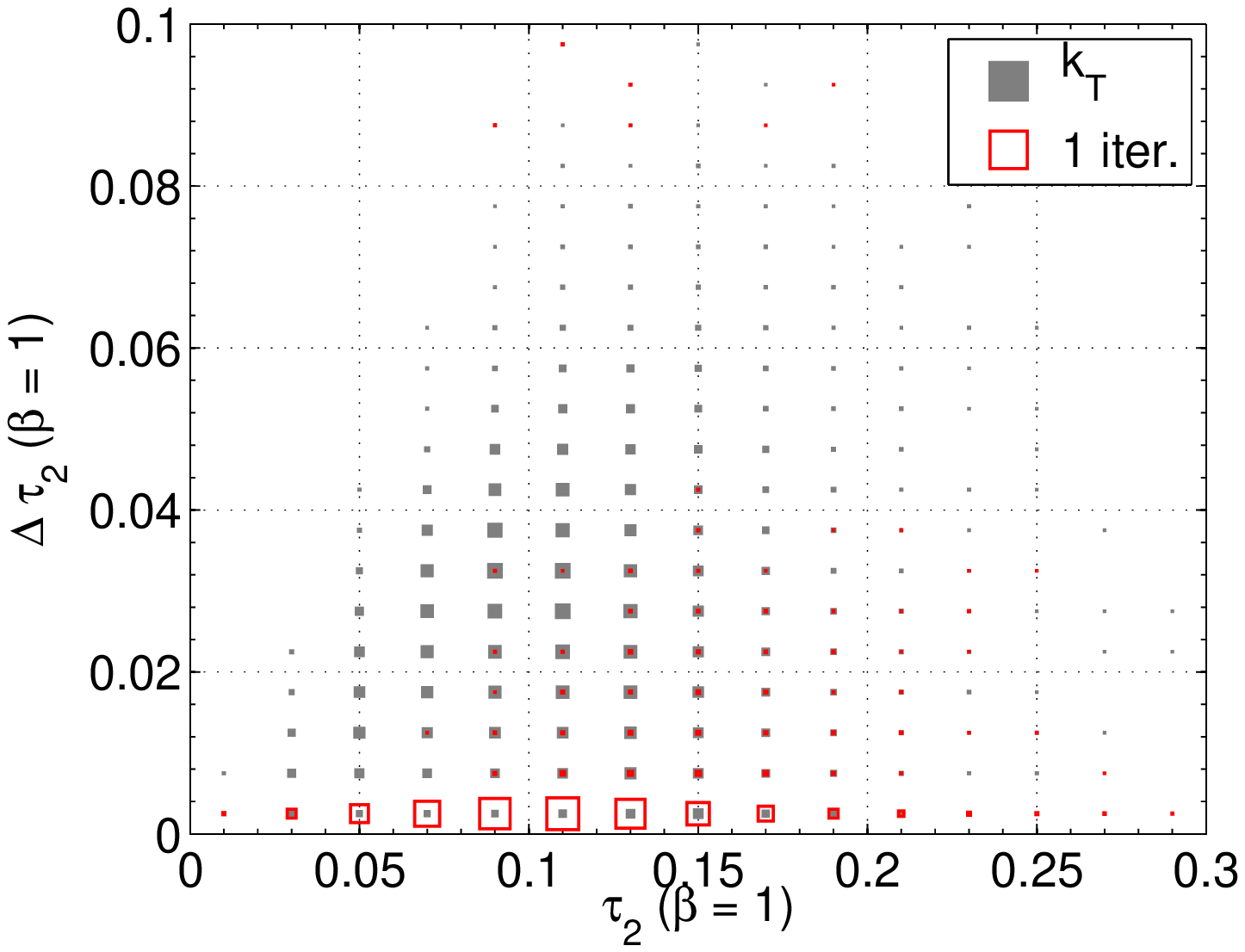}}\;\;
    \subfigure[]{\label{fig:t3CompBeta1}\includegraphics[trim = 0mm 0mm 0mm 0mm, clip, width = 0.31\textwidth]{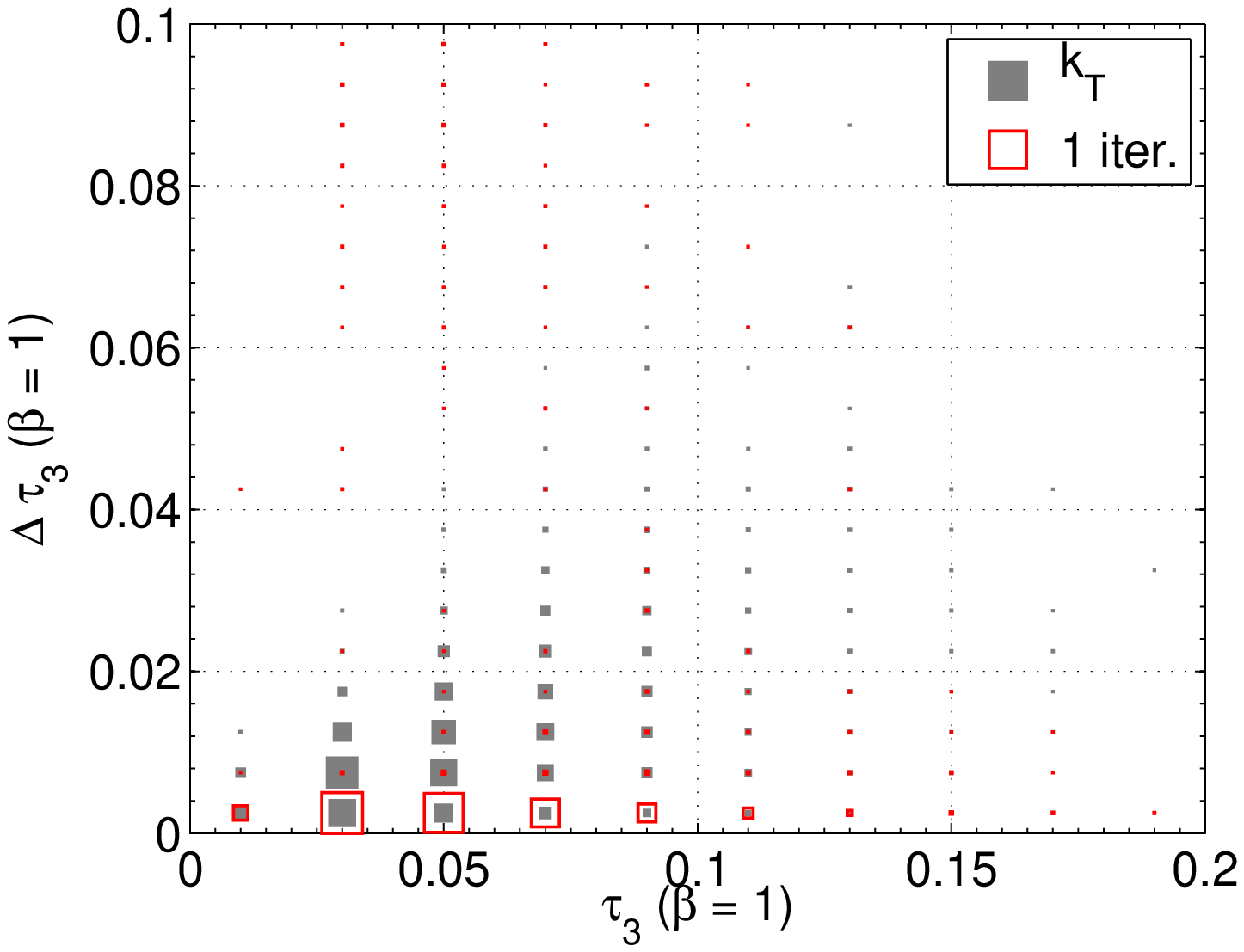}}\\
    \subfigure[]{\label{fig:t1CompBeta2}\includegraphics[trim = 0mm 0mm 0mm 0mm, clip, width = 0.31\textwidth]{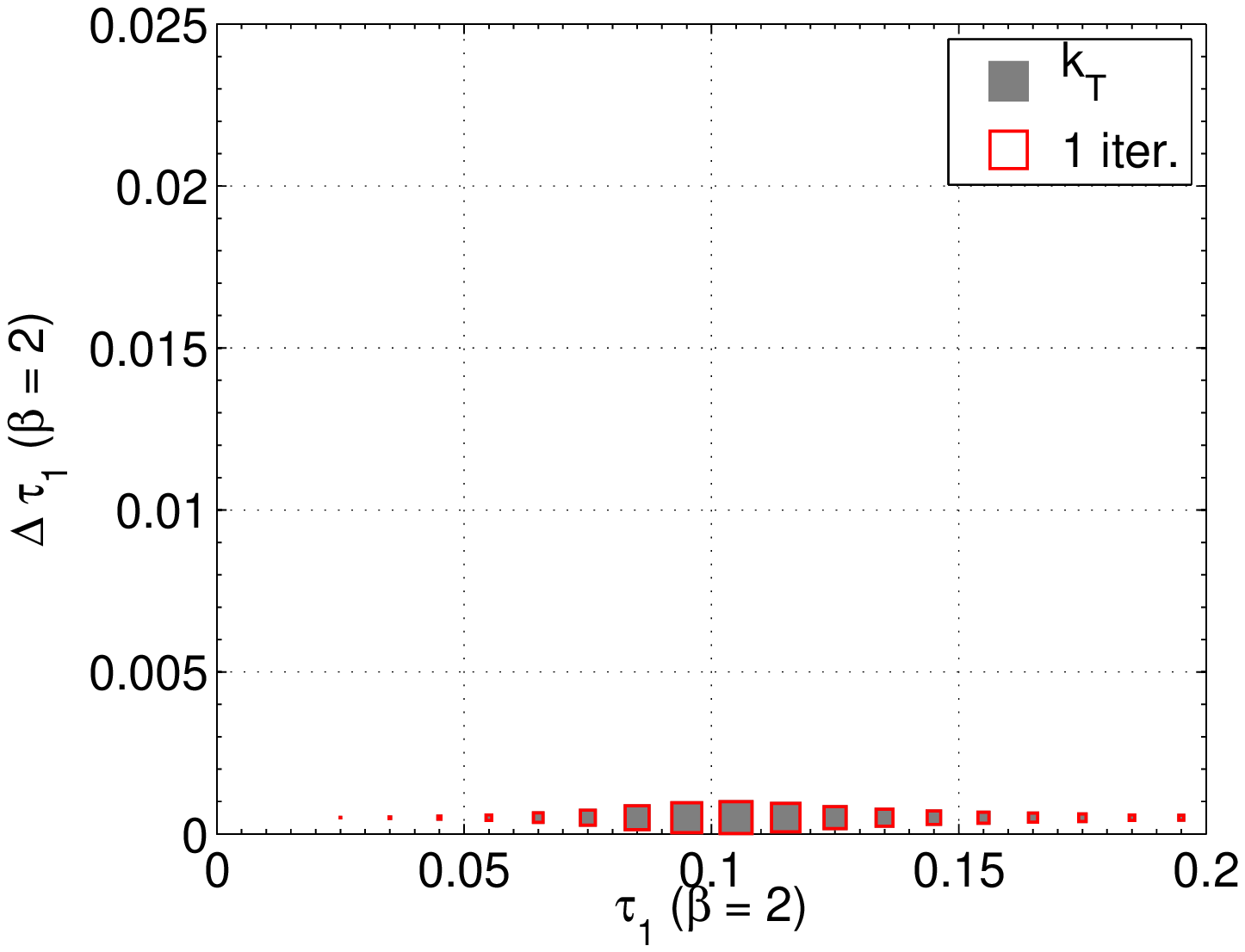}}\;\;
    \subfigure[]{\label{fig:t2CompBeta2}\includegraphics[trim = 0mm 0mm 0mm 0mm, clip, width = 0.31\textwidth]{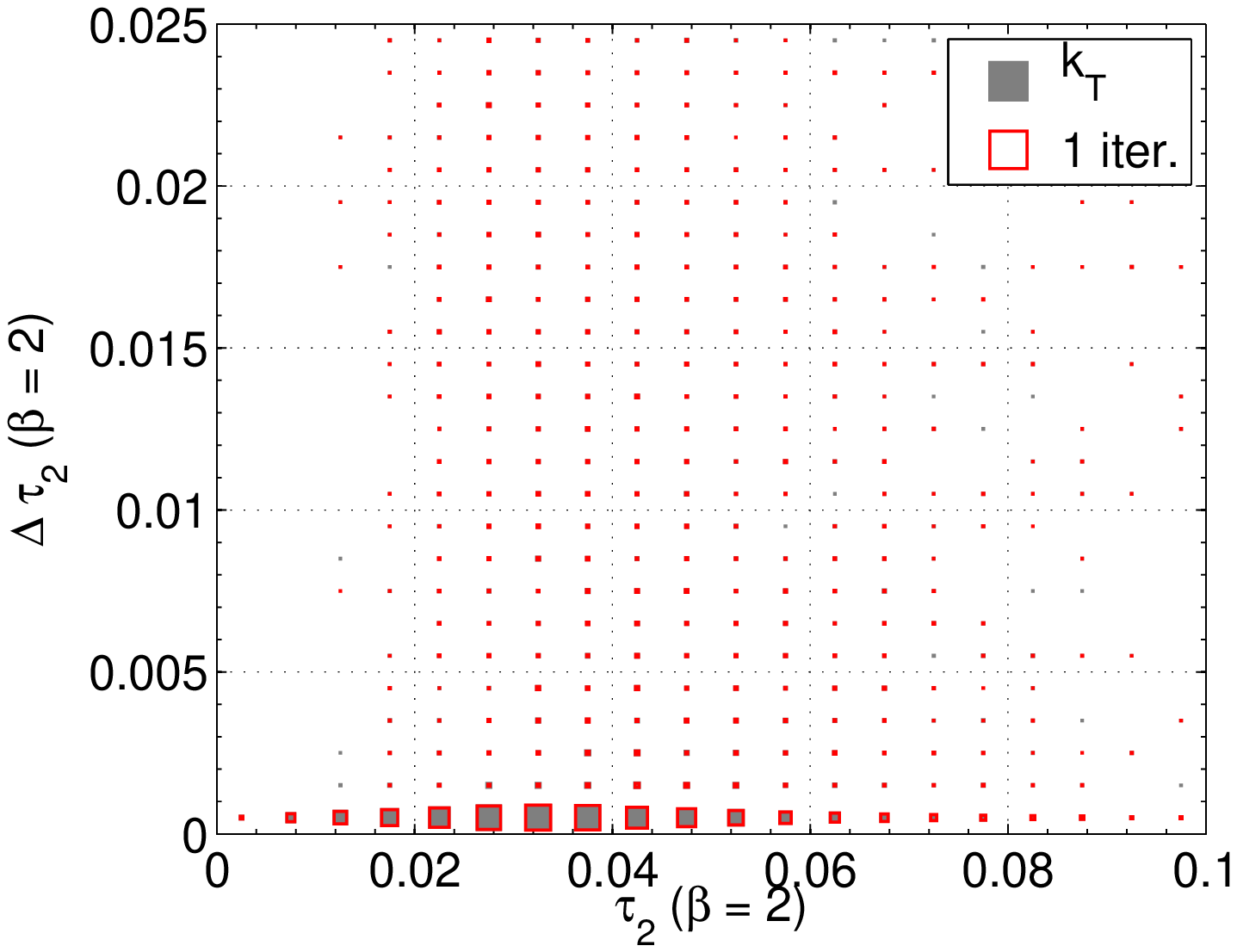}}\;\;
    \subfigure[]{\label{fig:t3CompBeta2}\includegraphics[trim = 0mm 0mm 0mm 0mm, clip, width = 0.31\textwidth]{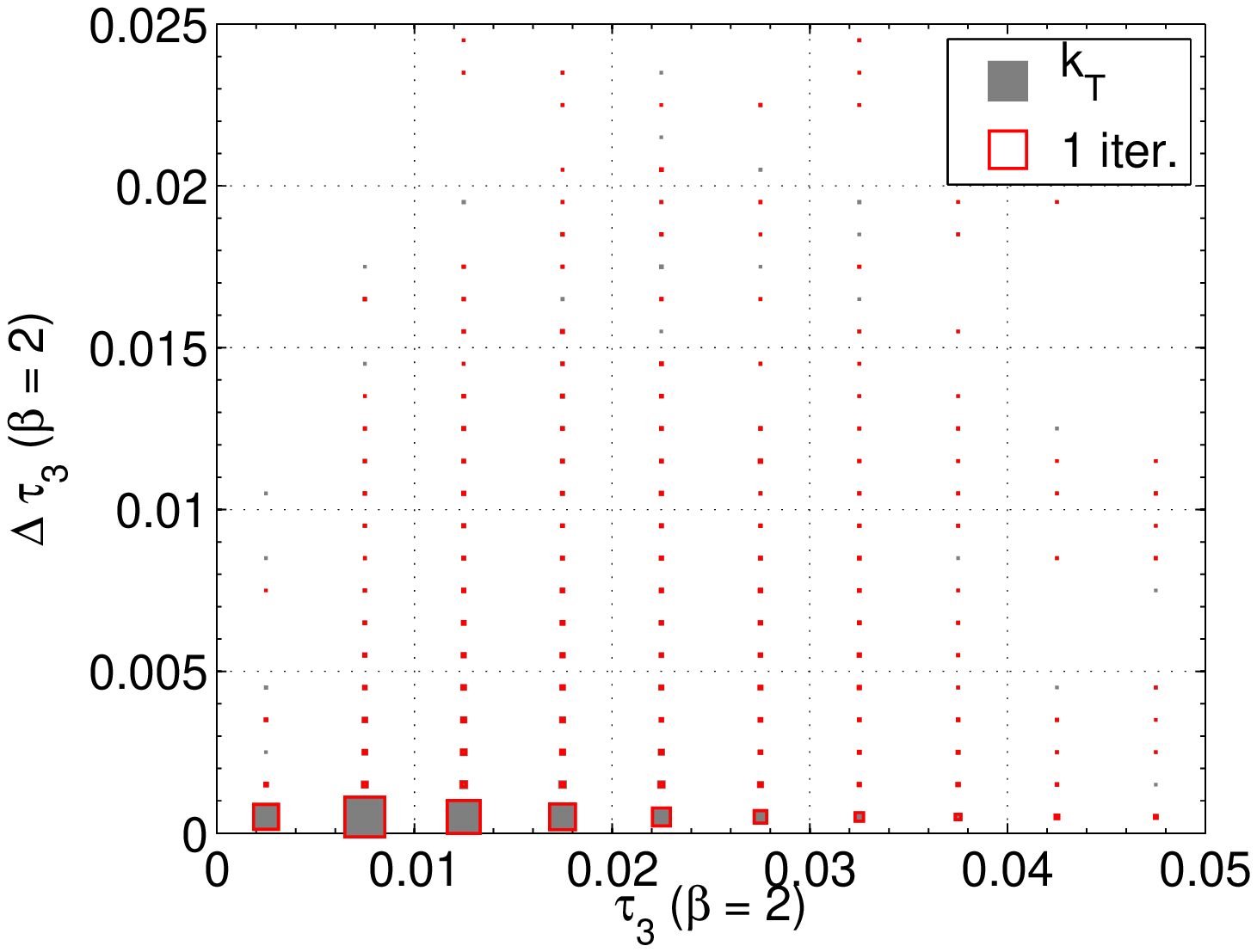}}
   \end{center}
   \vsh
  \caption{Difference between the minimum value of $\tau_N$ and the exclusive $k_T$ $\tilde{\tau}_N$.  The event sample is the 500-600 GeV $t\bar{t}$ sample detailed in \Sec{sec:analysisOverview}, with the same event selection as \Fig{fig:tauHistograms2Axes}.  The top row is $\beta = 1$, the bottom row is $\beta = 2$, and the columns are $\tau_1^{(\beta)}$, $\tau_2^{(\beta)}$, and $\tau_3^{(\beta)}$.  For $\beta = 1$, the difference between the minimum $\tau_N$ and the exclusive $k_T$ $\tilde{\tau}_N$ can be of order 50\%, though this difference is ameliorated by doing a single pass of the minimization procedure using the exclusive $k_T$ axes as a seed.  For $\beta = 2$, the values of $N$-subjettiness are typically different by less than 10\%, except for rare cases where the exclusive $k_T$ axes are near a local minimum of $\tilde{\tau}_N$, such that even doing a single pass of the minimization procedure does not help much.}
  \label{fig:tComp}
\end{figure}

In \Fig{fig:tComp}, we show the difference between the minimum value (after 100 seeds) of $N$-subjettiness compared to the value using the exclusive $k_T$ axes using the $t \bar{t}$ sample described in \Sec{sec:analysisOverview}.  We also show the (local) minimum value obtained by doing a single pass at minimization starting from the exclusive $k_T$ axes (without added noise).  We see that for $\beta = 2$, the minimum axes and the $k_T$ axes are quite similar, as expected from the discussion below \Eq{eq:minfinderbeta2}.  For $\beta =1$, there can be a 50\% shift between the minimum $\tau_N$ and the exclusive $k_T$ $\tilde{\tau}_N$, though this difference is quickly diminished by one pass of the minimization procedure.

\subsection{Uniqueness of 1-subjettiness Minima}
\label{sec:uniqueness}

The fact that a relatively small number of seeds are needed to find the global minimum of $N$-subjettiness is due in part to an interesting property of 1-subjettiness, which is that $\tilde{\tau}_1$ has a unique minimum for $\beta \ge 1$ (i.e.\ the global minimum).  One can think of finding $\tau^{(\beta)}_N$ for $N = k \ge 2$ and $\beta \ge 1$ as being separated into two tasks:  first partitioning the jet constituents into $N$ subclusters $C_J$ which together yield the lowest sum of subcluster $\tilde{\tau}^{(\beta)}_1$ values, and then finding the unique minimum of each $\tilde{\tau}^{(\beta)}_1$.  Of course, the algorithm in \Sec{sec:minimizationAlgorithm} tackles partitioning and minimization at the same time.

To see why 1-subjettiness has a unique minimum, note that \Eq{eq:1subjettiness} is a sum of continuous ``potential'' functions of $(y_0, \phi_0)$, one for each particle.  For $\beta > 1$, these potential functions are strictly convex since they behave like $\Delta R^\beta$.   The sum of strictly convex functions is also strictly convex, and strictly convex functions (and thus $\tilde{\tau}_1$) have a unique minimum.  For the special case $\beta = 1$, $\tilde{\tau}_1$ is convex but not strictly so, which means that the minimum value of $\tilde{\tau}_1$ is unique, but this minimum could be obtained at multiple positions $(y_0, \phi_0)$.  

When we talk about $N$-jettiness as a jet algorithm in \Sec{sec:jetAlgorithm}, we will encounter the potential function for iterative cone algorithms in \Eq{eq:1subR0}.  That potential function scales like $\min\{\Delta R^2, R^2_0\}$, where $R_0$ is a fixed parameter, and thus the potential function is not convex.  This leads to a proliferation of local minima for iterative cone algorithms, and the related problems of infrared safety in seeded cone algorithms.

\section{Top Tagging Performance}
\label{sec:topTaggingPerformance}

In this section, we investigate the tagging efficiencies for top jets and the mistag rates for QCD jets using $N$-subjettiness.  Compared to the preliminary study in \Ref{Thaler:2010tr}, we will use the top tagging benchmark samples from the BOOST2010 report \cite{Abdesselam:2010pt}.  This will enable an apples-to-apples comparison to common top tagging methods in the literature.\footnote{In \Ref{Thaler:2010tr}, $N$-subjettiness was compared to simplistic implementations of the Johns Hopkins Top Tagger \cite{Kaplan:2008ie} and the ATLAS YSplitter method \cite{Brooijmans:1077731} on event samples from the default tune of \texttt{Pythia 8.135} \cite{Sjostrand:2006za,Sjostrand:2007gs}.}  Note, however, that the BOOST2010 samples are particle level samples and do not include realistic detector resolutions, efficiencies, or acceptances.  

\subsection{Analysis Overview}
\label{sec:analysisOverview}

For our tagging performance study, we use benchmark samples from the BOOST2010 report \cite{Abdesselam:2010pt}.  These event samples are publicly available at:
\begin{itemize}
\item \url{http://www.lpthe.jussieu.fr/~salam/projects/boost2010-events/}
\item \url{http://tev4.phys.washington.edu/TeraScale/boost2010/}
\end{itemize}
We will utilize samples from two different benchmark Monte Carlo programs which simulate proton-proton collisions at a center-of-mass energy of 7 TeV.  The primary benchmark is \texttt{HERWIG 6.510}~\cite{Corcella:2002jc} with a description of the underlying event from \texttt{JIMMY}~\cite{Butterworth:1996zw} using an ATLAS tune~\cite{ATL-PHYS-PUB-2010-002}.  We will also do one comparison study to \texttt{PYTHIA 6.4} \cite{Sjostrand:2006za} with a $p_T$-ordered shower using the Perugia0 tune \cite{Skands:2010ak}.

\begin{figure}[tp]
  \begin{center}
    \subfigure[]{\label{fig:perp}\includegraphics[trim = 0mm 0mm 00mm 0mm, clip, width = 0.48\textwidth]{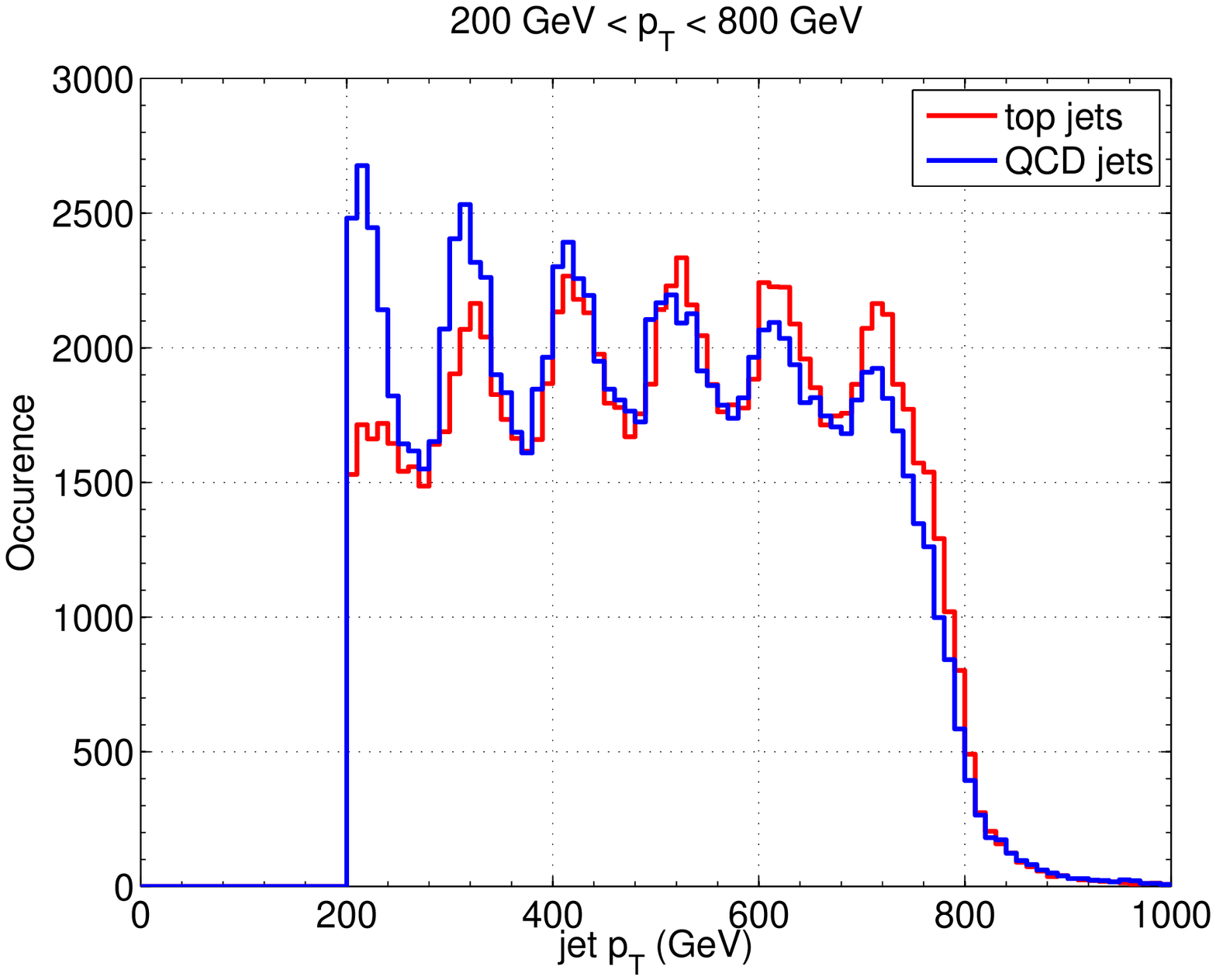}} \;\;\;
    \subfigure[]{\label{fig:mass}\includegraphics[trim = 0mm 0mm 00mm 0mm, clip, width = 0.48\textwidth]{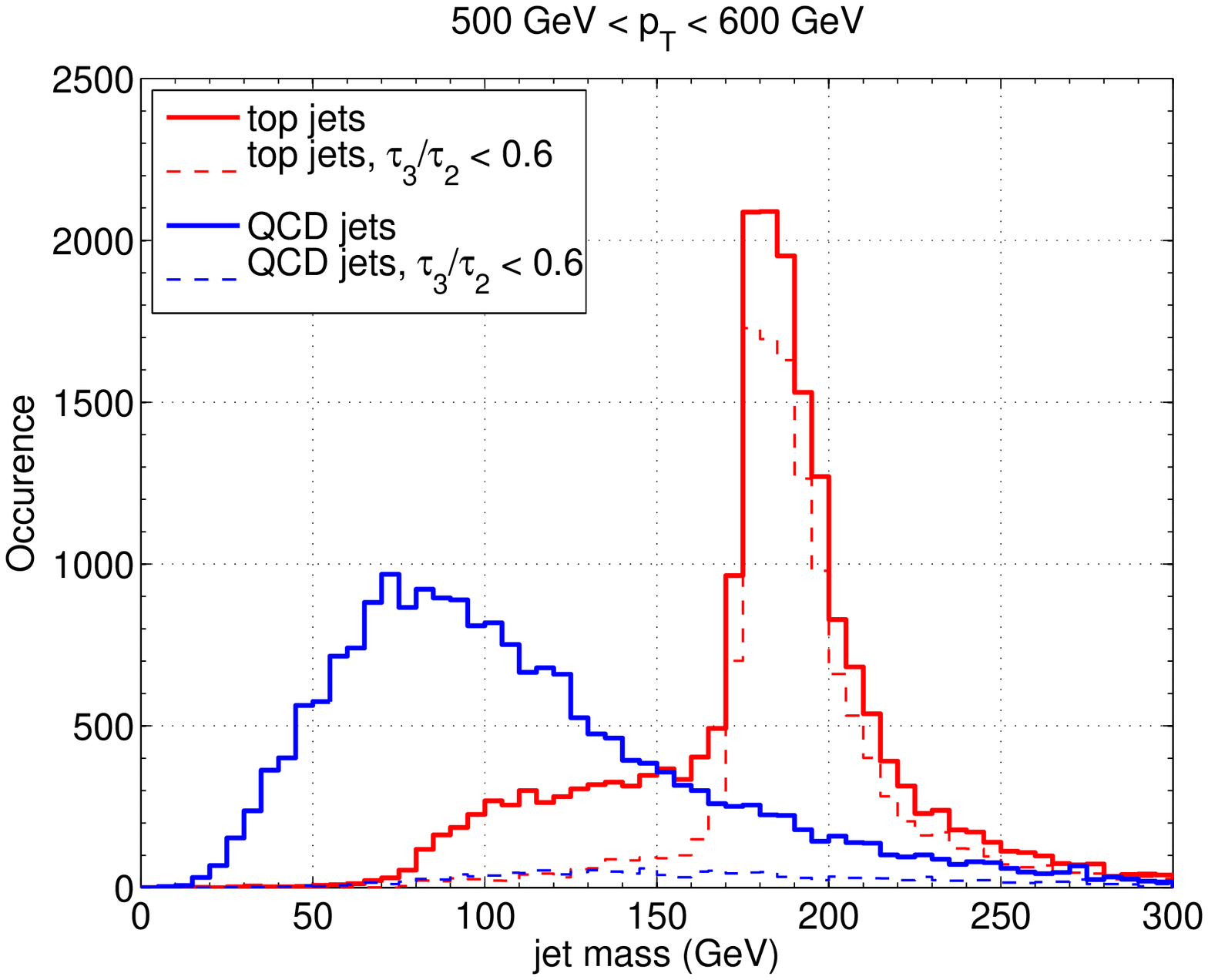}}
  \end{center}
  \vsh
  \caption{Basic kinematics of the $t\bar{t}$ and dijet BOOST2010 samples, after clustering with $R = 1.0$ anti-$k_T$ jets. (a) Jet transverse momentum in the combined sample with parton-level $p_T$ between 200 and 800 GeV.  Note that this is an unphysical $p_T$ distribution, but serves as a useful testing ground for the various top tagging methods. (b) Jet invariant mass in the $500~\GeV < p_T < 600~\GeV$ sample.  An $N$-subjettiness cut $\tau_3/\tau_2 < 0.6$ eliminates the bulk of the QCD jets as well as top jets with a mass much smaller than $m_{\rm top}$, but leaves most of the top resonance peak intact.}
  \label{fig:samples}
\end{figure}

The signal sample is hadronically-decaying $t\bar{t}$ production, and the background sample is QCD dijet production.   They are divided in subsamples of equal size, with parton $p_T$ ranges from 200--300 GeV, 300--400 GeV, \dots, 700--800 \GeV, as shown in \Fig{fig:perp}.  Together they yield an approximately flat jet transverse momentum distribution in a kinematic regime that is interesting for new physics searches at the LHC.\footnote{This flat distribution in transverse momentum is of course artificial, since physical cross-sections fall off with $p_T$, but it is helpful for testing the performance of tagging methods across a wide kinematic range.}   

In accordance with the BOOST2010 report, jets are defined with the anti-$k_T$ algorithm \cite{Cacciari:2008gp} with a jet radius parameter of $R = 1.0$ using \texttt{FastJet 2.4.4} \cite{FastJet,Cacciari:2005hq}.  No simulation of detector effects is performed, but only final state particles with pseudorapidity $|\eta| < 5.0$ (except neutrinos and muons) are considered in the jet clustering.  Only the two hardest jets with $p_T > 200~\GeV$ are considered from each event, and efficiencies and fake rates are determined on a per jet basis. 

The various top tagging algorithms studied in the BOOST2010 report are summarized in \Ref{Abdesselam:2010pt} and described in more detail in the original papers.  The five algorithms shown in  \Fig{fig:executiveSummary} are referred to as ``Hopkins'' \cite{Kaplan:2008ie}, ``CMS'' \cite{CMS-PAS-JME-09-001,CMS-PAS-EXO-09-002,Rappoccio:1358770}, ``Pruning'' \cite{Ellis:2009su,Ellis:2009me}, ``ATLAS'' \cite{ATL-PHYS-PUB-2010-008,ATL-PHYS-PUB-2009-081,Brooijmans:1077731}, and ``Thaler/Wang''  \cite{Thaler:2008ju}.  After the BOOST2010 report, two other top tagging methods were applied to this sample \cite{Hook:2011cq,Jankowiak:2011qa}, though the comparisons in \Fig{fig:executiveSummary} and later in \Tab{tab:taggersummary} only include the originally tested algorithms.  

The basic criterion for tagging a boosted top quark is that the jet mass should fall near $m_{\rm top} \simeq 171~\GeV$.  In \Fig{fig:mass}, we show the jet invariant mass distribution from the 500--600 GeV sub-sample, where one can clearly see the top resonance.   One can also see that an $N$-subjettiness cut of $\tau^{(1)}_3/\tau^{(1)}_2 < 0.6$ substantially decreases the background in the top peak region without adversely affecting the signal much.  For concreteness, we will consider the mass window $160~\GeV < m_{\rm jet} < 240~\GeV$ for top jets in \Sec{sec:results}.  The upper limit of this mass range is relatively high compared to the lower limit, because boosted top jets often acquire additional mass from the underlying event.  We consider possible optimizations of the mass window in \Sec{sec:multivariateMethods}.  In addition to the invariant mass cut, we will apply a cut on the ratio $\tau_3 / \tau_2$, where the cut is adjusted to change the relative signal tagging efficiency and background mistag rate.  

\subsection{N-subjettiness Performance}
\label{sec:results}

\begin{figure}[tp]
  \begin{center}
    \subfigure[]{\label{fig:tau1_pt500_a100}\includegraphics[trim = 0mm 0mm 0mm 0mm, clip, width = 0.47\textwidth]{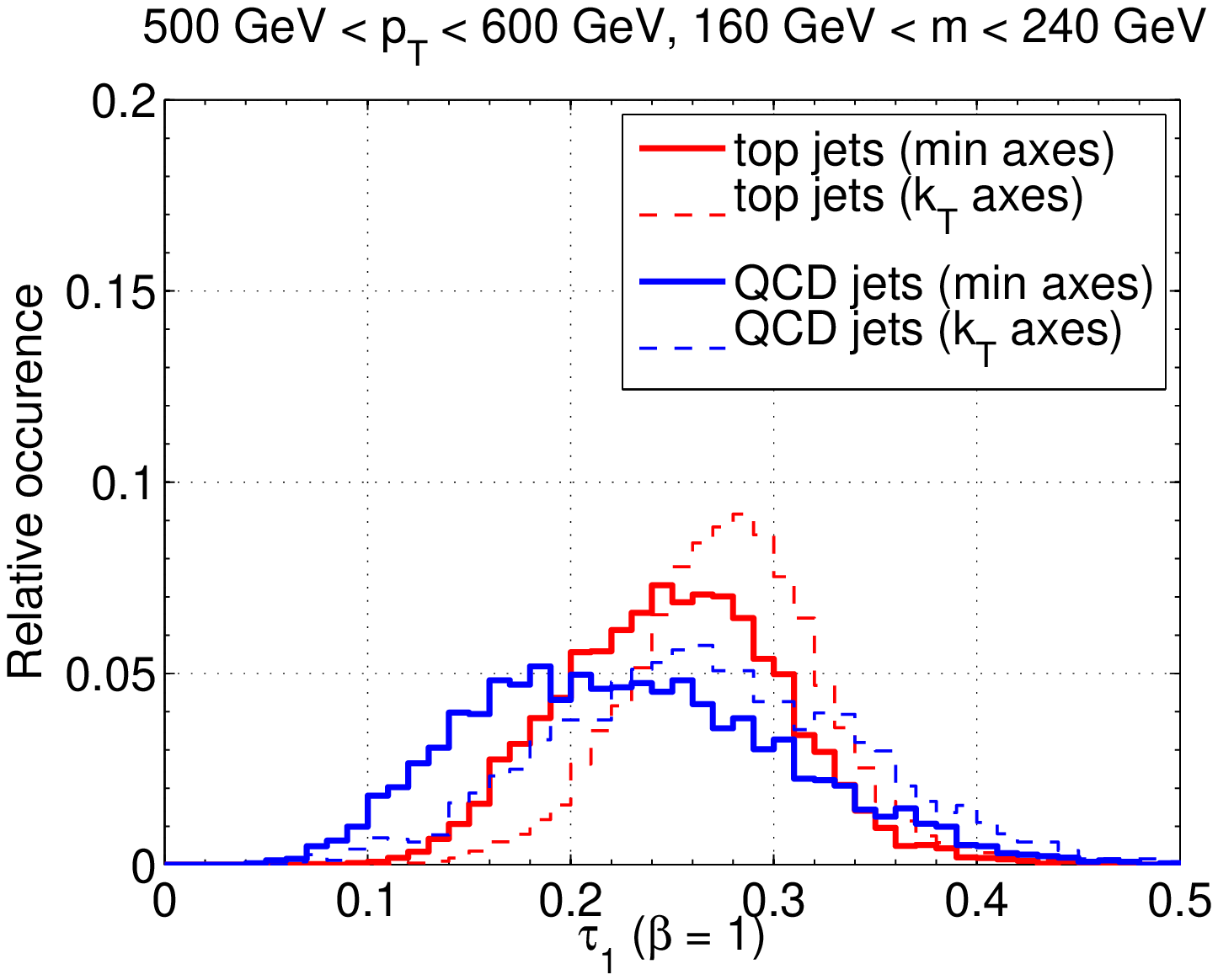}} \;\;\;
    \subfigure[]{\label{fig:tau1_pt500_a200}\includegraphics[trim = 0mm 0mm 0mm 0mm, clip, width = 0.47\textwidth]{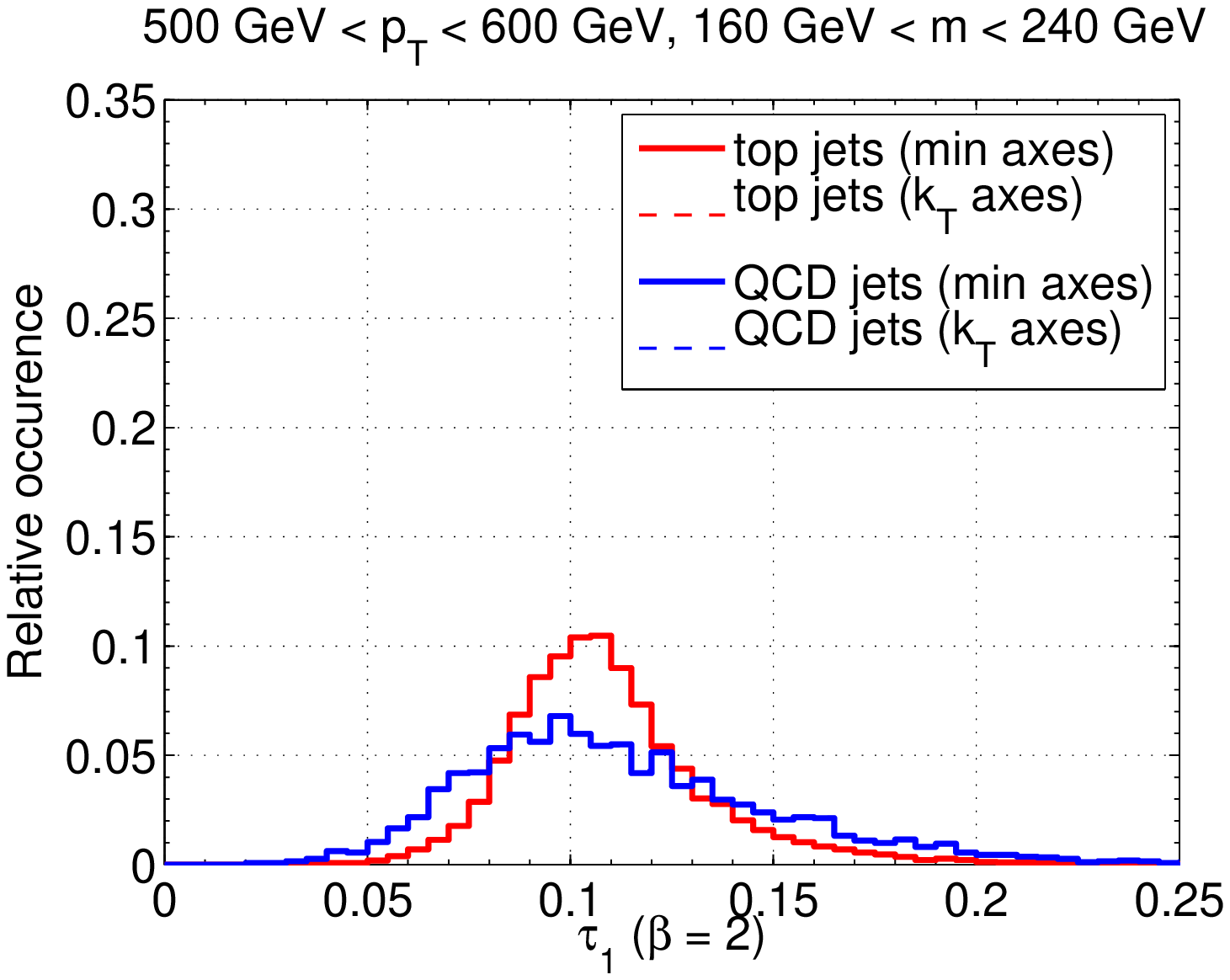}}
    \subfigure[]{\label{fig:tau2_pt500_a100}\includegraphics[trim = 0mm 0mm 0mm 0mm, clip, width = 0.47\textwidth]{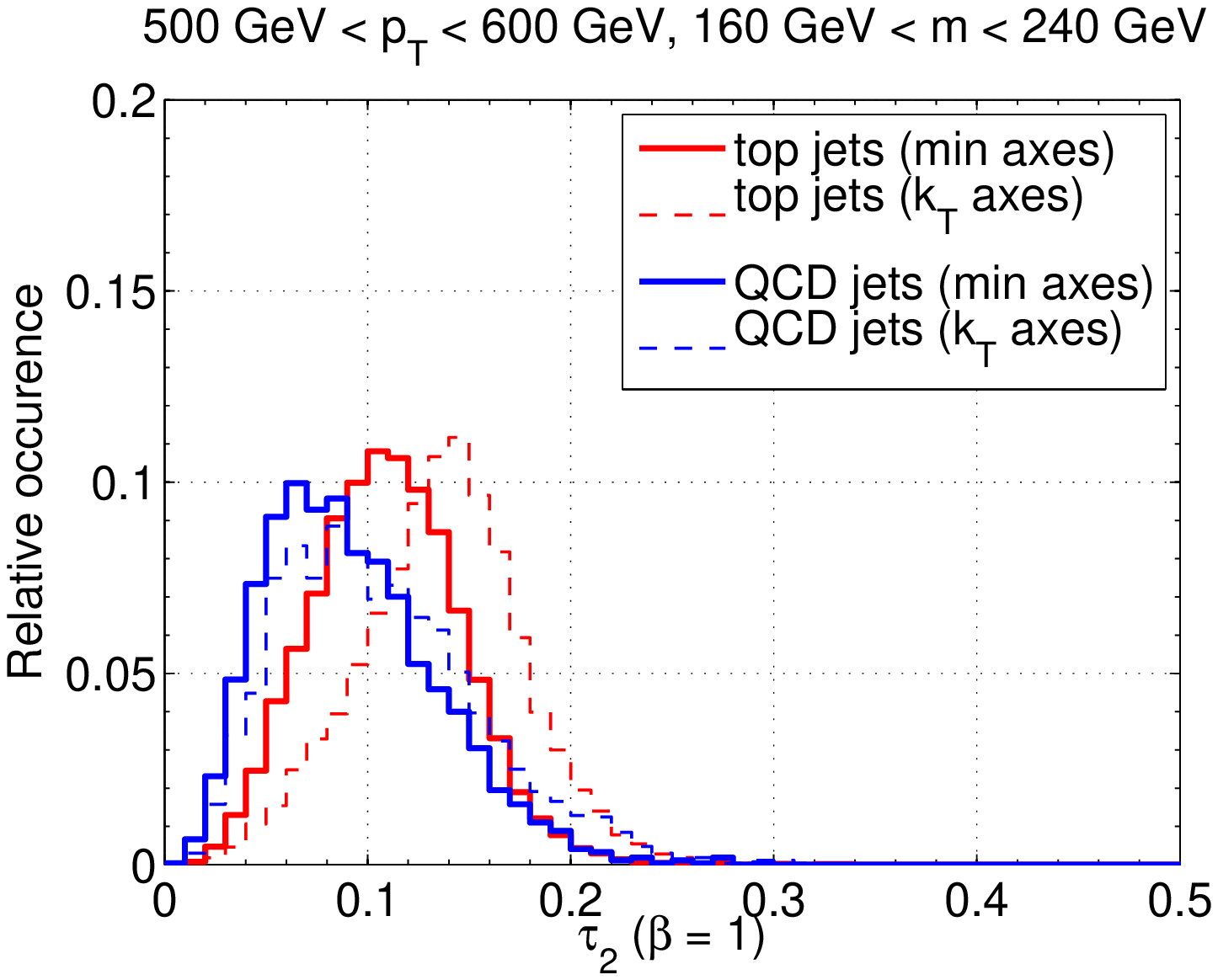}} \;\;\;
    \subfigure[]{\label{fig:tau2_pt500_a200}\includegraphics[trim = 0mm 0mm 0mm 0mm, clip, width = 0.47\textwidth]{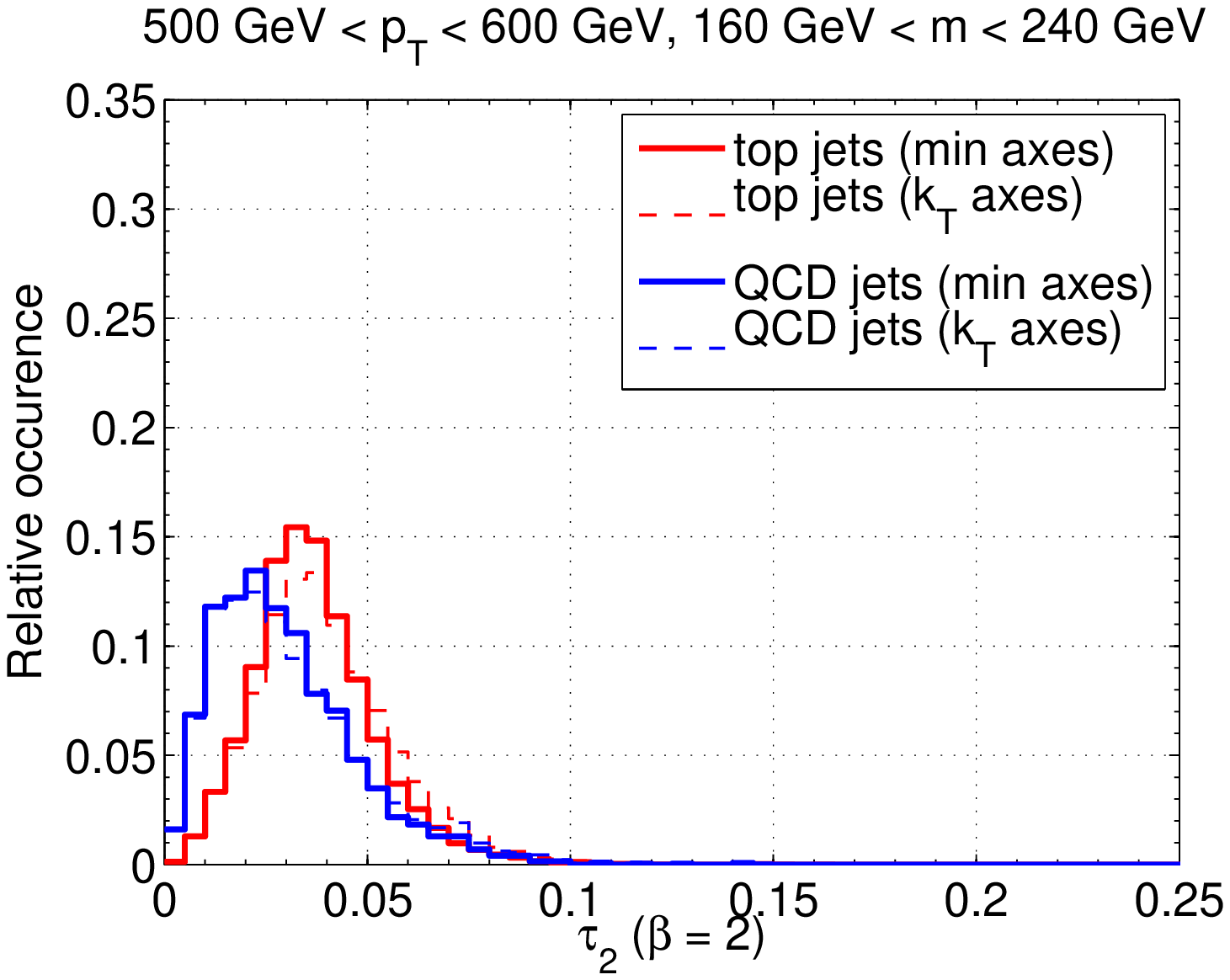}}
    \subfigure[]{\label{fig:tau3_pt500_a100}\includegraphics[trim = 0mm 0mm 0mm 0mm, clip, width = 0.47\textwidth]{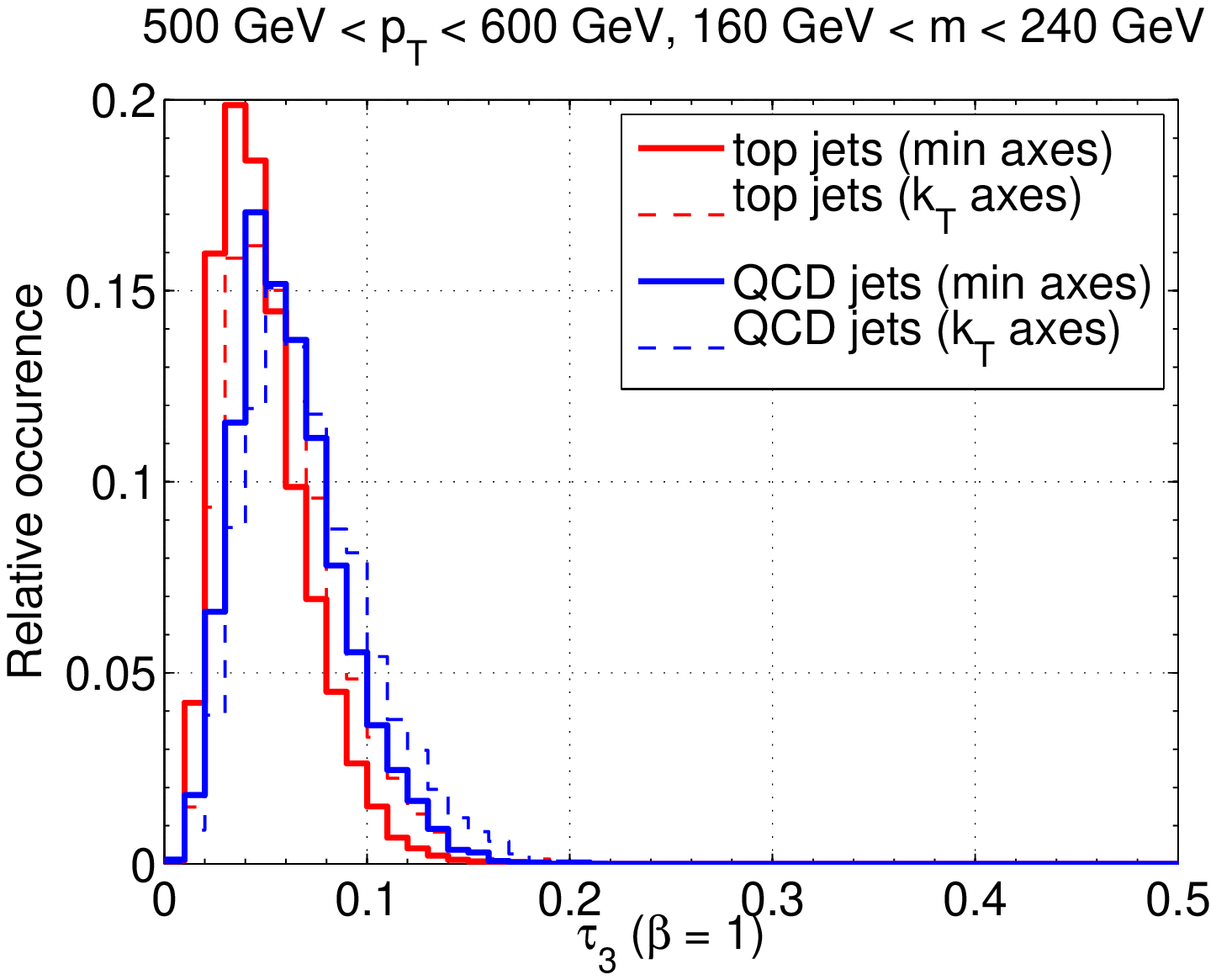}} \;\;\;
    \subfigure[]{\label{fig:tau3_pt500_a200}\includegraphics[trim = 0mm 0mm 0mm 0mm, clip, width = 0.47\textwidth]{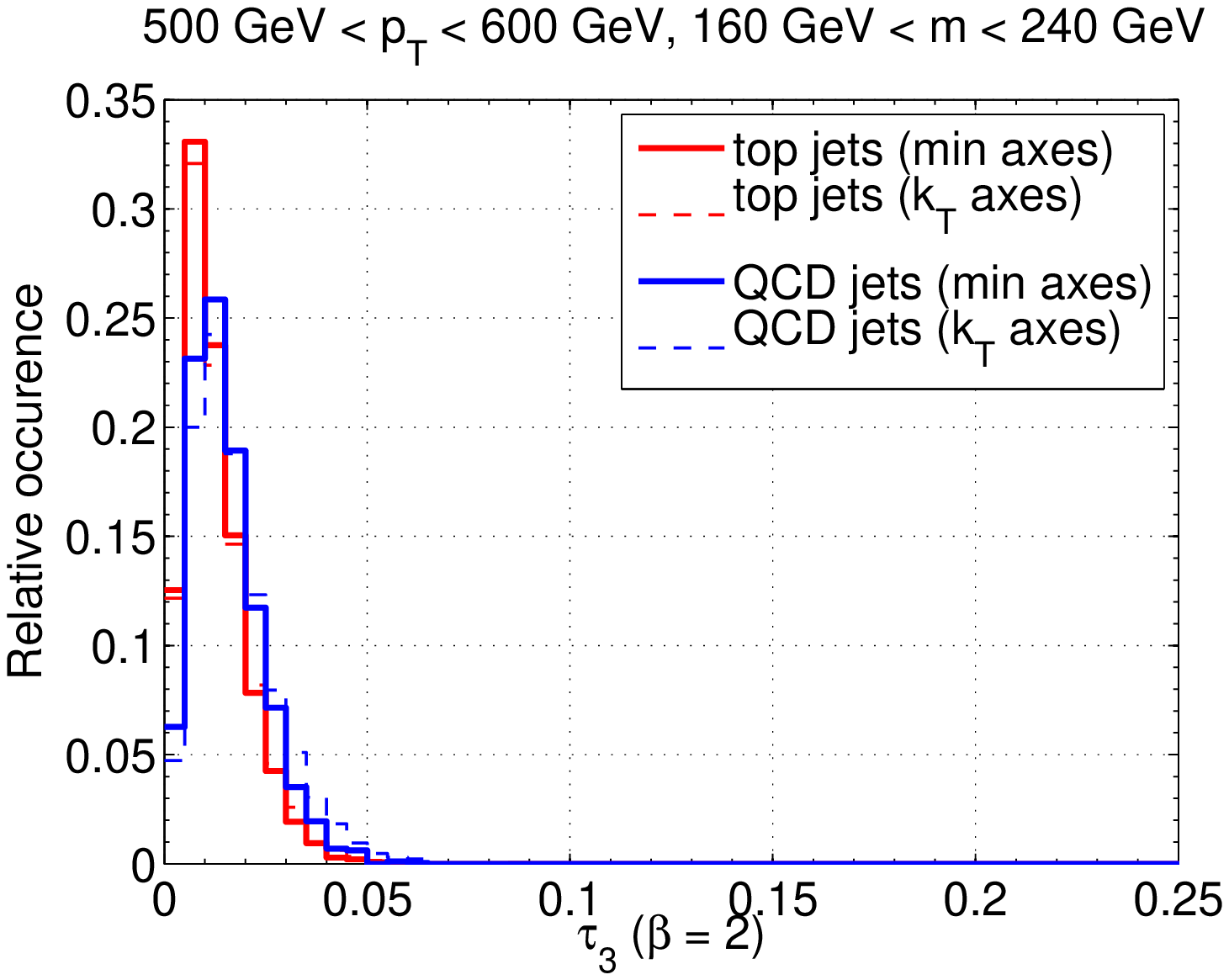}}
 \end{center}
 \vsh
  \caption{Left column: distributions of (a) $\tau_1$, (c) $\tau_2$ and (e) $\tau_3$  for $\beta = 1$ comparing boosted top and QCD jets.  For these plots, we impose an invariant mass window of $160~\GeV < m_{\rm jet} < 240~\GeV $ on jets with $R = 1.0$ and $500~\GeV < p_T < 600~\GeV$.  The solid bold lines are for the $\beta = 1$ minimization axes, while the dashed thin lines are for the exclusive $k_T$ axes.  As expected, the minimization axes yield smaller values of $\tau_N$ than the exclusive $k_T$ axes.   Right column: equivalent plots for $\beta = 2$.}
  \label{fig:tauHistograms2Axes}
\end{figure}

We first show the effect of the minimization procedure on the raw $N$-subjettiness distributions in the BOOST2010 sample.  Plots of $\tau_1$, $\tau_2$, and $\tau_3$ comparing top jets and QCD jets are shown in \Fig{fig:tauHistograms2Axes}, after imposing the $160~\GeV < m_{\rm jet} < 240~\GeV$ criterion.  As expected, the average value of $\tau_N$ is smaller using the minimum axes compared to the exclusive $k_T$ axes, though the shift is not as pronounced for $\beta = 2$, as expected from the discussion below \Eq{eq:minfinderbeta2}.  

As argued in \Ref{Thaler:2010tr}, $\tau_N$ is by itself not a very good discriminant for identifying boosted top quarks.  While one might naively expect that an event with small $\tau_3$ would be more likely to be a top jet, a quark- or gluon-initiated jet can also have small $\tau_3$, as shown in \Fig{fig:tau3_pt500_a100}.  Though top jets are likely to have large $\tau_1$ and $\tau_2$, QCD jets with a diffuse spray of large-angle radiation can also have large $\tau_1$ and $\tau_2$, as shown in \Figs{fig:tau1_pt500_a100}{fig:tau2_pt500_a100}.  However, those QCD jets with large $\tau_2$ typically have large values of $\tau_3$ as well, so it is in fact the \emph{ratio} $\tau_3/\tau_2$ which is the preferred discriminating variable.  

\begin{figure}[tp]
  \begin{center}
    \subfigure[]{\label{fig:t2t1_pt0500_a100}\includegraphics[trim = 0mm 0mm 0mm 0mm, clip,width = 0.48\textwidth]{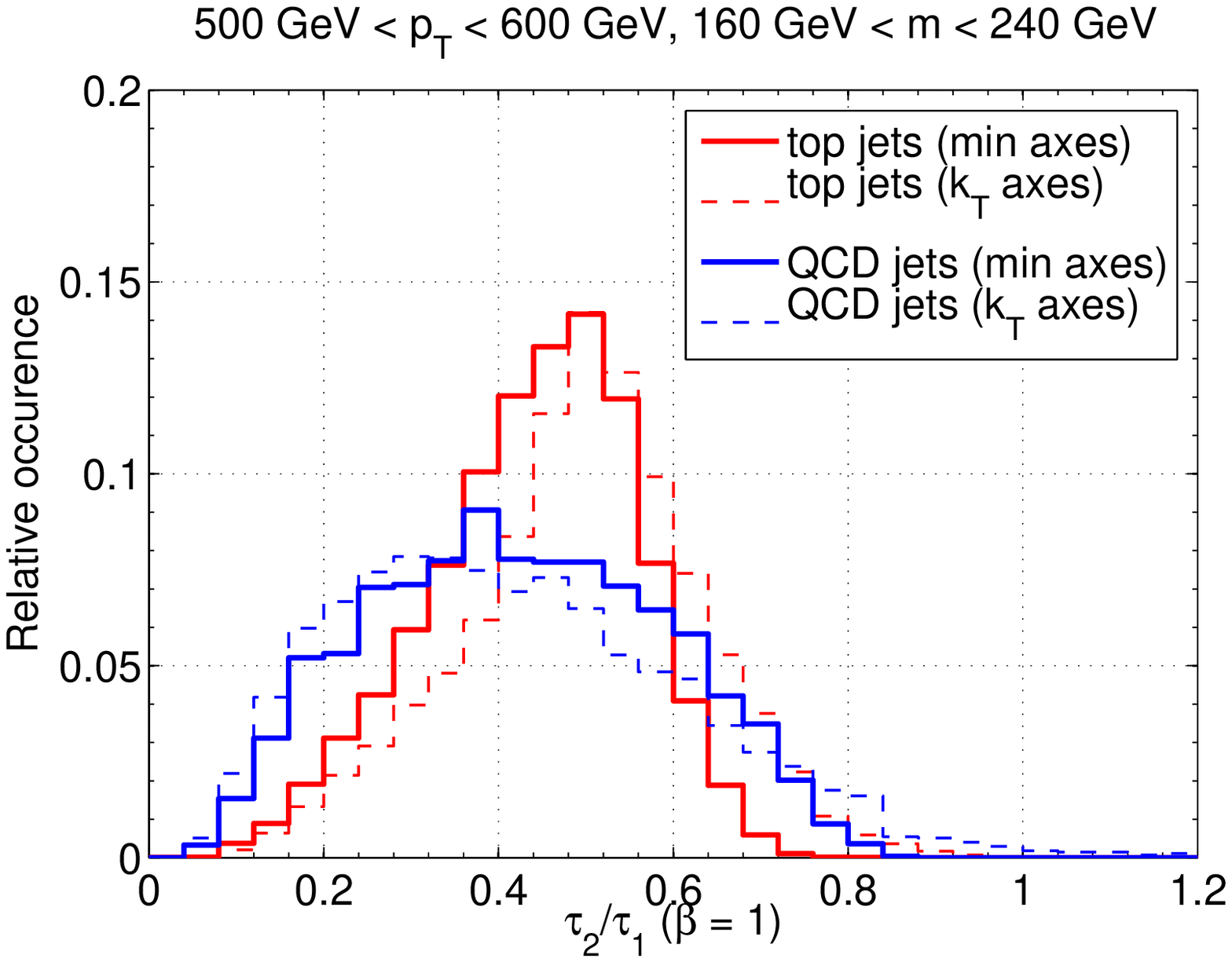}} \;\;\;
    \subfigure[]{\label{fig:t2t1_pt0500_a200}\includegraphics[trim = 0mm 0mm 0mm 0mm, clip,width = 0.48\textwidth]{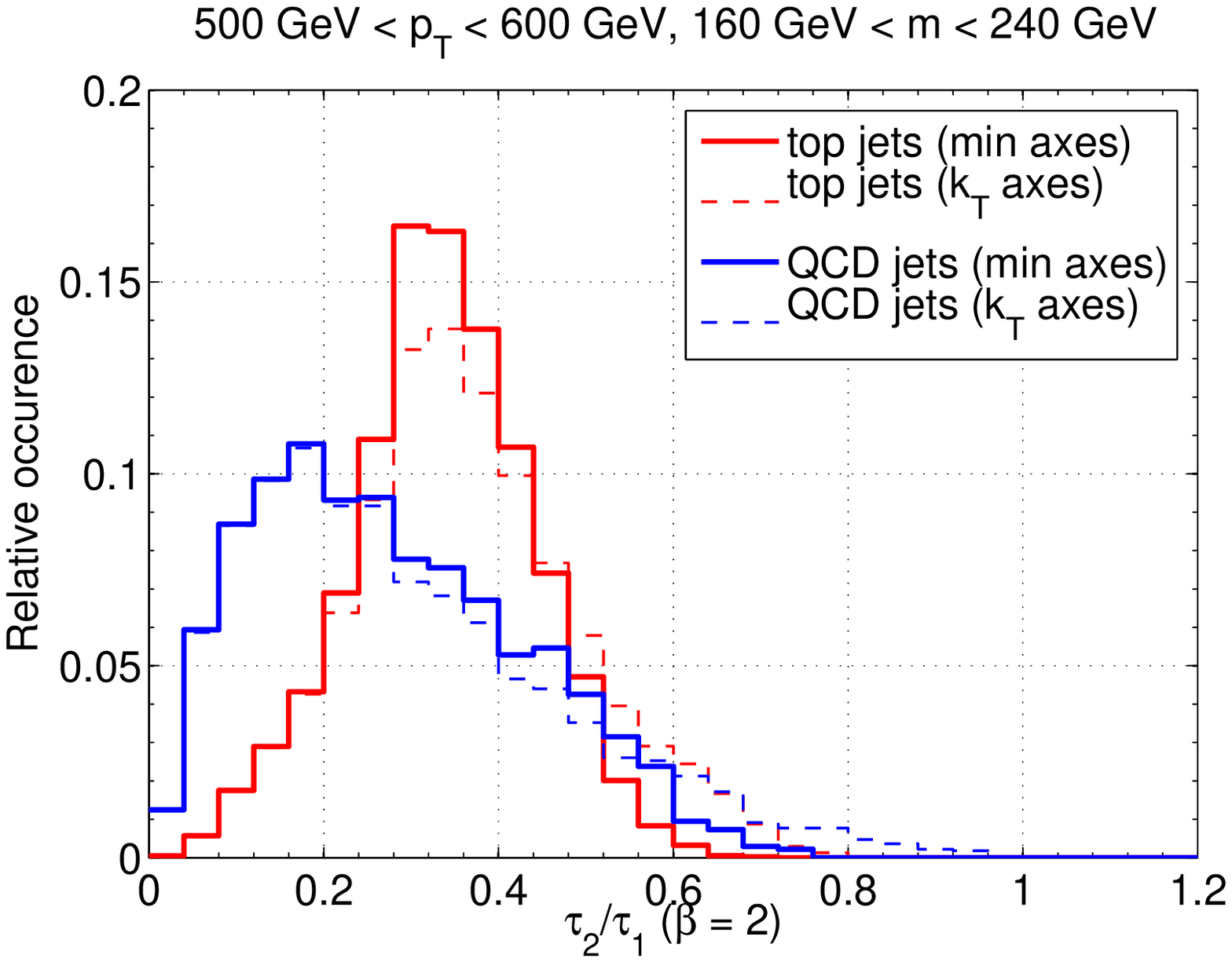}} 
    \subfigure[]{\label{fig:t3t2_pt0500_a100}\includegraphics[trim = 0mm 0mm 0mm 0mm, clip,width = 0.48\textwidth]{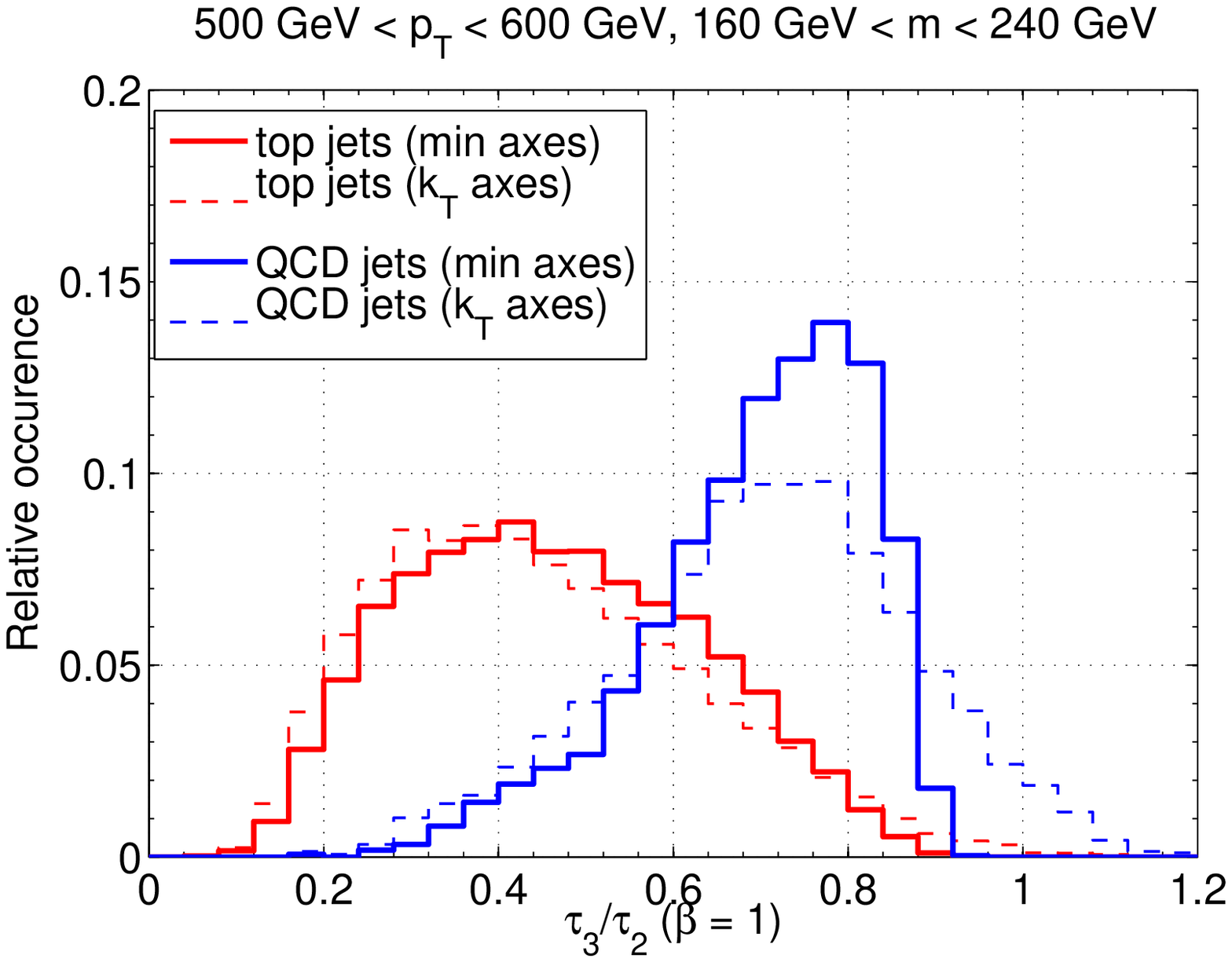}}\;\;\;
    \subfigure[]{\label{fig:t3t2_pt0500_a200}\includegraphics[trim = 0mm 0mm 0mm 0mm, clip,width = 0.48\textwidth]{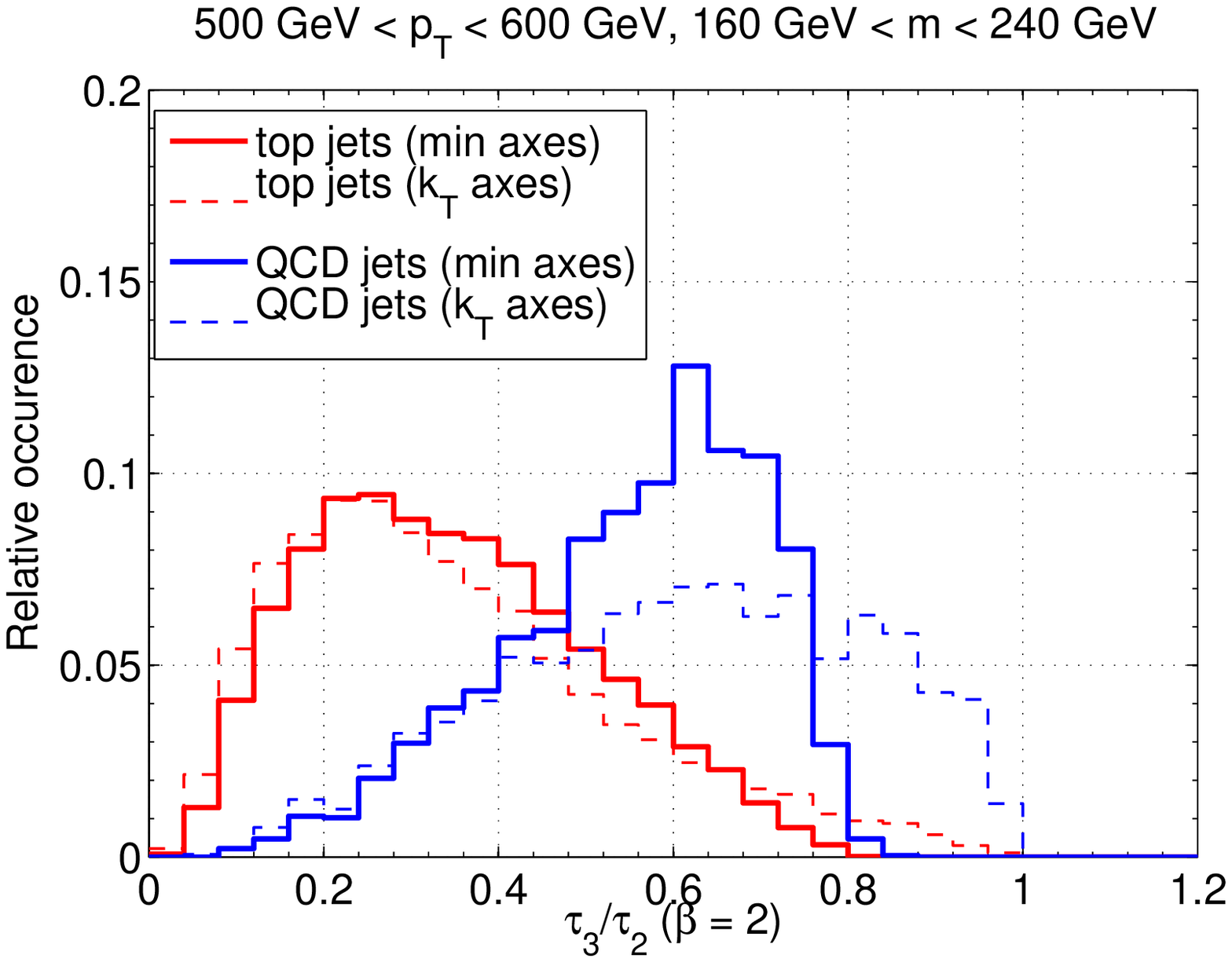}}
  \end{center}
  \vsh
    \caption{Distributions of (a) $\tau^{(1)}_2/\tau^{(1)}_1$, (b) $\tau^{(2)}_2/\tau^{(2)}_1$, (c) $\tau^{(1)}_3/\tau^{(1)}_2$, (d) $\tau^{(2)}_3/\tau^{(2)}_2$ for boosted top and QCD jets, using the same formatting and event selection as \Fig{fig:tauHistograms2Axes}.  Note that after applying the minimization procedure, all of these ratios are strictly less than 1.  For boosted top identification, the best individual discriminating variable is (c) $\tau^{(1)}_3/\tau^{(1)}_2$, though especially (b) $\tau^{(2)}_2/\tau^{(2)}_1$ contains some additional information.}
      \label{fig:tauRatioHistograms2Axes}  
\end{figure}

Plots of the $\tau_N/\tau_{N-1}$ ratios are shown in \Fig{fig:tauRatioHistograms2Axes} for $\beta =1$ and $\beta =2$.  Notice that with the minimization procedure, we have $\tau_N/\tau_{N-1} < 1$, as expected.  By eye, there is better top/QCD separation using the minimized $\tau_N$ values, and $\tau_3/\tau_2$ with $\beta = 1$ appears to be the best single variable for discrimination.   Both of these observations will be confirmed below.  There is additional distinguishing power in $\tau_2/\tau_1$ and (to a smaller extent) raw $N$-subjettiness values, which will be explored in \Sec{sec:multivariateMethods}.

 \begin{figure}[tp]
  \begin{center}
    \subfigure[]{\label{fig:eff_500_axis}\includegraphics[trim = 0mm 0mm 0mm 0mm, clip, width = 0.49\textwidth]{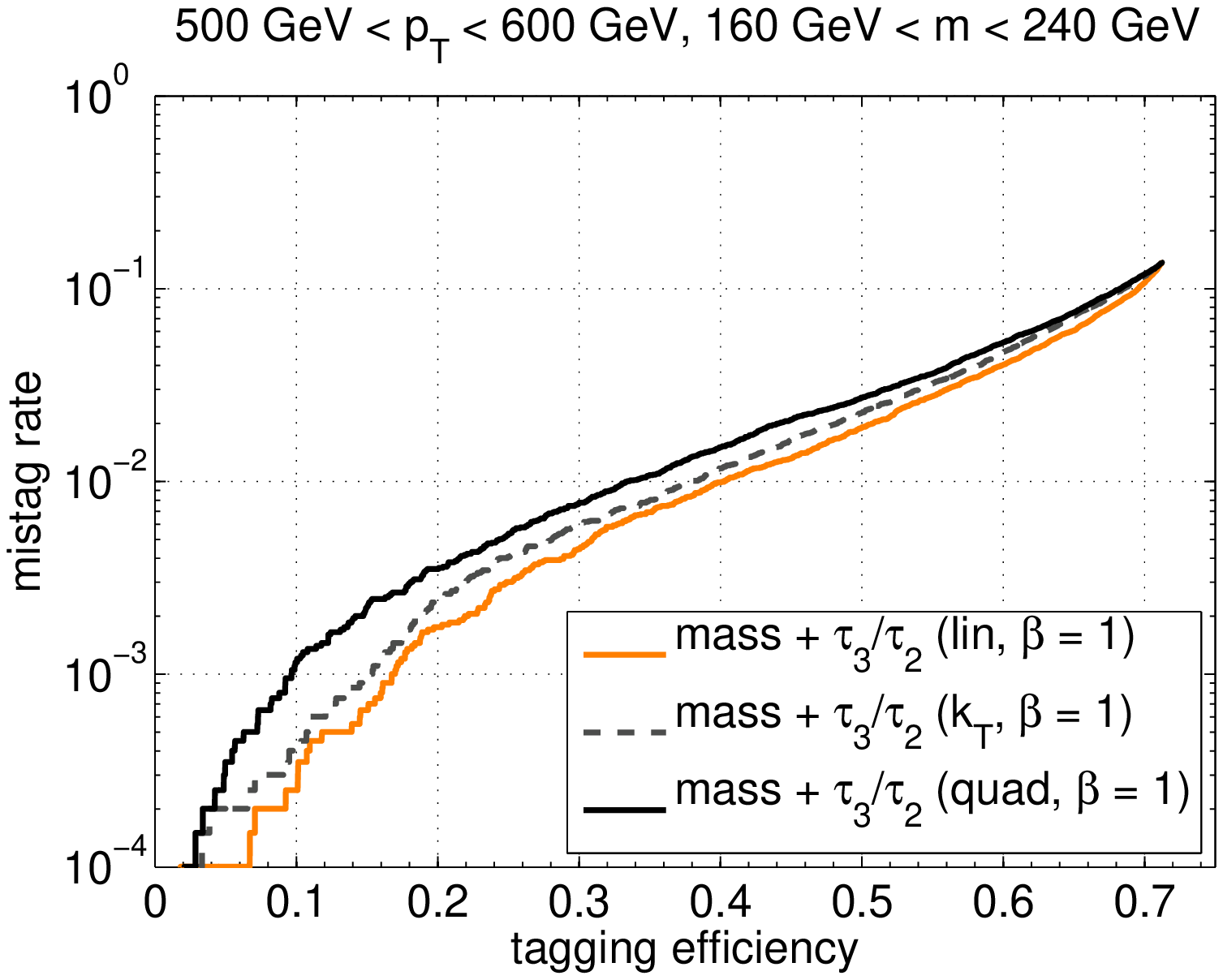}}
    \subfigure[]{\label{fig:eff_all_axis}\includegraphics[trim = 0mm 0mm 0mm 0mm, clip, width = 0.49\textwidth]{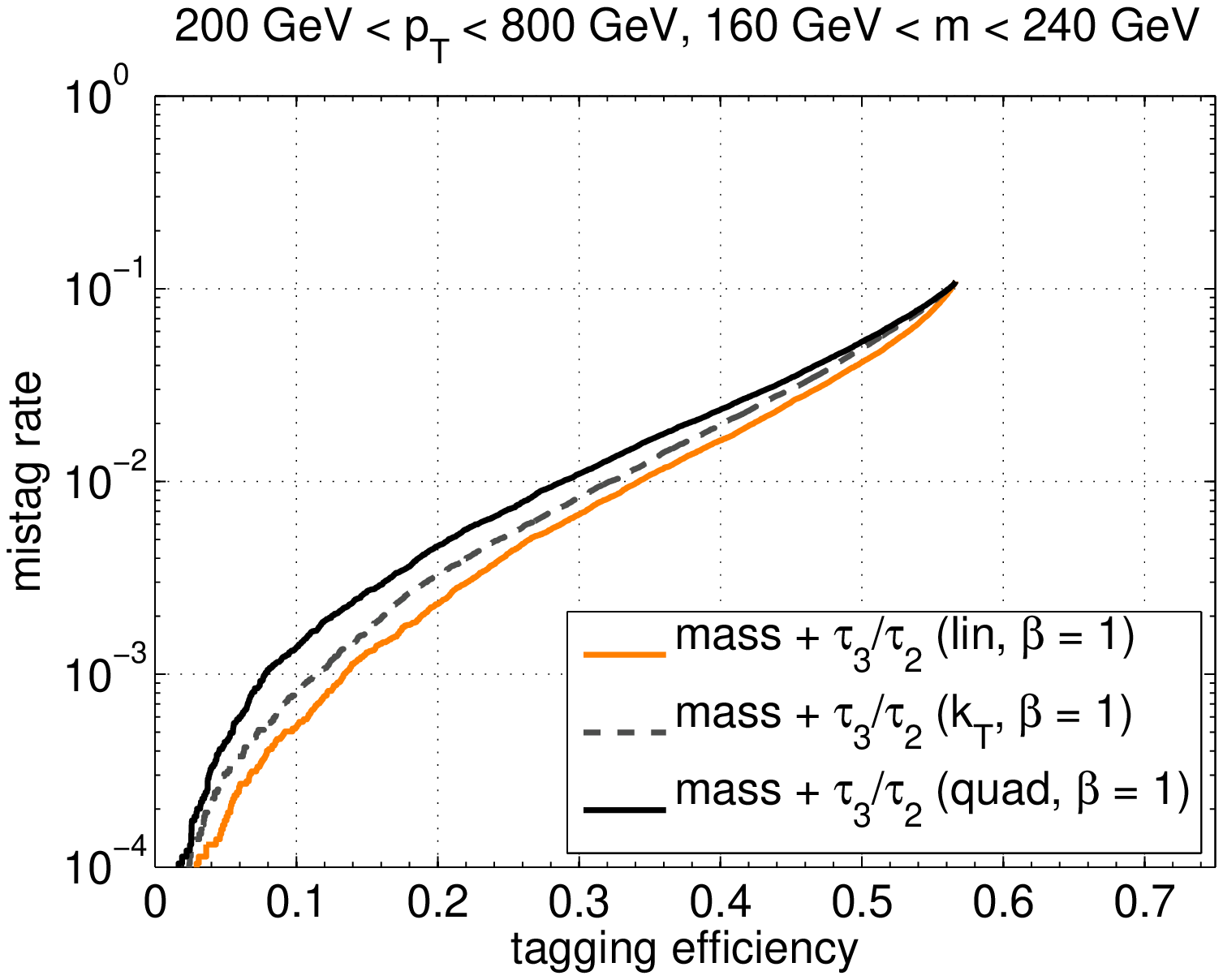}}
    \subfigure[]{\label{fig:eff_500_beta}\includegraphics[trim = 0mm 0mm 0mm 0mm, clip, width = 0.49\textwidth]{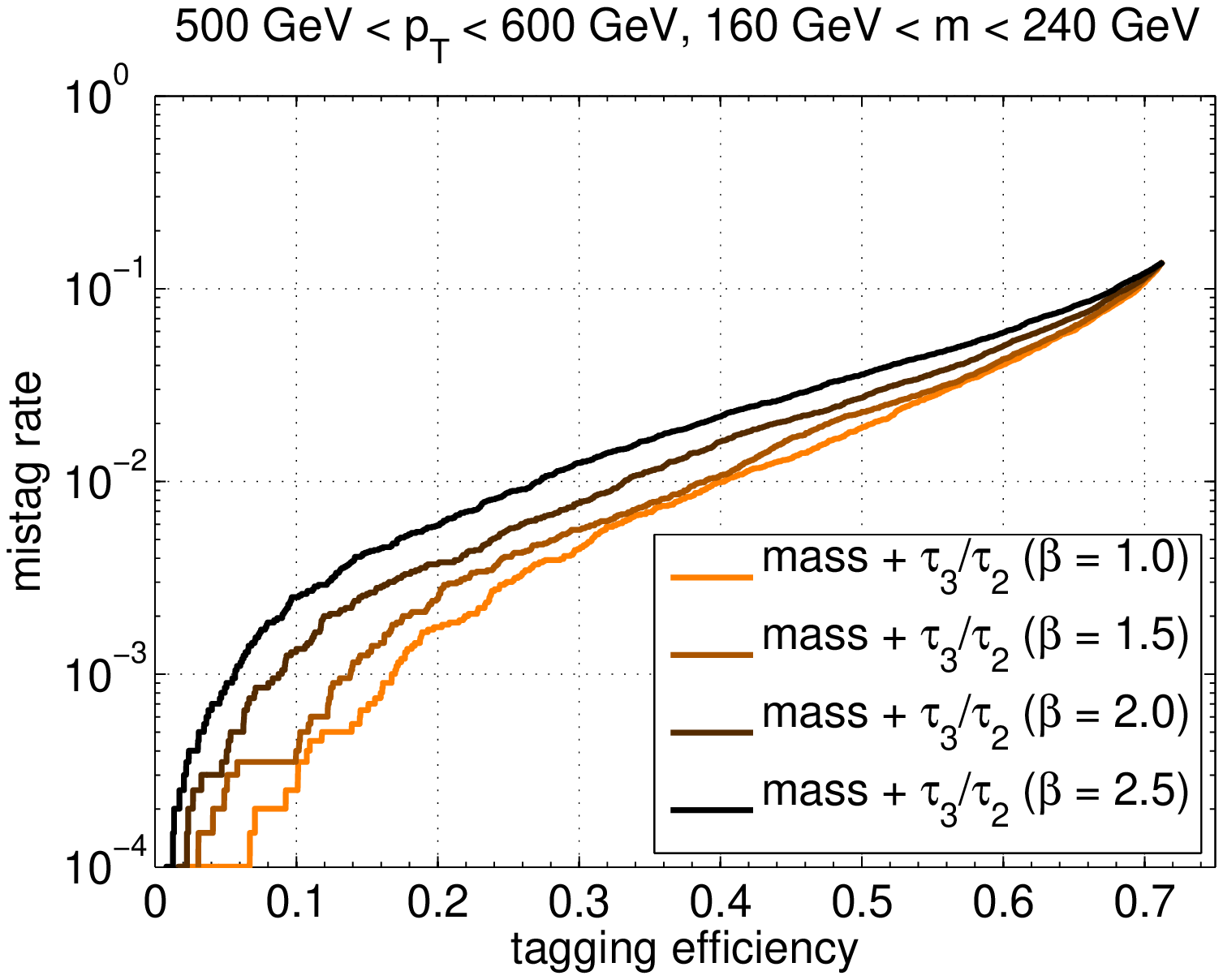}} 
    \subfigure[]{\label{fig:eff_all_beta}\includegraphics[trim = 0mm 0mm 00mm 0mm, clip, width = 0.49\textwidth]{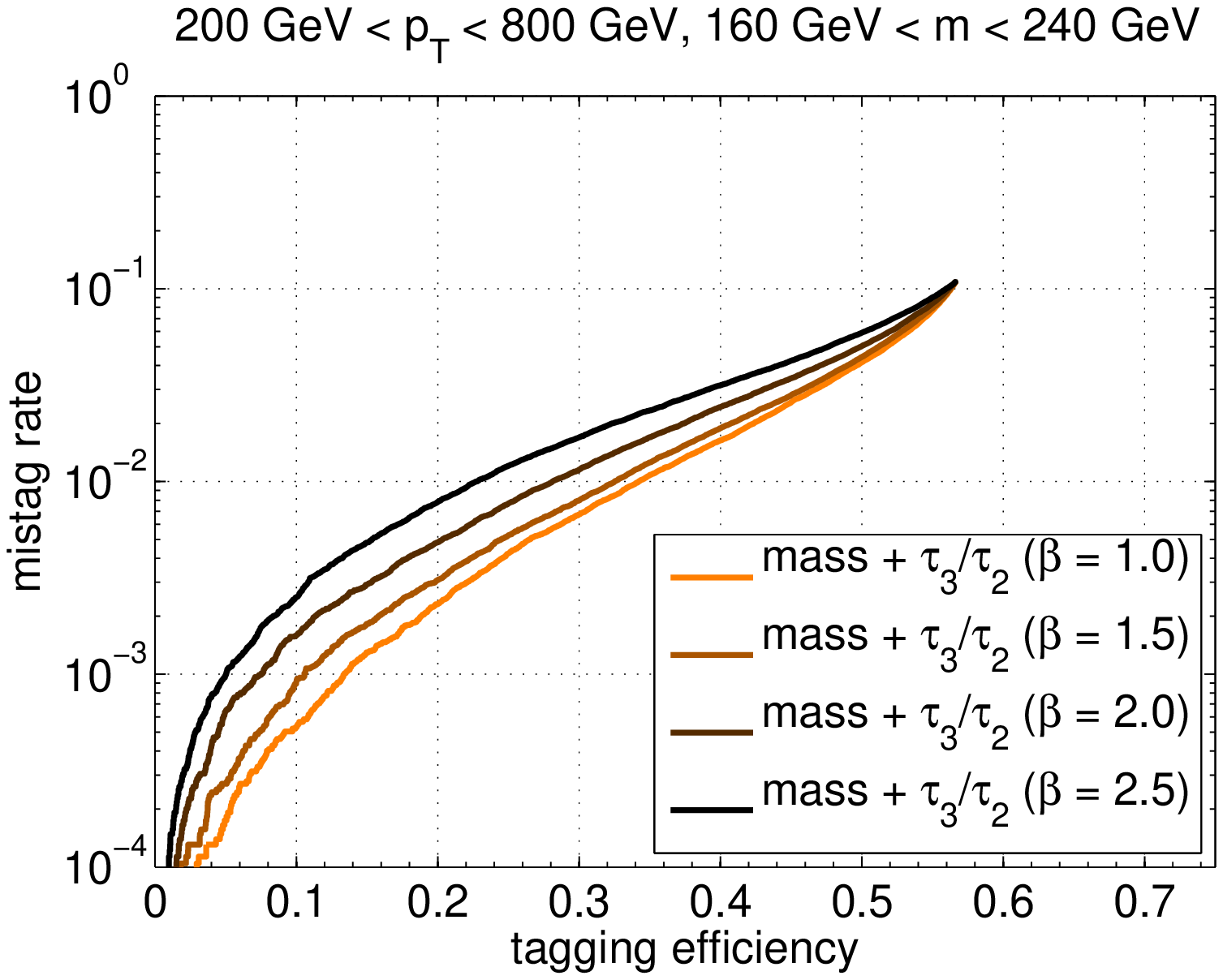}}
  \end{center}
  \vsh
  \caption{Efficiency/mistag curves for a variable cut on $\tau_3/\tau_2$ and a fixed top mass window $160~\GeV < m_{\rm jet} < 240~\GeV $.  Top row:  Fixing $\beta = 1$, but calculating $\tilde{\tau}^{(1)}_N$ using different axes:  the axes that minimize the $\beta = 1$ measure (``lin''), the axes from exclusive $k_T$ (``$k_T$''), and axes that minimize the $\beta = 2$ measure (``quad'').  The left panel is for the $500~\GeV < p_T < 600~\GeV$ sample while the right panel is for the entire $p_T$ range.  We see that using $\tilde{\tau}^{(1)}_N$ with the corresponding minimization axes (``lin'') gives the best performance.  Bottom row:  Changing $\beta$, but always using the axes that minimize $\tilde{\tau}^{(\beta)}_N$.  The jet broadening measure ($\beta = 1$) gives the best performance.}
  \label{fig:TopSigEff}
\end{figure}
 
We now quantify the performance of $\tau_3/\tau_2$ as a boosted top tagger.  In \Fig{fig:TopSigEff}, we show the effect of varying a cut on $\tau_3/\tau_2$ on the signal efficiency and background mistag rate using the BOOST2010 samples.  \Figs{fig:eff_500_axis}{fig:eff_all_axis} show curves for the $\beta = 1$ measure using different choices for the subjet axes, and the best performance is obtained for the axes that minimize $\tau^{(1)}_N$.  \Figs{fig:eff_500_beta}{fig:eff_all_beta} show the effect of changing the angular weighting exponent $\beta$, using the axes that minimize  $\tilde{\tau}^{(\beta)}_N$, and the best performance is obtained for $\beta = 1$.  Thus, the best tagging performance for $\tau_3/\tau_2$ is achieved by using the minimum axes with $\beta = 1$ (the jet broadening measure).\footnote{Though not shown, it is indeed the case that all other axes/measure combinations perform worse than the minimum $\beta = 1$ axes with the $\beta =1$ measure.}  When compared to other tagging methods in \Fig{fig:executiveSummary}, $N$-subjettiness shines as a boosted top tagger (at least on the BOOST2010 benchmark samples).

\begin{table}[tp]
  \begin{center}
 
    \begin{tabular}{|r ||r@{ : }lr@{ : }lr@{ : }l|}
     \hline
     $p_T$ range (GeV) & \multicolumn{2}{c}{200--300} & \multicolumn{2}{c}{300--400} & \multicolumn{2}{c|}{400--500}  \\ \hline \hline
     No $\tau^{(1)}_3/\tau^{(1)}_2$ cut &.069 &.0054&.34 &.034 &.59 &.092    \\
     $\tau^{(1)}_3/\tau^{(1)}_2 < 0.8$  &.065 &.0049&.32 &.027 &.58 &.073   \\
     $\tau^{(1)}_3/\tau^{(1)}_2 < 0.6$  &.049 &.0016&.26 &.0085&.47 &.022  \\
     $\tau^{(1)}_3/\tau^{(1)}_2 < 0.4$  &.026 &.0003&.15 &.0011&.25 &.0033   \\
     $\tau^{(1)}_3/\tau^{(1)}_2 < 0.3$  &.012 &.0000&.079&.0003&.12 &.0006 \\
     \hline
    \multicolumn{7}{c}{}\\
    \hline
     $p_T$ range (GeV) & \multicolumn{2}{c}{500--600} & \multicolumn{2}{c}{600--700} & \multicolumn{2}{c|}{700--800} \\ \hline \hline
     No $\tau^{(1)}_3/\tau^{(1)}_2$ cut &.71 &.14  &.75 &.17  &.75 &.19   \\
     $\tau^{(1)}_3/\tau^{(1)}_2 < 0.8$  &.70 &.11  &.74 &.13  &.74 &.14   \\
     $\tau^{(1)}_3/\tau^{(1)}_2 < 0.6$  &.55 &.028 &.57 &.034 &.57 &.035  \\
     $\tau^{(1)}_3/\tau^{(1)}_2 < 0.4$  &.28 &.0039&.27 &.0040&.27 &.0044  \\
     $\tau^{(1)}_3/\tau^{(1)}_2 < 0.3$  &.13 &.0005&.13 &.0011&.12 &.0008 \\
     \hline
    \end{tabular}
      \end{center}
  \caption{
Efficiencies vs.\ mistag rates for top jets : QCD jets for each of the \texttt{HERWIG} parton $p_T$ subsamples.  The top row corresponds to just applying the $m_{\rm top}$ invariant mass window (160 GeV to 240 GeV), and the subsequent rows include additional $\tau^{(1)}_3/\tau^{(1)}_2$ cuts.  Once the top quarks have sufficient $p_T$ for their decay products to be collimated, $N$-subjettiness exhibits fairly uniform performance as a function of $p_T$.}
  \label{tab:slices}
\end{table}

In \Tab{tab:slices}, we show the top tagging efficiency versus QCD mistag rate as a function of jet $p_T$ for different cuts on $\tau^{(1)}_3/\tau^{(1)}_2$.  At low $p_T$ (200--400 GeV), the efficiency for finding a jet within the top mass window is quite small, as an $R=1.0$ jet is unlikely to capture all of the top decay products.  For higher $p_T$ ranges (400--800 GeV), the efficiency is remarkably stable as a function of the jet $p_T$.  The cut $\tau^{(1)}_3/\tau^{(1)}_2 < 0.6$ yields approximately a 50\% efficiency operating point, while $\tau^{(1)}_3/\tau^{(1)}_2 < 0.4$ yields approximately a 25\% efficiency operating point.

\begin{table}[!t]
\begin{center}
\begin{tabular}{|c||c|c|c|c|}
\hline
        \multicolumn{5}{|c|}{{\sc HERWIG} results}  \\ \hline
               &  eff.  & mistag & eff.  & mistag \\
Tagger         &  ( \% ) & rate ( \% ) & ( \% ) & rate ( \% ) \\ \hline \hline
Hopkins \cite{Kaplan:2008ie}      & 20  &  0.4 $\pm$ 0.02  & 50  &  4.9 $\pm$ 0.06 \\ 
CMS \cite{CMS-PAS-JME-09-001,CMS-PAS-EXO-09-002,Rappoccio:1358770}         & 20  &  0.4 $\pm$ 0.02  & 50  & 5.2 $\pm$ 0.06  \\ 
Pruning \cite{Ellis:2009su,Ellis:2009me}     & 20  &  0.3 $\pm$ 0.02  & 50  &  7.6 $\pm$ 0.08  \\
ATLAS \cite{ATL-PHYS-PUB-2010-008,ATL-PHYS-PUB-2009-081,Brooijmans:1077731}       & 20  &  0.7 $\pm$ 0.02  & 50  &  4.6 $\pm$ 0.06  \\   
T/W  \cite{Thaler:2008ju} & 20  &  1.5 $\pm$ 0.04  & 50 & 6.0 $\pm$ 0.07 
\\ \hline
$\tau_3/\tau_2$ & 20 & 0.23 $\pm$ 0.01 & 50 & 4.12 $\pm$ 0.06 \\
Multivariate $\tau_N$ & 20 & 0.18 $\pm$ 0.01 & 50 & 2.96 $\pm$ 0.05 \\
     \hline
    \multicolumn{5}{c}{}\\
    \hline
                 \multicolumn{5}{|c|}{{\sc PYTHIA} results}  \\  \hline
              &  eff.  & mistag & eff.  & mistag \\
Tagger         &  ( \% ) & rate ( \% ) & ( \% ) & rate ( \% ) \\ \hline \hline
Hopkins      &  20       &   0.2  $\pm$ 0.01    &  47     &   3.2  $\pm$ 0.05    \\ 
CMS          &  22   &  0.3 $\pm$ 0.01 & 49  &  3.5 $\pm$ 0.05 \\
Pruning      &  19   & 0.2  $\pm$ 0.01  &   49    & 4.5   $\pm$ 0.06     \\
ATLAS          &  18   &  0.5 $\pm$ 0.02 & 49 &  3.1 $\pm$ 0.05 \\     
T/W  &  18   &  0.8 $\pm$ 0.02 & 57  &  7.0 $\pm$ 0.08 \\ 
\hline
$\tau_3/\tau_2$ & 18 & 0.14 $\pm$ 0.01 & 49 & 2.63 $\pm$ 0.05 \\
Multivariate $\tau_N$ & 18 & 0.11 $\pm$ 0.01 & 48 & 1.84 $\pm$ 0.04 \\
\hline
\end{tabular}
\end{center}
\caption{Summary of tagging efficiencies vs.\ mistag rates at different working points for a number of top taggers, including the $\tau_3/\tau_2$ cut in \Sec{sec:results} and the multivariate $\tau_N$ method in \Sec{sec:multivariateMethods}. The performance numbers for the other taggers are taken from \Ref{Abdesselam:2010pt} and described in more detail there.  The parameters are chosen such that all taggers run at 20\% and 50\% efficiency for the \texttt{HERWIG} samples, and the same parameters are then applied to the \texttt{PYTHIA} sample. Statistical errors on the mistag rate are indicated, and the efficiency numbers have uncertainties of 0.1\%.  There are systematic shifts in the tagging efficiencies and mistag rates between the two Monte Carlo programs, though the discrepancies for $N$-subjettiness tagging are similar in magnitude as those for other tagging methods. \label{tab:taggersummary}}
\end{table}

Finally, since boosted top identification depends on the precise radiation pattern within a jet, it is subject to potentially large uncertainties from Monte Carlo modeling of the parton shower, hadronization, and underlying event.   An indication of these uncertainties can be seen in \Tab{tab:taggersummary}, which compares the various top tagging algorithms on the \texttt{HERWIG} and \texttt{PYTHIA} samples.  The relative performance between different algorithms is consistent between the programs, though the absolute fake rates in the \texttt{PYTHIA} sample are significantly smaller.  We conclude that while the tagging performance of $N$-subjettiness does have Monte Carlo modeling dependence, it is no worse than for other proposed tagging methods.

\subsection{Multivariate Methods}
\label{sec:multivariateMethods}

In the previous subsection, we saw that a simple cut on the ratio $\tau_3^{(1)}/\tau_2^{(1)}$ and a fixed top mass window $160~\GeV < m_{\rm jet} < 240~\GeV$ yielded impressive top tagging performance.  Here, we will explore whether multivariate classification might be able to further optimize tagging performance.  One feature of $N$-subjettiness is that the entire set of $\tau_N^{(\beta)}$ values for various choices of $N$ and $\beta$ can be calculated on a jet-by-jet basis and then used as inputs to multivariate methods.\footnote{In addition, $\tau_N$ is a sum over sub-$\tau_1$ values for each Voronoi region.   These sub-$\tau_1$ values could be used as well, though we did not find any particular gain when including them in our multivariate studies.}  In principle, $N$-subjettiness could be used in tandem with some of the other top tagging methods, though we have not done a systematic study of that possibility.  

\begin{figure}[tp]
  \begin{center}
    \subfigure[]{\label{fig:densTau32}\includegraphics[trim = 0mm 0mm 0mm 0mm, clip,width = 0.48\textwidth]{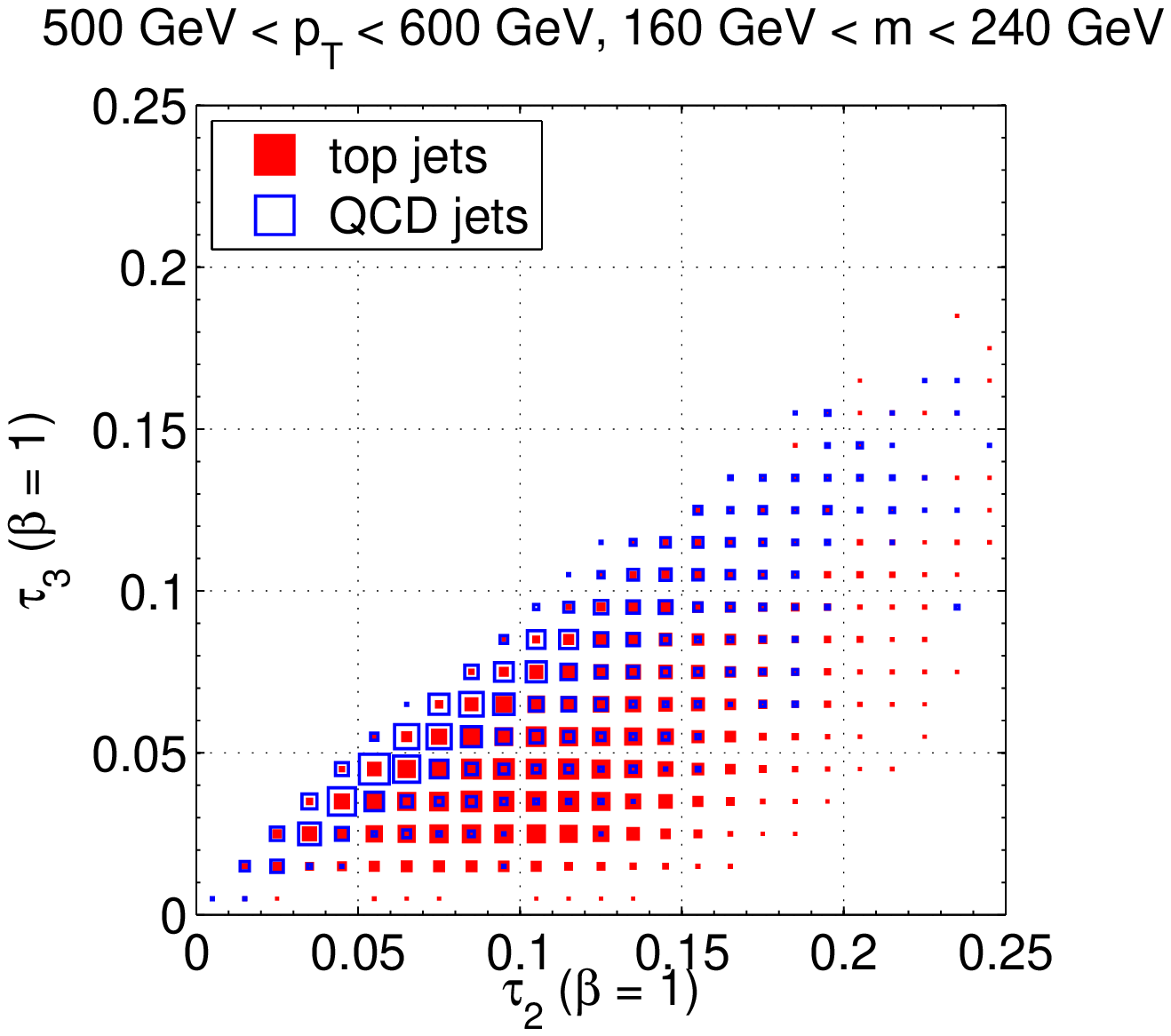}} \;\;\;
    \subfigure[]{\label{fig:densTau3221}\includegraphics[trim = 0mm 0mm 0mm 0mm, clip,width = 0.48\textwidth]{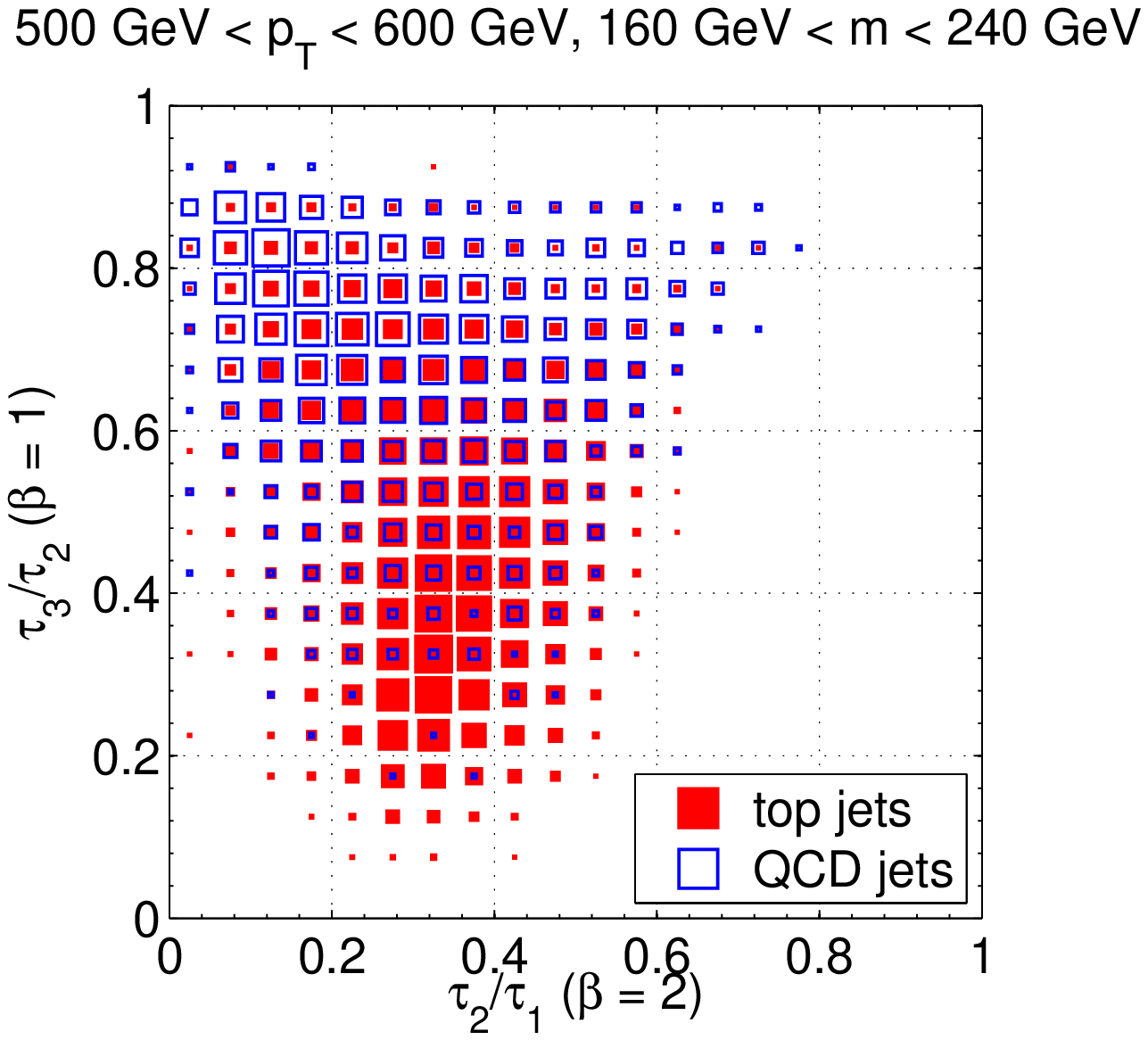}}
  \end{center}
  \vsh
  \caption{Density plots in the (a) $\tau^{(1)}_2$ vs.\ $\tau^{(1)}_3$ plane and (b) $\tau^{(2)}_2 / \tau^{(2)}_1$ vs. $\tau^{(1)}_3 / \tau^{(1)}_2$ plane for boosted top and QCD jets.  The selection criteria are the same as in \Fig{fig:tauHistograms2Axes}.  While a linear cut on the $\tau^{(1)}_3 / \tau^{(1)}_2$ ratio is clearly an effective boosted top tagger, there is additional information in other $N$-subjettiness variables that can be used in a multivariate method.}
  \label{fig:Top2Dtau123}
\end{figure}

To motivate the potential power of multivariate methods, consider \Fig{fig:Top2Dtau123}.  In \Fig{fig:densTau32}, we show distributions on the $\tau_3^{(1)}$ vs.\ $\tau_2^{(1)}$ plane, which demonstrate (as expected) that $\tau_3^{(1)}/\tau_2^{(1)}$ is a better discriminant variable than $\tau_3^{(1)}$ or $\tau_2^{(1)}$ alone.  We saw in \Fig{fig:t2t1_pt0500_a200} that the ratio $\tau^{(2)}_2/\tau^{(2)}_1$ might also have discriminating power, but square cuts on that ratio did not yield much improvement over using $\tau_3^{(1)}/\tau_2^{(1)}$ alone.  However, in \Fig{fig:densTau3221} we see that top jets and QCD jets are fairly well-separated in the $\tau^{(2)}_2/\tau^{(2)}_1$ vs.\ $\tau^{(1)}_3/\tau^{(1)}_2$ plane, and multivariate analyses can capitalize on these kinds of correlations.

Though there are a variety of multivariate classification methods one could study, we will focus on a (modified) linear Fisher discriminant \cite{fisher,anderson_bahadur}, since it is a straightforward extension of the one-dimensional one-sided cut $\tau_3^{(1)}/\tau_2^{(1)} < c$.  The Fisher discriminant uses multi-dimensional information to form a linear one-sided cut:
\be
\vec{L} \cdot \vec{X} < c,
\ee
where $\vec{X}$ is a vector of jet characteristics, such as $N$-subjettiness values and jet mass, while $\vec{L}$ encodes the linear weights of each of these attributes.  In a geometric language, the goal of a Fisher discriminant is to find the set of parallel hyperplanes (each one encoded by their own $c$ parameter) defined by one normal vector $\vec{L}$ which best separate the top jet signal from the QCD jet background in the space $\mathbb{R}^{\dim(X)}$.  

The standard (but not necessarily optimal) way to choose $\vec{L}$ is to take the $\vec{L} \cdot \vec{X}$ distributions for two different classes, and calculate the $\vec{L}$ which maximizes the variance between the two classes compared to the variance within each class.  We will instead use a modified Fisher discriminant \cite{anderson_bahadur}, which is better suited for defining efficiency/rejection curves:  
\be
\label{eq:fishersolution}
\vec{L} = (\Sigma_{\mathrm{QCD}} + \gamma \Sigma_{\mathrm{top}})^{-1} (\vec{\mu}_{\mathrm{QCD}} -  \vec{\mu}_{\mathrm{top}}),
\ee
where $\Sigma$ are the covariance matrices and $\vec{\mu}$ are the mean vectors for the variables in $\vec{X}$.  The standard Fisher discriminant takes $\gamma = 1$, but we found that lower values of $\gamma$ were preferred for tagging performance, especially if $\tau_3^{(1)}/\tau_2^{(1)}$ is one of the variables in $\vec{X}$.

As an instructive example, we include the following $N$-subjettiness variables and ratios to define a discriminant:
\be
\tau_1^{(1)}, \quad \tau_2^{(1)}, \quad \tau_3^{(1)}, \quad \frac{\tau_2^{(1)}}{\tau_1^{(1)}}, \quad \frac{\tau_3^{(1)}}{\tau_2^{(1)}},  \quad \frac{\tau_2^{(2)}}{\tau_1^{(2)}}, \quad \frac{\tau_3^{(2)}}{\tau_2^{(2)}}.
\ee
 In addition, we include jet mass information in the variables
 \be
\left(\frac{\Delta m}{m}\right)_{+} \equiv \max \left\{ \frac{m_{\mathrm{jet}} - m_{\mathrm{top}}}{m_{\mathrm{top}}},0 \right\}, \qquad \left(\frac{\Delta m}{m}\right)_{-} \equiv \min \left\{ \frac{m_{\mathrm{jet}} - m_{\mathrm{top}}}{m_{\mathrm{top}}},0 \right\},
 \ee
 where we have separated out jet mass values above and below the top mass in order to effectively adjust the top mass window.  Note that if the goal is to obtain a function $\vec{L} \cdot \vec{X}$ which works for a variety of operating efficiencies, the Fisher discriminant is not always improved by adding additional variables.  In particular, we found more uniform performance across the whole efficiency/rejection curve by not including $\tau^{(2)}_N$ values nor jet $p_T$.  We also found that performance was improved by first applying a uniform cut on jet mass $160~\GeV < m_{\mathrm{jet}} < 280~\GeV$.\footnote{Including jets with masses below 160 GeV threw off the determination of $\vec{L}$ and thus the tagging efficiency (even though we use jet mass information in $\vec{L}$).  We suspect that QCD jets at lower jet masses have a distinctly different covariance matrix, making it difficult to obtain high-purity signal separation.}

 \begin{table}[tp]
\begin{center}
\begin{tabular}{|c||c|c|}
\hline
Tagger   & \hspace{.4in}20\% Efficiency  \hspace{.4in} &  \hspace{.4in}    50\% Efficiency \hspace{.4in}  \\
\hline
\hline
\multirow{3}{*} {$\tau_3$/$\tau_2$}   & \multicolumn{2}{|c|}{$160~\GeV < m_{\mathrm{jet}} < 240~\GeV$}  \\ 
                                                          & \multicolumn{2}{|c|}{$\tau_3/\tau_2 < c$}  \\ 
\cline{2-3}
  					& c = 0.384                                      & c = 0.680 \\ 
\hline
\hline
\multirow{5}{*} {multivariate $\tau_N$}   & \multicolumn{2}{|c|}{$160~\GeV < m_{\mathrm{jet}} < 280~\GeV$}        \\ 
                  & \multicolumn{2}{|c|}{$\vec{L} \cdot \vec{X} < c$}        \\ 
					& \multicolumn{2}{|c|}{$\vec{L}$ = [2.30, -5.85, -1.89, 6.21, 7.25, -5.35, -0.86, 1.61, -14.07]} \\ 
					& \multicolumn{2}{|c|}{$\vec{X} = \left[\tau_1^{(1)}, \tau_2^{(1)} , \tau_3^{(1)} , \frac{\tau_2^{(1)}}{\tau_1^{(1)}} , \frac{\tau_3^{(1)}}{\tau_2^{(1)}} , \frac{\tau_2^{(2)}}{\tau_1^{(2)}} , \frac{\tau_3^{(2)}}{\tau_2^{(2)}} , \left(\frac{\Delta m}{m}\right)_{+} , \left(\frac{\Delta m}{m}\right)_{-} \right]$} \\ 
\cline{2-3}
                                        & c = 3.51                                      			   & c = 5.29           \\
\hline
\end{tabular}
\caption{Optimized parameters for the $N$-subjettiness taggers at different working points.  These parameters are used for the results in \Tab{tab:taggersummary}.  Both the $\tau_3$/$\tau_2$ method and the multivariate method make use of a linear one-sided cut.  The parameters for the modified Fisher discriminant were obtained from \Eq{eq:fishersolution} with $\gamma = 0.7$.
\label{tab:taggerpars}}
\end{center}
\end{table}

 \begin{figure}[tp]
  \begin{center}
      \includegraphics[trim = 0mm 0mm 0mm 0mm, clip,width = 0.65\textwidth]{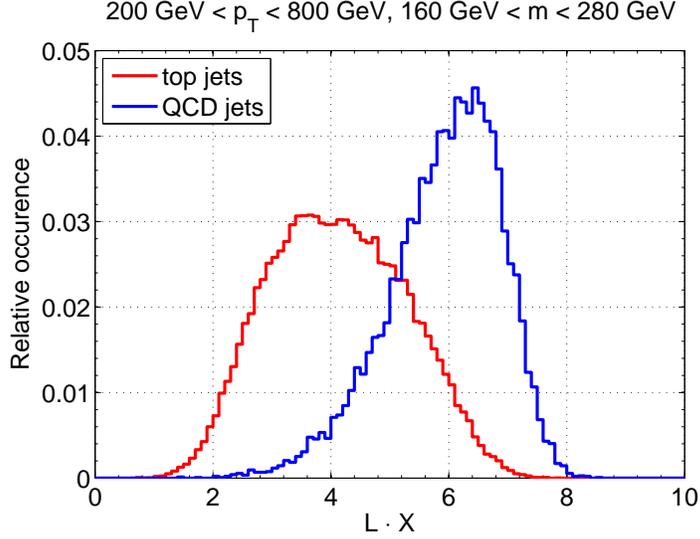}
  \end{center}
  \vsh
  \caption{Plot of the linear discriminant $\vec{L} \cdot \vec{X}$ for boosted top and QCD jets.  The value of $\vec{L}$ is given in \Tab{tab:taggerpars}, and a top mass window $160~\GeV < m_{\rm jet} < 280~\GeV$ has already been applied.  This linear discriminant has more separation power than the $\tau^{(1)}_{3}/\tau^{(1)}_{2}$ variable in \Fig{fig:t3t2_pt0500_a100}.}
  \label{fig:FisherDistribution}
\end{figure}
 
 \begin{figure}[tp]
  \begin{center}
    \subfigure[]{\label{fig:fisher_eff_allpt}\includegraphics[trim = 0mm 0mm 0mm 0mm, clip,width = 0.49\textwidth]{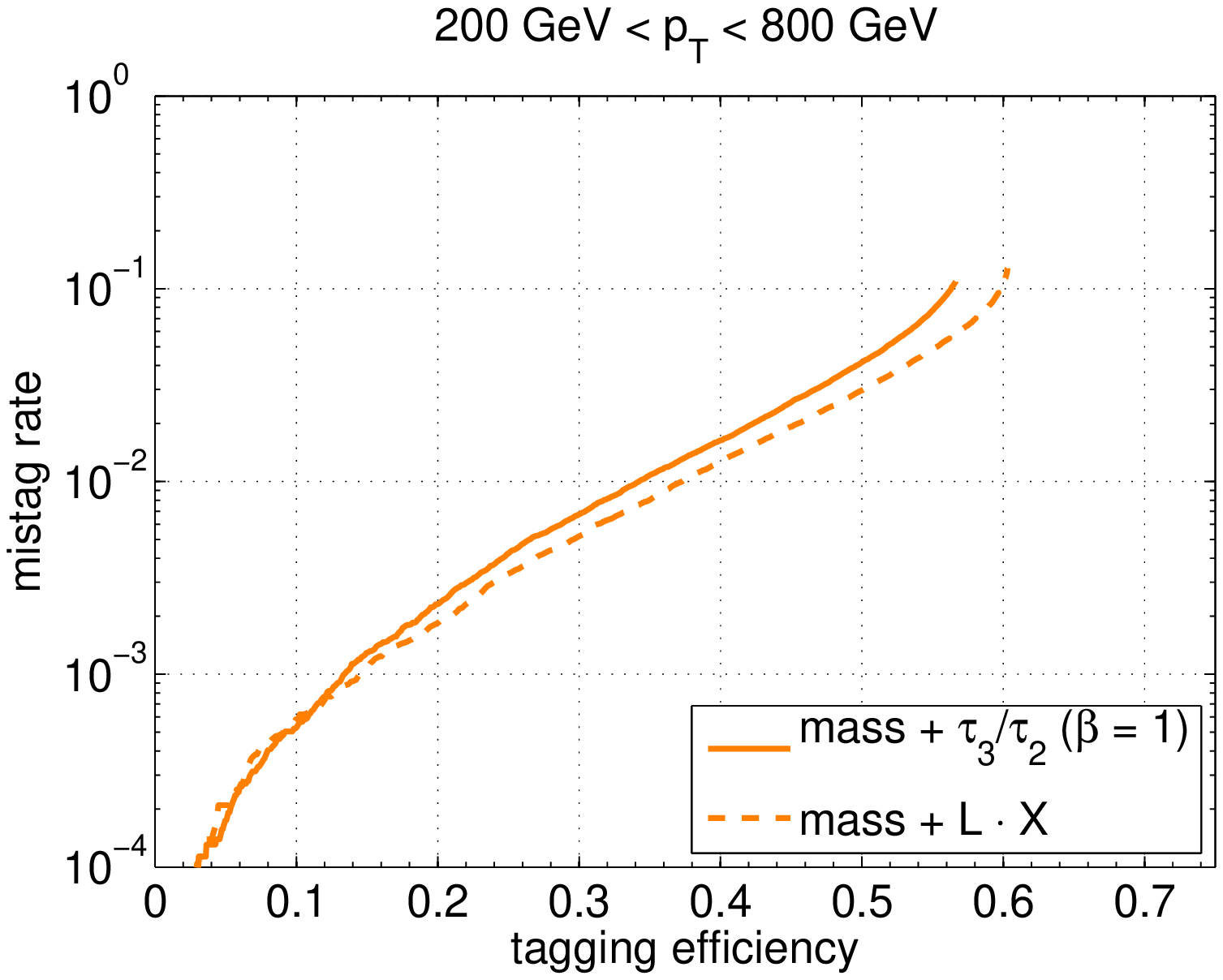}}
    \subfigure[]{\label{fig:fisher_eff_500}\includegraphics[trim = 0mm 0mm 0mm 0mm, clip,width = 0.49\textwidth]{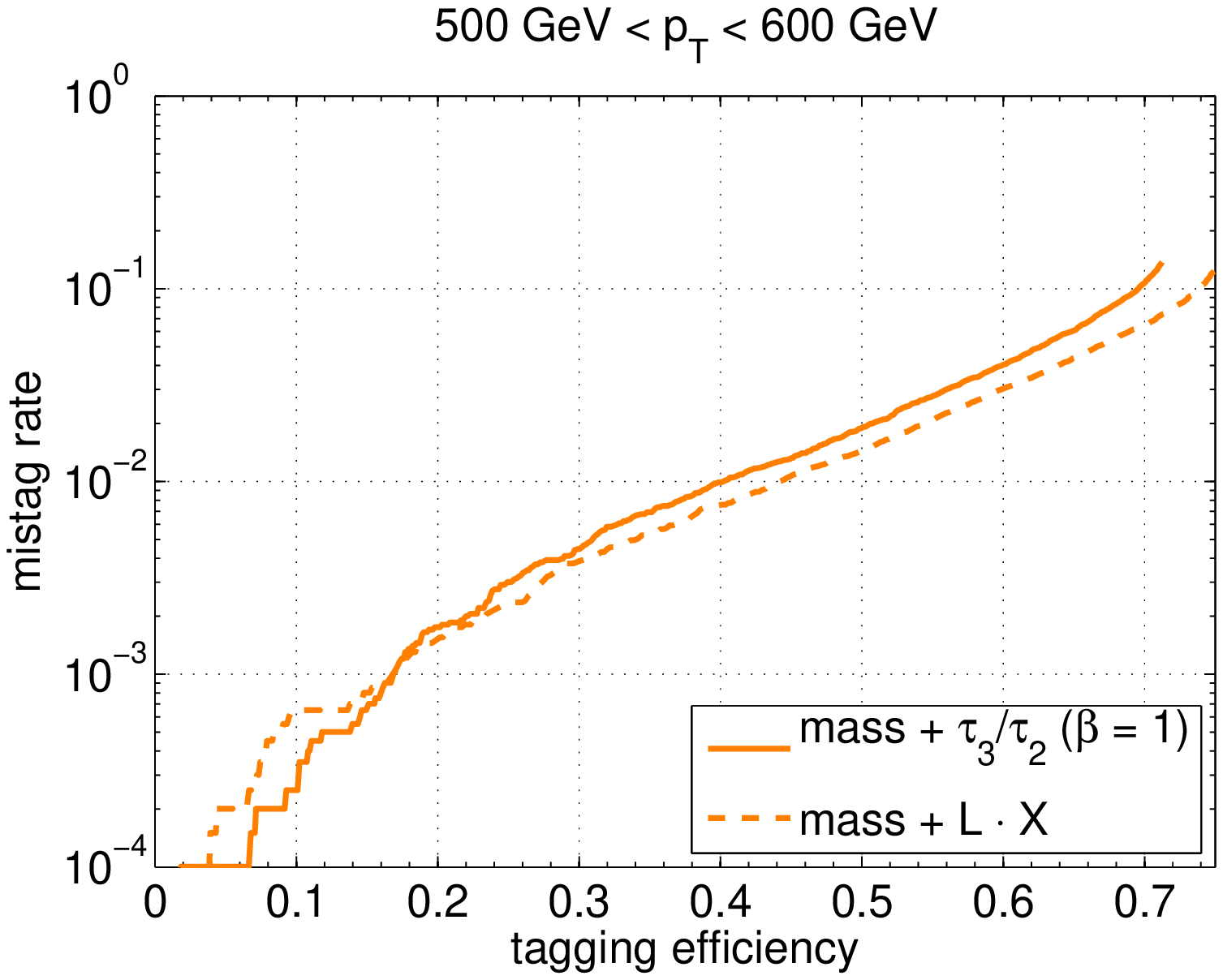}}
  \end{center}
  \vsh
  \caption{Efficiency/mistag curves for the linear discriminant.  Here, we compare a cut on $\tau^{(1)}_{3}/\tau^{(1)}_{2}$ to a cut on $\vec{L} \cdot \vec{X}$, and find roughly a 20\% improvement in the mistag rate for fixed efficiency, though the simple $\tau^{(1)}_{3}/\tau^{(1)}_{2}$ cut performs better as very small mistag rate.  These are the same curves that appear in \Fig{fig:executiveSummary}, albeit with a different range on the vertical axis to highlight the small mistag rate region.}
  \label{fig:FisherEfficiency}
\end{figure}

Applying \Eq{eq:fishersolution} with $\gamma = 0.7$, we found the linear coefficients $\vec{L}$ listed in \Tab{tab:taggerpars}.  A plot of $\vec{L} \cdot \vec{X}$ appears in \Fig{fig:FisherDistribution} which shows excellent signal/background separation.  With these parameters, a sliding cut on $\vec{L} \cdot \vec{X}$ defines the efficiency curves in \Fig{fig:FisherEfficiency}.  Compared to the simple $\tau_3^{(1)}/\tau_2^{(1)}$ cut in \Sec{sec:results}, we see about a 20\% decrease in the mistag rate for fixed efficiency.  \Tab{tab:taggersummary} tests the performance of the Fisher discriminant between \texttt{HERWIG} and \texttt{PYTHIA}, and we see that the relative improvement in using a multivariate method is consistent between the programs.  Finally, we compare the multivariate $\tau_N$ selection to other top taggers in \Fig{fig:executiveSummary}, where we again see fantastic performance.

\section{N-jettiness as a Jet Algorithm}
\label{sec:jetAlgorithm}

The minimization procedure in \Sec{sec:minimizationProcedure} for individual jets can be extended to the full event, thus minimizing (some version of) $N$-jettiness \cite{Stewart:2010tn}.  The axes and regions determined by this procedure would then define a jet algorithm, with the important caveat that the number of jets is fixed at $N$.  Here, we briefly sketch how such an algorithm might be used at a hadron collider, leaving a more complete study to future work.

\subsection{Previous Literature}
\label{sec:previousliterature}

The idea of using minimization to define jets is not new, and we will briefly mention some of the previous literature.  Cluster optimization is a rich area of study in computer science, but has only been used in limited cases for jet physics.

The most basic example is in $e^+ e^-$  collisions, where the axis which minimizes thrust is used to define a hemisphere jet algorithm.  To our knowledge, the first use of the $k$-means clustering algorithm as a jet finder was given in \Ref{Chekanov:2005cq}, where two all-hadronic channels with fixed jet multiplicity were studied, $e^+ e^- \rightarrow t \bar{t}$ ($N=6$) and $e^+ e^- \rightarrow W^+W^-$ ($N = 4$).  Jet finding can be seen more generally as an optimization problem, with different optimization measures proposed in \Refs{Berger:2002jt,Angelini:2002et,Angelini:2004ac,Grigoriev:2003yc,Grigoriev:2003tn,Lai:2008zp,Volobouev:2009rv}.

The most well-known application of minimization for jets is in iterative cone finding.  Stable cones are cones for which the jet axis and the jet 3-momentum are aligned.  The procedure outlined in \Ref{Ellis:2001aa} for finding stable cones (in the $p_T$ scheme) is equivalent to minimizing
\be
\label{eq:1subR0}
\tau_1^{(2)}(R_0) = \sum_i p_{T, i} \min \{ \Delta R_{1,i}^2, R_0^2 \},
\ee
which the reader will recognize as (unnormalized) 1-subjettiness with the thrust measure ($\beta = 2$) and an additional $R_0$ cutoff.  The choice $\beta = 2$ is important since only then does the minimization criterion for $\tau_1(R_0)$ enforce jet axis/momentum alignment (see \Eq{eq:minfinderbeta2}).  Of course, stable cone finding by itself is not sufficient for defining a jet algorithm because cones generically overlap, so iterative cone finding is usually augmented with a split-merge procedure.

Note that $\tau^{(2)}_1(R_0)$ does not have a monotonically increasing first derivative in radial directions, so it generically has many local minima (see \Sec{sec:uniqueness}), leading to the famous problems of infrared safety of seeded cone algorithms (see \Refs{Ellis:2007ib,Salam:2009jx} for a review).  Interestingly, the anti-$k_T$ algorithm \cite{Cacciari:2008gp} acts like an idealized cone algorithm when applied to only the hardest jet in an event, and it tends to do an excellent job of minimizing $\tau^{(2)}_1(R_0)$ without any seeding problems.

\subsection{Extension to N-jettiness}

It is now straightforward to extend \Eq{eq:1subR0} to $N$-jettiness, and thereby use minimization to define a fixed $N$ jet cone algorithm for hadronic collisions.  Including both a jet radius cutoff $R_0$ and a beam pseudorapidity cut $\eta_0$, one possible definition of $N$-jettiness is: 
\be
\tau^{(\beta,\gamma)}_N(R_0, \eta_0) = \sum_i p_{T,i} \min \left\{  \left(\exp \frac{-\eta_i}{\eta_0} \right)^{\gamma}, \left(\exp \frac{\eta_i}{\eta_0} \right)^{\gamma}, \left(\frac{\Delta R_{1,i}}{R_0}\right)^\beta, \ldots,  \left(\frac{\Delta R_{N,i}}{R_0}\right)^\beta, 1\right\}.
\ee
The first two entries in the minimum are ``beam measures'', the next $N$ entries are ``jet measures'', and the final entry defines unclustered momentum.  The exponents $\gamma$ and $\beta$ are angular weighting exponents for the beam measure and the jet measure, respectively, and the choice $\beta = 2$ and $\gamma = \infty$ is similar in spirit to traditional iterative cone finding with hard cutoffs on $R_0$ and $\eta_0$.  

While there are a number of different distance measures that could be used to define $N$-jettiness, this one is well-suited for hadronic collisions, since it is boost invariant along the beam axis and yields circular cones in rapidity/azimuth.  The quantity $\tau_N(R_0, \eta_0)$ corresponds roughly to unclustered $p_T$, so minimizing $\tau_N(R_0, \eta_0)$ is essentially maximizing the amount of radiation contained in $N$ cones.  Unlike iterative cone algorithms which require a split-merge procedure, minimizing $N$-jettiness automatically splits overlapping cones at the Voronoi edges.\footnote{It is also possible to further generalize the definition of $N$-subjettiness (and the minimization algorithm) to include ``fuzzy edges'' through partial assignment of particles to clusters.  Instead of using absolute Voronoi assignment, one could assign a particle to all clusters but with normalized weight factors that are negatively correlated with the distance to the respective cluster centers, similar to \Ref{Grigoriev:2003tn}.}  Of course, one could define the jets entirely by the Voronoi regions by taking $R_0$ to be very large.

The minimization procedure for $\tau_N(R_0, \eta_0)$ is nearly identical to \Sec{sec:minimizationAlgorithm} with one important change.  At each stage of the iteration, the only particles which  participate in the axes update step are those for which the jet measure is smallest.  In this way, the beam measure and $R_0$ affect which particles can be clustered into jets, but not the way in which they are clustered.  As in \Sec{sec:minimizationAlgorithm}, different values for $\beta$ require different update steps.

\begin{figure}[tp]
  \begin{center}
  \includegraphics[trim = 0mm 0mm 0mm 0mm, clip,width = 0.95\textwidth]{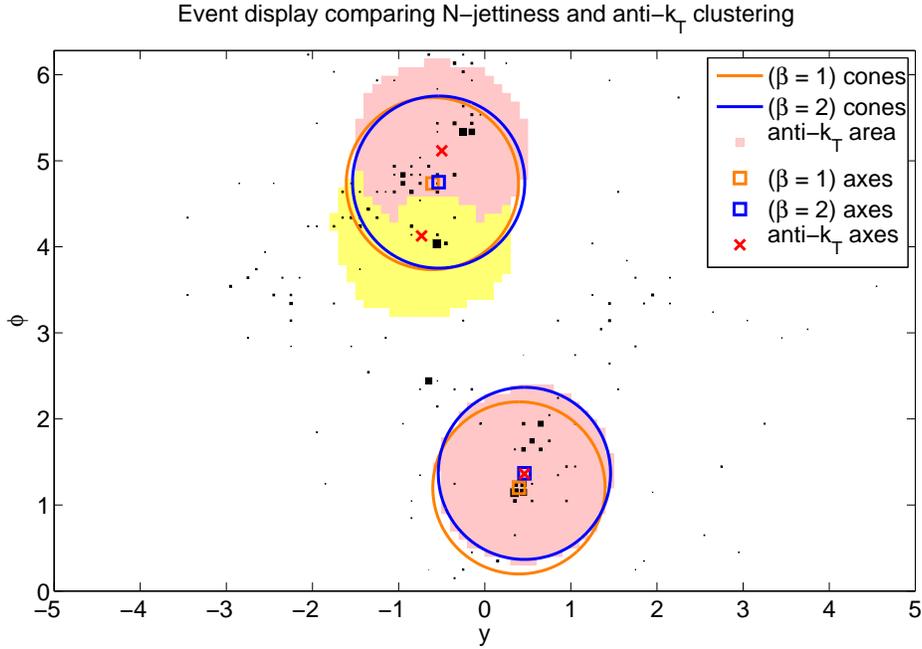}
\end{center}
\vsh
  \caption{A boosted $t \bar{t}$ event display comparing the 2-jettiness minimization procedure for $\beta = 1$ and $\beta = 2$ to the anti-$k_T$ jet algorithm.  All three methods use $R = 1.0$. 2-jettiness yields perfectly circular cones, while the two hardest anti-$k_T$ jets can be modified by the presence of a third jet.  The cluster of particles in the lower half of the figure is arranged into jets of $p_T = 233/231/231~\GeV$ for the three respective methods ($\beta = 1$/$\beta = 2$/anti-$k_T$).  The anti-$k_T$ and $\beta = 2$ axes are well-aligned for this jet, while the $\beta = 1$ axis is offset from the former two.  The cluster of particles in the top half of the figure has jets of $p_T = 235/226~\GeV$ for the $\beta = 1$/$\beta = 2$ cones and is split into two jets of $p_T = 167~\GeV$ (red) and $p_T = 103~\GeV$ (yellow) with the anti-$k_T$ algorithm.}
  \label{fig:comparisonAKTevent}
\end{figure}

\begin{figure}[tp]
  \begin{center}
      \subfigure[]{\label{fig:akt_Njet1_comp}\includegraphics[trim = 0mm 0mm 0mm 0mm, clip,width = 0.48\textwidth]{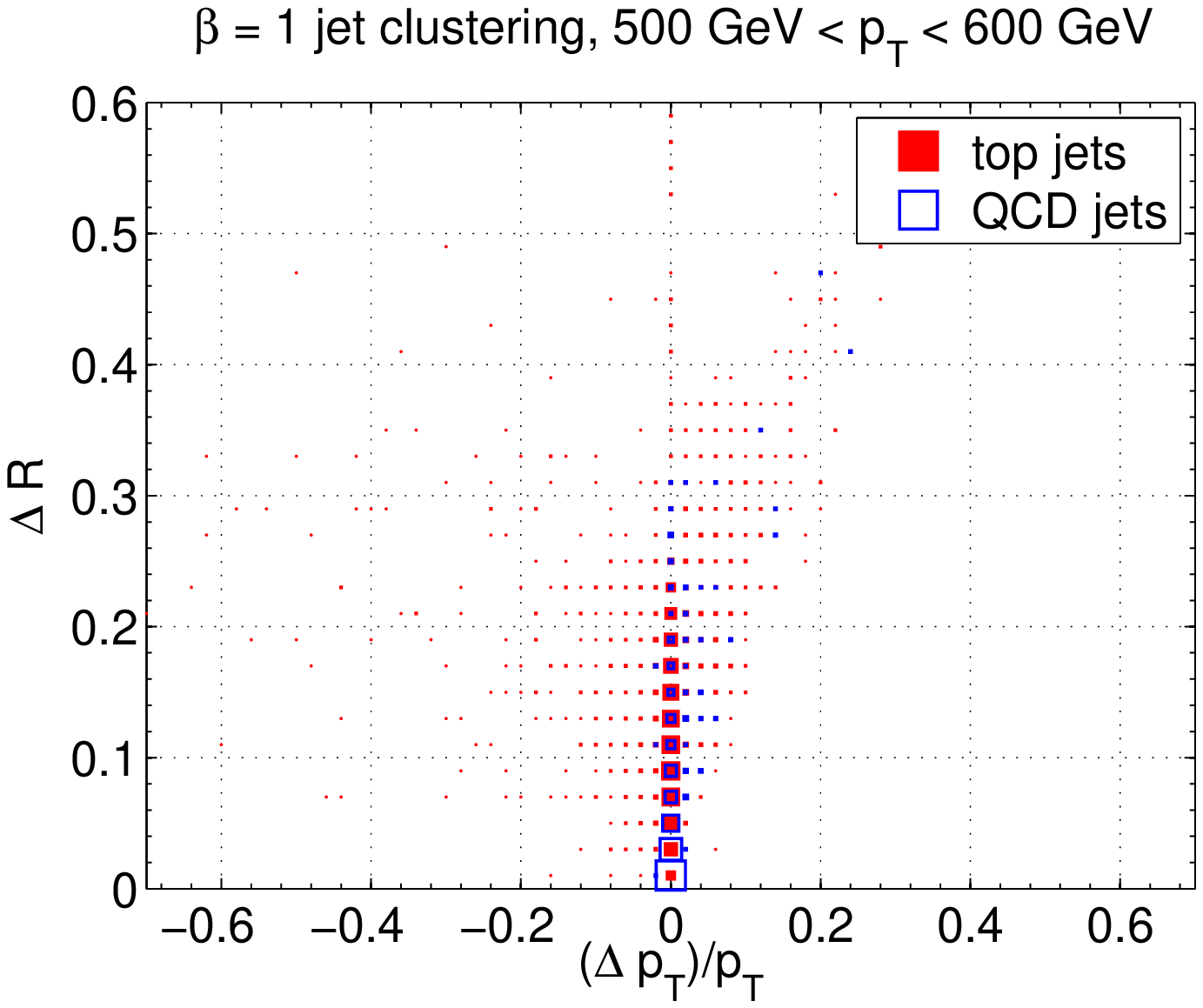}} \;\;\;
      \subfigure[]{\label{fig:akt_Njet2_comp}\includegraphics[trim = 0mm 0mm 0mm 0mm, clip,width = 0.48\textwidth]{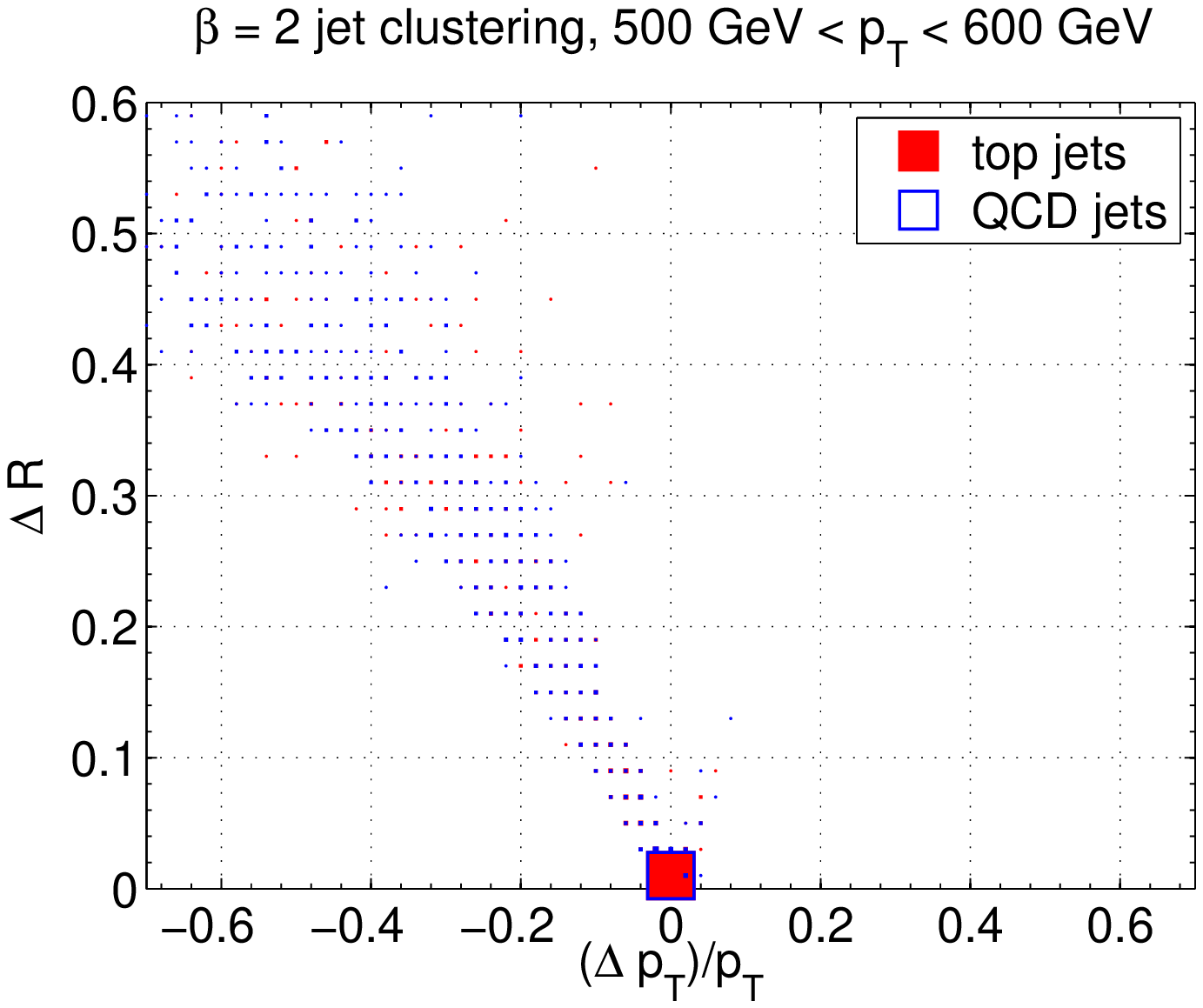}}
  \end{center}
  \vsh
  \caption{Comparison of the two hardest jets found with anti-$k_T$ to the jets found with $2$-jettiness minimization. The $500~\GeV < p_T < 600~\GeV$ event samples are used without any cut on the jet mass.  Shown is the $\Delta R$ difference in the jet axes compared to the fractional difference in $p_T$ (anti-$k_T$ minus $N$-jettiness, divided by anti-$k_T$).  Left:  $\tau_N$ minimization with $\beta = 1$.  The jet broadening measure does not require the jet axis to align with the momentum axis, but the resulting jets have comparable $p_T$.  Right:  $\tau_N$ minimization with $\beta = 2$.  Both the thrust measure and the anti-$k_T$ algorithm enforce jet axis/momentum alignment, yielding small $\Delta R$ separation.  There is, however, a tail region where a third jet is identified by anti-$k_T$, decreasing the $p_T$ of the second hardest jet.}
  \label{fig:comparisonAKT}
\end{figure}

\begin{figure}[tp]
  \begin{center}
      \subfigure[]{\label{fig:newtopmass200}\includegraphics[trim = 0mm 0mm 0mm 0mm, clip,width = 0.48\textwidth]{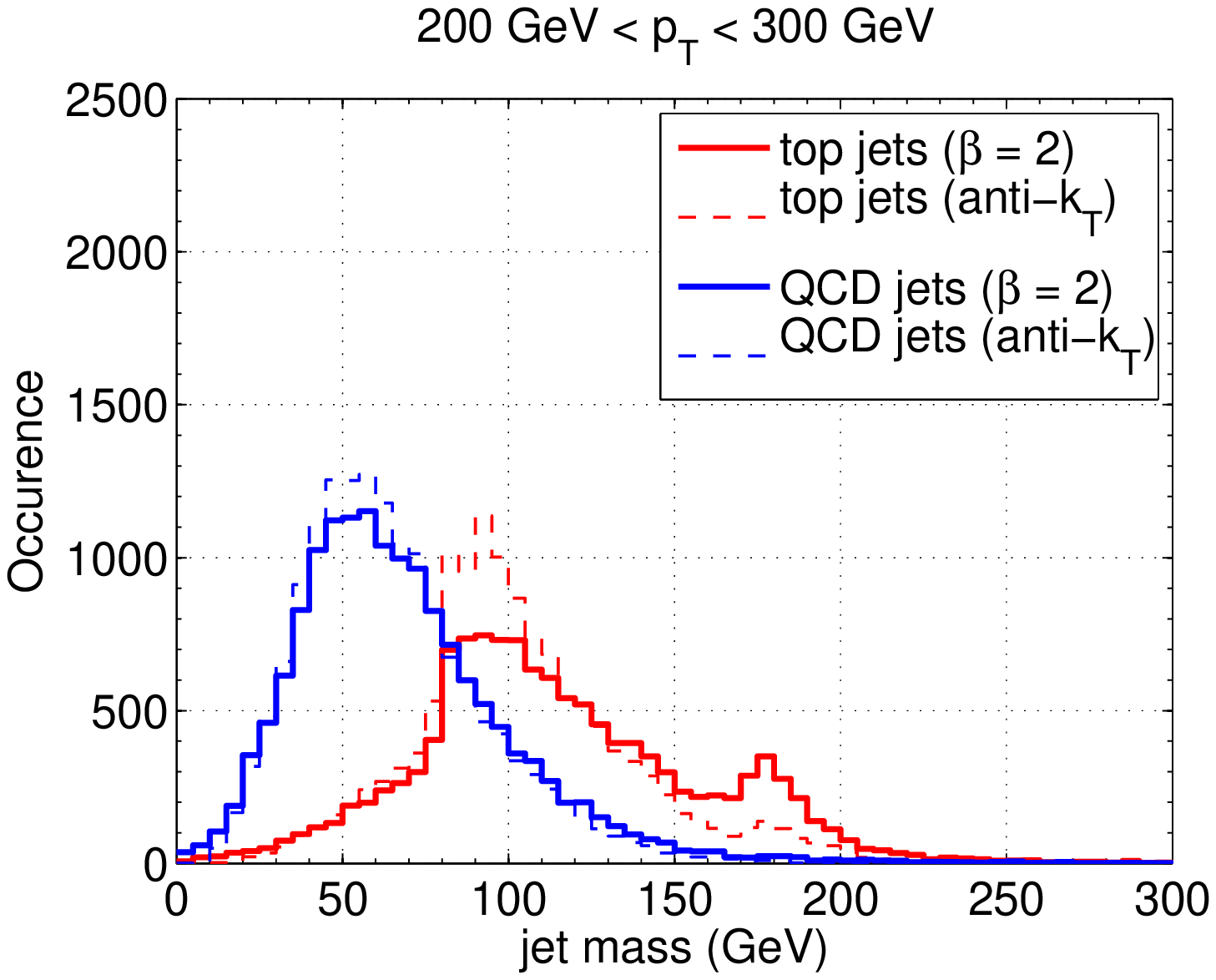}} \;\;\;
      \subfigure[]{\label{fig:newtopmass300}\includegraphics[trim = 0mm 0mm 0mm 0mm, clip,width = 0.48\textwidth]{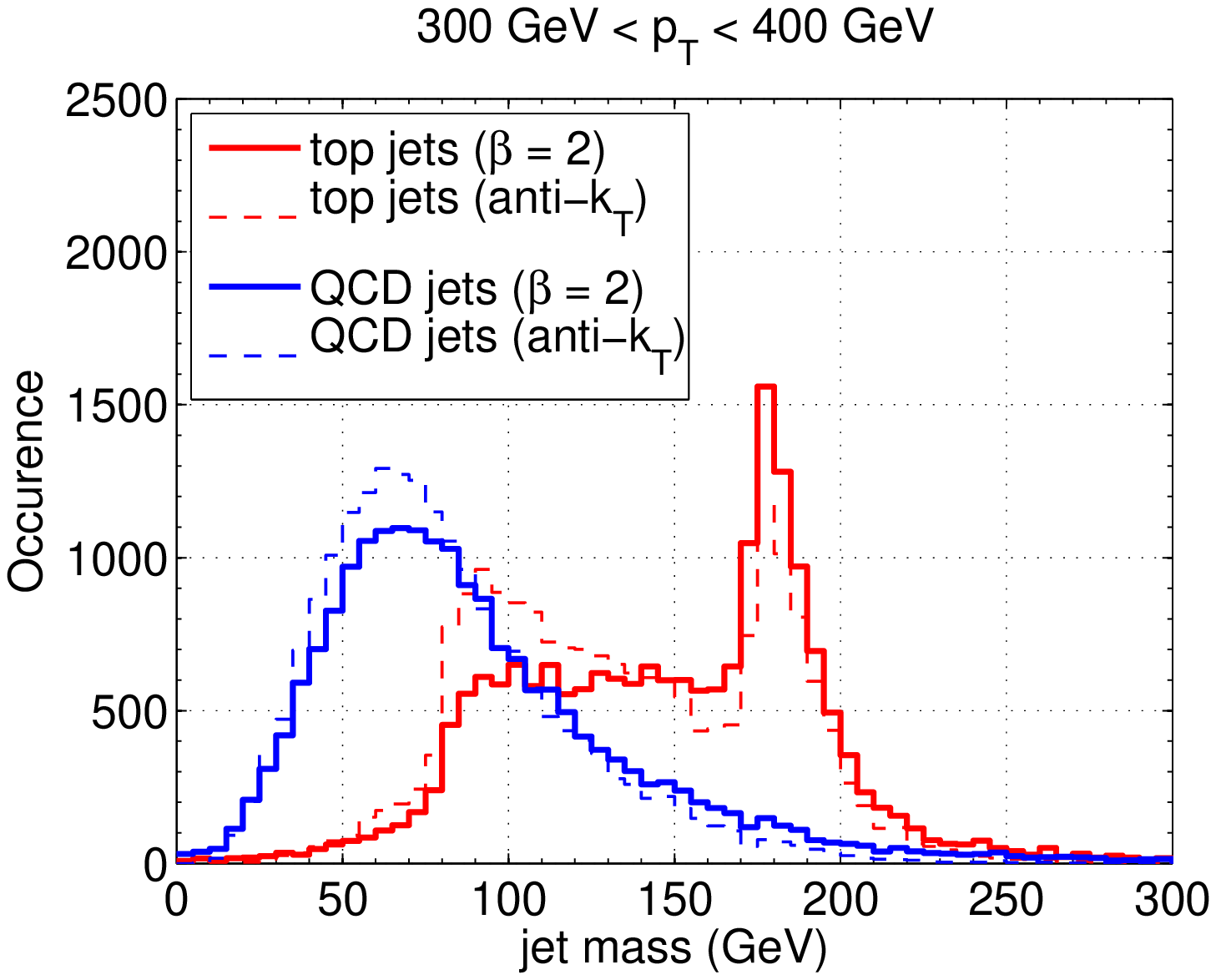}}
  \end{center}
  \vsh
  \caption{Invariant mass of jets found with 2-jettiness minimization and the anti-$k_T$ algorithm for low to moderate parton $p_T$.  For $200~\GeV < p_T < 300~\GeV$ in (a), the top mass peak is more than twice as prominent for 2-jettiness jets than for anti-$k_T$ jets. For $300~\GeV < p_T < 400~\GeV$ in (b), the beneficial effect on top mass reconstruction is less pronounced though still significant.  For parton $p_T$ greater than 400 GeV, where the decay products are more collimated, the effect disappears.  The 2-jettiness jets were found with $\beta = 2$ minimization with the two hardest anti-$k_T$ jets used as seeds (no noise added to the coordinates).}
  \label{fig:coneMass}
\end{figure}

In preliminary studies, we find that the jet regions determined by $N$-jettiness are very similar to the $N$ hardest jets returned by the anti-$k_T$ algorithm.\footnote{We in fact use the anti-$k_T$ jets (plus noise) as the seed axes for the minimization procedure.}  \Fig{fig:comparisonAKTevent} shows an event display where the anti-$k_T$ region for $R = 1.0$ is closely aligned with the Voronoi regions defined by $\tau^{(2,\infty)}_2$ with $R_0 = 1.0$ and $\eta_0 = 5.0$.  However, there is a crucial difference:  for any process with well-separated jets, 2-jettiness yields two perfect cones by definition, whereas the anti-$k_T$ jet areas can be modified by the presence of a nearby third jet (even if only two jets are studied).  For $\beta=1$, the jet axis can move substantially, though the actual jet constituents are quite similar.

We can quantify the difference between the anti-$k_T$ jets and the $N$-jettiness jets using the BOOST2010 samples.   As demonstrated in \Fig{fig:comparisonAKT}, the two hardest jets determined by anti-$k_T$ are closely aligned with the axes found by $2$-jettiness minimization with $\beta = 2$ ($\Delta R \lesssim 0.02$), and the $p_T$ of the resulting jets are quite similar ($|\Delta p_T| / p_T \lesssim 0.05$).  However, there is a tail to the distribution where the anti-$k_T$ jets have smaller $p_T$ than the $N$-jettiness jets, due to the presence of a nearby third jet.  As expected, there is a much larger change in the jet direction for  $\beta = 1$ (though the actual $p_T$ of the jet is rather stable), and this difference may be useful for studying jet systematics.  In particular, note that $\Delta p_T$ between anti-$k_T$ and $\beta = 1$ jets is roughly symmetric about zero.  For identifying moderately boosted tops, \Fig{fig:coneMass} shows how the top decay products are more likely to be clustered into the same jet with 2-jettiness minimization compared to the anti-$k_T$ algorithm.

\subsection{Discussion}

There are a number of potential benefits with using $N$-jettiness as a jet algorithm.  First, as advocated in \Ref{Stewart:2010tn}, $N$-jettiness is a way to define exclusive $N$-jet samples, and there is a growing interest in calculating (and resumming) $N$-jettiness distributions \cite{Stewart:2010pd,Berger:2010xi,Jouttenus:2011wh,Bauer:2011hj,Bauer:2011uc}.  Second, for inclusive $N$-jet samples, minimizing $\tau_N$ simultaneously determines the jet regions and gives a quality measure for the jet reconstruction (namely $\tau_N$ itself, corresponding roughly to unclustered $p_T$).  Third, unlike traditional iterative cone finding, $N$-jettiness automatically incorporates a ``split-merge step'' into the cone finding.  In particular, the stable cones found in $N$-jettiness include ``recoil'', meaning that (for $\beta = 2$) the jet axis and the jet 3-momentum are aligned even when cones collide, a behavior that is similar to anti-$k_T$.\footnote{Note that anti-$k_T$ and $N$-jettiness have different ways of dealing with overlapping cones, with $N$-jettiness splitting jets democratically in area while anti-$k_T$ preferentially allowing the harder jets to remain circular.}  Finally, the angular exponents $\beta$ and $\gamma$ have no known analog in traditional jet algorithms, and adjusting their values may be useful to test the robustness of jet finding.  The jet broadening measure ($\beta = 1$) is particularly intriguing given its power for boosted object tagging and its novelty relative to standard jet algorithms.  

There are also a number of challenges for using $N$-jettiness as a jet algorithm.  From an algorithmic point of view, $N$-jettiness minimization yields well-defined jets as long as the global minimum for $\tau_N$ is found, but just like for the $k$-means algorithm, finding the global minimum can be challenging.   The iterative procedure in \Sec{sec:minimizationAlgorithm} often converges to a local minimum for poorly chosen seeds, especially with the $R_0$ cut.   For practical purposes, it may be necessary to use an infrared/collinear safe method (such as anti-$k_T$) to determine seed axes, and then be satisfied with finding a local minimum.  Though not a major concern with modern computers, $N$-jettiness minimization is significantly slower than recursive clustering, especially with a large number of particles.   From a physics point of view, the fact that $N$ is fixed means that this algorithm does not define non-overlapping $N$-jet samples.  For analyses where the number of jets is known ahead of time, this is not an issue, but for more general searches this may be a liability.  

In the context of boosted hadronic objects, the fact that the number of cones is fixed at $N$ suggests a way to smoothly interpolate between traditional jet studies and jet substructure.  For example, minimizing $6$-jettiness on a boosted $t\bar{t}$ sample could in principle find all of the top constituents in both the boosted and non-boosted regimes.  In practice, this procedure is complicated by initial state radiation (ISR), since when minimizing $N$-jettiness, there is competition between splitting a fat jet into smaller jets and trying to minimize unclustered $p_T$ by identifying ISR jets.  Such competition could be alleviated by using $(>\!N)$-jettiness, at the expense of complicating the analysis.  

\section{Conclusions}
\label{sec:conclusion}

Jets are an important probe of short distance physics, as they offer a window to phenomena beyond the standard model.  The goal of jet substructure techniques is to maximize the physics reach for jets, and these ``fat jet'' methods are helpful for exploring extreme kinematic regimes with boosted hadronic resonances.  

In this context, $N$-subjettiness is a particularly interesting jet shape, since it directly measures the $N$-prong nature of a jet.  As originally defined in \Ref{Thaler:2010tr}, $N$-subjettiness required external input to determine the $N$ candidate subjet directions, and therefore had residual algorithmic dependence.  In this paper, we have shown how a modified version of the $k$-means clustering algorithm can be generalized to minimize $\tilde{\tau}_N$, and the minimum value of $N$-subjettiness is then a true jet shape.  

Using the BOOST2010  benchmarks, we have shown that $N$-subjettiness is a successful boosted top tagger, validating the preliminary study in \Ref{Thaler:2010tr}.  The ratio $\tau_3/\tau_2$ is an effective discriminant between top jets and QCD jets, especially if one uses the jet broadening measure ($\beta = 1$) and our minimization technique.  Additional discrimination power is possible using multivariate techniques, and a modified Fisher discriminant incorporating jet mass, $\tau_N$, and $\tau_N/\tau_{N-1}$ is particularly promising.  It would be interesting to study whether other top tagging methods could be improved with $\tau_N$ information, and to test the performance of $\tilde{\tau}_N$ minimization on boosted 2-prong objects like $W$, $Z$, or Higgs bosons.

Finally, the procedure to minimize $N$-subjettiness on a single jet can be used to minimize $N$-jettiness on an entire event.  This allows $N$-jettiness to define a fixed $N$ cone jet algorithm.  While there have been a few attempts in the past to define jets in terms of minimization or optimization, $N$-jettiness has the benefit that it is closely related to well-understood iterative cone algorithms, but does not suffer from the ambiguities of split-merge procedures.  The fact that $N$-jettiness includes an adjustable angular weighting exponent may prove useful, as one could interpolate between standard $\beta = 2$ weighting and more exotic $\beta = 1$ weighting.  As the LHC continues to explore new (and old) physics with jets, we are encouraged by this interesting connection between jet substructure observables and jet finding algorithms.

\acknowledgments  We thank Richard Cavanaugh, Nicholas Dunn, Stephen Ellis, Christopher Lee, Elliot Lipeles, David Miller, Salvatore Rappoccio, Iain Stewart, Frank Tackmann, and Christopher Vermilion for helpful physics conversations.  We also thank Christian Therkelsen, Colin Hom, and David Witmer for helpful advice on algorithms.  Portions of the work are based on the MIT senior thesis of K.V.T., which was selected for a 2011 Joel Matthew Orloff Award by the MIT Department of Physics.  This work was supported in part by the National Science Foundation under Grant No.~1066293 and the hospitality of the Aspen Center for Physics.   J.T. is supported by the U.S. Department of Energy under the Early Career research program DE-FG02-11ER-41741 as well as under cooperative research agreement DE-FG02-05ER-41360.

\bibliography{NsubFollowup}

\providecommand{\href}[2]{#2}\begingroup\raggedright\begin{thebibliography}{10}

\bibitem{Cacciari:2008gp}
M.~Cacciari, G.~P. Salam, and G.~Soyez, {\it {The Anti-k(t) jet clustering
  algorithm}},  {\em JHEP} {\bf 0804} (2008) 063,
  [\href{http://xxx.lanl.gov/abs/0802.1189}{{\tt arXiv:0802.1189}}].

\bibitem{Salam:2009jx}
G.~P. Salam, {\it {Towards Jetography}},  {\em Eur.Phys.J.} {\bf C67} (2010)
  637--686, [\href{http://xxx.lanl.gov/abs/0906.1833}{{\tt arXiv:0906.1833}}].

\bibitem{Aad:2011xw}
{\bf ATLAS} Collaboration, G.~Aad {\em et.~al.}, {\it {Search for new phenomena
  with the monojet and missing transverse momentum signature using the ATLAS
  detector in sqrt(s) = 7 TeV proton-proton collisions}},
  \href{http://xxx.lanl.gov/abs/1106.5327}{{\tt arXiv:1106.5327}}.

\bibitem{Chatrchyan:2011nd}
{\bf CMS} Collaboration, S.~Chatrchyan {\em et.~al.}, {\it {Search for New
  Physics with a Mono-Jet and Missing Transverse Energy in pp Collisions at
  sqrt(s) = 7 TeV}},  \href{http://xxx.lanl.gov/abs/1106.4775}{{\tt
  arXiv:1106.4775}}.

\bibitem{Aad:2011aj}
{\bf ATLAS} Collaboration, G.~Aad {\em et.~al.}, {\it {Search for New Physics
  in Dijet Mass and Angular Distributions in pp Collisions at $\sqrt{s} = 7$
  TeV Measured with the ATLAS Detector}},  {\em New J.Phys.} {\bf 13} (2011)
  053044, [\href{http://xxx.lanl.gov/abs/1103.3864}{{\tt arXiv:1103.3864}}].

\bibitem{Collaboration:2011ns}
{\bf CMS} Collaboration, S.~Chatrchyan {\em et.~al.}, {\it {Search for
  Resonances in the Dijet Mass Spectrum from 7 TeV pp Collisions at CMS}},
  \href{http://xxx.lanl.gov/abs/1107.4771}{{\tt arXiv:1107.4771}}.

\bibitem{daCosta:2011qk}
{\bf ATLAS} Collaboration, G.~Aad {\em et.~al.}, {\it {Search for squarks and
  gluinos using final states with jets and missing transverse momentum with the
  ATLAS detector in sqrt(s) = 7 TeV proton-proton collisions}},  {\em
  Phys.Lett.} {\bf B701} (2011) 186--203,
  [\href{http://xxx.lanl.gov/abs/1102.5290}{{\tt arXiv:1102.5290}}].

\bibitem{Collaboration:2011ida}
{\bf CMS} Collaboration, S.~Chatrchyan {\em et.~al.}, {\it {Search for New
  Physics with Jets and Missing Transverse Momentum in pp collisions at sqrt(s)
  = 7 TeV}},  \href{http://xxx.lanl.gov/abs/1106.4503}{{\tt arXiv:1106.4503}}.

\bibitem{CMS-PAS-JME-10-013}
{\bf CMS} Collaboration, {\it Jet substructure algorithms},  Tech. Rep.
  CMS-PAS-JME-10-013, 2011.

\bibitem{CMS-PAS-EXO-11-006}
{\bf CMS} Collaboration, {\it Search for bsm $t \bar{t}$ production in the
  boosted all-hadronic final state},  tech. rep., 2011.

\bibitem{ATLAS-CONF-2011-073}
{\bf ATLAS} Collaboration, {\it Measurement of jet mass and substructure for
  inclusive jets in Ãs = 7 tev pp collisions with the atlas experiment},  Tech.
  Rep. ATLAS-CONF-2011-073, CERN, Geneva, May, 2011.

\bibitem{Abazov:2011vh}
{\bf D0} Collaboration, V.~M. Abazov {\em et.~al.}, {\it {Measurement of color
  flow in $\mathbf{t\bar{t}}$ events from $\mathbf{p\bar{p}}$ collisions at
  $\mathbf{\sqrt{s}=1.96}$ TeV}},  {\em Phys.Rev.} {\bf D83} (2011) 092002,
  [\href{http://xxx.lanl.gov/abs/1101.0648}{{\tt arXiv:1101.0648}}].

\bibitem{Aaltonen:2011pg}
{\bf CDF} Collaboration, T.~Aaltonen {\em et.~al.}, {\it {Study of Substructure
  of High Transverse Momentum Jets Produced in Proton-Antiproton Collisions at
  $\sqrt{s}=1.96$ TeV}},  \href{http://xxx.lanl.gov/abs/1106.5952}{{\tt
  arXiv:1106.5952}}.

\bibitem{Abdesselam:2010pt}
A.~Abdesselam, E.~Kuutmann, U.~Bitenc, G.~Brooijmans, J.~Butterworth, {\em
  et.~al.}, {\it {Boosted objects: A Probe of beyond the Standard Model
  physics}},  {\em Eur.Phys.J.} {\bf C71} (2011) 1661,
  [\href{http://xxx.lanl.gov/abs/1012.5412}{{\tt arXiv:1012.5412}}]. Long
  author list - awaiting processing.

\bibitem{Seymour:1993mx}
M.~H. Seymour, {\it {Searches for new particles using cone and cluster jet
  algorithms: A Comparative study}},  {\em Z.Phys.} {\bf C62} (1994) 127--138.

\bibitem{Butterworth:2002tt}
J.~Butterworth, B.~Cox, and J.~R. Forshaw, {\it {$W W$ scattering at the CERN
  LHC}},  {\em Phys.Rev.} {\bf D65} (2002) 096014,
  [\href{http://xxx.lanl.gov/abs/hep-ph/0201098}{{\tt hep-ph/0201098}}].

\bibitem{Brooijmans:1077731}
G.~Brooijmans, {\it High pt hadronic top quark identification},  Tech. Rep.
  ATL-PHYS-CONF-2008-008. ATL-COM-PHYS-2008-001, CERN, Geneva, Jan, 2008.

\bibitem{Thaler:2008ju}
J.~Thaler and L.-T. Wang, {\it {Strategies to Identify Boosted Tops}},  {\em
  JHEP} {\bf 0807} (2008) 092, [\href{http://xxx.lanl.gov/abs/0806.0023}{{\tt
  arXiv:0806.0023}}].

\bibitem{Kaplan:2008ie}
D.~E. Kaplan, K.~Rehermann, M.~D. Schwartz, and B.~Tweedie, {\it {Top Tagging:
  A Method for Identifying Boosted Hadronically Decaying Top Quarks}},  {\em
  Phys.Rev.Lett.} {\bf 101} (2008) 142001,
  [\href{http://xxx.lanl.gov/abs/0806.0848}{{\tt arXiv:0806.0848}}].

\bibitem{Plehn:2009rk}
T.~Plehn, G.~P. Salam, and M.~Spannowsky, {\it {Fat Jets for a Light Higgs}},
  {\em Phys.Rev.Lett.} {\bf 104} (2010) 111801,
  [\href{http://xxx.lanl.gov/abs/0910.5472}{{\tt arXiv:0910.5472}}].

\bibitem{Plehn:2010st}
T.~Plehn, M.~Spannowsky, M.~Takeuchi, and D.~Zerwas, {\it {Stop Reconstruction
  with Tagged Tops}},  {\em JHEP} {\bf 1010} (2010) 078,
  [\href{http://xxx.lanl.gov/abs/1006.2833}{{\tt arXiv:1006.2833}}].

\bibitem{Almeida:2008yp}
L.~G. Almeida, S.~J. Lee, G.~Perez, G.~F. Sterman, I.~Sung, {\em et.~al.}, {\it
  {Substructure of high-$p_T$ Jets at the LHC}},  {\em Phys.Rev.} {\bf D79}
  (2009) 074017, [\href{http://xxx.lanl.gov/abs/0807.0234}{{\tt
  arXiv:0807.0234}}].

\bibitem{Gallicchio:2010sw}
J.~Gallicchio and M.~D. Schwartz, {\it {Seeing in Color: Jet Superstructure}},
  {\em Phys.Rev.Lett.} {\bf 105} (2010) 022001,
  [\href{http://xxx.lanl.gov/abs/1001.5027}{{\tt arXiv:1001.5027}}].

\bibitem{Hook:2011cq}
A.~Hook, M.~Jankowiak, and J.~G. Wacker, {\it {Jet Dipolarity: Top Tagging with
  Color Flow}},  \href{http://xxx.lanl.gov/abs/1102.1012}{{\tt
  arXiv:1102.1012}}.

\bibitem{Jankowiak:2011qa}
M.~Jankowiak and A.~J. Larkoski, {\it {Jet Substructure Without Trees}},  {\em
  JHEP} {\bf 1106} (2011) 057, [\href{http://xxx.lanl.gov/abs/1104.1646}{{\tt
  arXiv:1104.1646}}].

\bibitem{Butterworth:2008iy}
J.~M. Butterworth, A.~R. Davison, M.~Rubin, and G.~P. Salam, {\it {Jet
  substructure as a new Higgs search channel at the LHC}},  {\em
  Phys.Rev.Lett.} {\bf 100} (2008) 242001,
  [\href{http://xxx.lanl.gov/abs/0802.2470}{{\tt arXiv:0802.2470}}].

\bibitem{Ellis:2009su}
S.~D. Ellis, C.~K. Vermilion, and J.~R. Walsh, {\it {Techniques for improved
  heavy particle searches with jet substructure}},  {\em Phys.Rev.} {\bf D80}
  (2009) 051501, [\href{http://xxx.lanl.gov/abs/0903.5081}{{\tt
  arXiv:0903.5081}}].

\bibitem{Ellis:2009me}
S.~D. Ellis, C.~K. Vermilion, and J.~R. Walsh, {\it {Recombination Algorithms
  and Jet Substructure: Pruning as a Tool for Heavy Particle Searches}},  {\em
  Phys.Rev.} {\bf D81} (2010) 094023,
  [\href{http://xxx.lanl.gov/abs/0912.0033}{{\tt arXiv:0912.0033}}].

\bibitem{Krohn:2009th}
D.~Krohn, J.~Thaler, and L.-T. Wang, {\it {Jet Trimming}},  {\em JHEP} {\bf
  1002} (2010) 084, [\href{http://xxx.lanl.gov/abs/0912.1342}{{\tt
  arXiv:0912.1342}}].

\bibitem{Soper:2010xk}
D.~E. Soper and M.~Spannowsky, {\it {Combining subjet algorithms to enhance ZH
  detection at the LHC}},  {\em JHEP} {\bf 1008} (2010) 029,
  [\href{http://xxx.lanl.gov/abs/1005.0417}{{\tt arXiv:1005.0417}}].

\bibitem{Almeida:2010pa}
L.~G. Almeida, S.~J. Lee, G.~Perez, G.~Sterman, and I.~Sung, {\it {Template
  Overlap Method for Massive Jets}},  {\em Phys.Rev.} {\bf D82} (2010) 054034,
  [\href{http://xxx.lanl.gov/abs/1006.2035}{{\tt arXiv:1006.2035}}].

\bibitem{Soper:2011cr}
D.~E. Soper and M.~Spannowsky, {\it {Finding physics signals with shower
  deconstruction}},  \href{http://xxx.lanl.gov/abs/1102.3480}{{\tt
  arXiv:1102.3480}}.

\bibitem{Thaler:2010tr}
J.~Thaler and K.~Van~Tilburg, {\it {Identifying Boosted Objects with
  N-subjettiness}},  {\em JHEP} {\bf 1103} (2011) 015,
  [\href{http://xxx.lanl.gov/abs/1011.2268}{{\tt arXiv:1011.2268}}].

\bibitem{Stewart:2010tn}
I.~W. Stewart, F.~J. Tackmann, and W.~J. Waalewijn, {\it {N-Jettiness: An
  Inclusive Event Shape to Veto Jets}},  {\em Phys.Rev.Lett.} {\bf 105} (2010)
  092002, [\href{http://xxx.lanl.gov/abs/1004.2489}{{\tt arXiv:1004.2489}}].

\bibitem{Berger:2003iw}
C.~F. Berger, T.~Kucs, and G.~F. Sterman, {\it {Event shape / energy flow
  correlations}},  {\em Phys.Rev.} {\bf D68} (2003) 014012,
  [\href{http://xxx.lanl.gov/abs/hep-ph/0303051}{{\tt hep-ph/0303051}}].

\bibitem{Ellis:2010rwa}
S.~D. Ellis, C.~K. Vermilion, J.~R. Walsh, A.~Hornig, and C.~Lee, {\it {Jet
  Shapes and Jet Algorithms in SCET}},  {\em JHEP} {\bf 1011} (2010) 101,
  [\href{http://xxx.lanl.gov/abs/1001.0014}{{\tt arXiv:1001.0014}}].

\bibitem{Kim:2010uj}
J.-H. Kim, {\it {Rest Frame Subjet Algorithm With SISCone Jet For Fully
  Hadronic Decaying Higgs Search}},  {\em Phys.Rev.} {\bf D83} (2011) 011502,
  [\href{http://xxx.lanl.gov/abs/1011.1493}{{\tt arXiv:1011.1493}}].

\bibitem{Englert:2011iz}
C.~Englert, T.~S. Roy, and M.~Spannowsky, {\it {Ditau jets in Higgs searches}},
   \href{http://xxx.lanl.gov/abs/1106.4545}{{\tt arXiv:1106.4545}}.

\bibitem{Bai:2011mr}
Y.~Bai and J.~Shelton, {\it {Composite Octet Searches with Jet Substructure}},
  \href{http://xxx.lanl.gov/abs/1107.3563}{{\tt arXiv:1107.3563}}.

\bibitem{Catani:1993hr}
S.~Catani, Y.~L. Dokshitzer, M.~Seymour, and B.~Webber, {\it {Longitudinally
  invariant $K_t$ clustering algorithms for hadron hadron collisions}},  {\em
  Nucl.Phys.} {\bf B406} (1993) 187--224.

\bibitem{Ellis:1993tq}
S.~D. Ellis and D.~E. Soper, {\it {Successive combination jet algorithm for
  hadron collisions}},  {\em Phys.Rev.} {\bf D48} (1993) 3160--3166,
  [\href{http://xxx.lanl.gov/abs/hep-ph/9305266}{{\tt hep-ph/9305266}}].

\bibitem{Lloyd82leastsquares}
S.~P. Lloyd, {\it Least squares quantization in pcm},  {\em IEEE Transactions
  on Information Theory} {\bf 28} (1982) 129--137.

\bibitem{Farhi:1977sg}
E.~Farhi, {\it {A QCD Test for Jets}},  {\em Phys.Rev.Lett.} {\bf 39} (1977)
  1587--1588.

\bibitem{Catani:1992jc}
S.~Catani, G.~Turnock, and B.~Webber, {\it {Jet broadening measures in $e^{+}
  e^{-}$ annihilation}},  {\em Phys.Lett.} {\bf B295} (1992) 269--276.

\bibitem{Ding:2006:RPR:1143844.1143880}
C.~Ding, D.~Zhou, X.~He, and H.~Zha, {\it R1-pca: rotational invariant l1-norm
  principal component analysis for robust subspace factorization},  in {\em
  Proceedings of the 23rd international conference on Machine learning}, ICML
  '06, (New York, NY, USA), pp.~281--288, ACM, 2006.

\bibitem{FastJet}
M.~Cacciari, G.~P. Salam, and G.~Soyez. \url{http://www.fastjet.fr}.

\bibitem{Cacciari:2005hq}
M.~Cacciari and G.~P. Salam, {\it {Dispelling the $N^{3}$ myth for the $k_t$
  jet-finder}},  {\em Phys.Lett.} {\bf B641} (2006) 57--61,
  [\href{http://xxx.lanl.gov/abs/hep-ph/0512210}{{\tt hep-ph/0512210}}].

\bibitem{Gallicchio:2011xq}
J.~Gallicchio and M.~D. Schwartz, {\it {Quark and Gluon Tagging at the LHC}},
  \href{http://xxx.lanl.gov/abs/1106.3076}{{\tt arXiv:1106.3076}}.

\bibitem{Dokshitzer:1997in}
Y.~L. Dokshitzer, G.~Leder, S.~Moretti, and B.~Webber, {\it {Better jet
  clustering algorithms}},  {\em JHEP} {\bf 9708} (1997) 001,
  [\href{http://xxx.lanl.gov/abs/hep-ph/9707323}{{\tt hep-ph/9707323}}].

\bibitem{Wobisch:1998wt}
M.~Wobisch and T.~Wengler, {\it {Hadronization corrections to jet
  cross-sections in deep inelastic scattering}},
  \href{http://xxx.lanl.gov/abs/hep-ph/9907280}{{\tt hep-ph/9907280}}.

\bibitem{Butterworth:2002xg}
J.~M. Butterworth, J.~P. Couchman, B.~E. Cox, and B.~M. Waugh, {\it {KtJet: A
  C++ implementation of the K(T) clustering algorithm}},  {\em Comput. Phys.
  Commun.} {\bf 153} (2003) 85--96,
  [\href{http://xxx.lanl.gov/abs/hep-ph/0210022}{{\tt hep-ph/0210022}}].

\bibitem{Sjostrand:2006za}
T.~Sjostrand, S.~Mrenna, and P.~Z. Skands, {\it {PYTHIA 6.4 Physics and
  Manual}},  {\em JHEP} {\bf 0605} (2006) 026,
  [\href{http://xxx.lanl.gov/abs/hep-ph/0603175}{{\tt hep-ph/0603175}}].

\bibitem{Sjostrand:2007gs}
T.~Sjostrand, S.~Mrenna, and P.~Z. Skands, {\it {A Brief Introduction to PYTHIA
  8.1}},  {\em Comput.Phys.Commun.} {\bf 178} (2008) 852--867,
  [\href{http://xxx.lanl.gov/abs/0710.3820}{{\tt arXiv:0710.3820}}].

\bibitem{Corcella:2002jc}
G.~Corcella, I.~Knowles, G.~Marchesini, S.~Moretti, K.~Odagiri, {\em et.~al.},
  {\it {HERWIG 6.5 release note}},
  \href{http://xxx.lanl.gov/abs/hep-ph/0210213}{{\tt hep-ph/0210213}}.

\bibitem{Butterworth:1996zw}
J.~Butterworth, J.~R. Forshaw, and M.~Seymour, {\it {Multiparton interactions
  in photoproduction at HERA}},  {\em Z.Phys.} {\bf C72} (1996) 637--646,
  [\href{http://xxx.lanl.gov/abs/hep-ph/9601371}{{\tt hep-ph/9601371}}].

\bibitem{ATL-PHYS-PUB-2010-002}
{\bf ATLAS} Collaboration, {\it Atlas monte carlo tunes for mc09},  Tech. Rep.
  ATL-PHYS-PUB-2010-002, CERN, Geneva, Mar, 2010.

\bibitem{Skands:2010ak}
P.~Z. Skands, {\it {Tuning Monte Carlo Generators: The Perugia Tunes}},  {\em
  Phys.Rev.} {\bf D82} (2010) 074018,
  [\href{http://xxx.lanl.gov/abs/1005.3457}{{\tt arXiv:1005.3457}}].

\bibitem{CMS-PAS-JME-09-001}
{\bf CMS} Collaboration, {\it A cambridge-aachen (c-a) based jet algorithm for
  boosted top-jet tagging},  Tech. Rep. CMS-PAS-JME-09-001, Jul, 2009.

\bibitem{CMS-PAS-EXO-09-002}
{\bf CMS} Collaboration, {\it Search for high mass tt resonances in the
  all-hadronic mode},  Tech. Rep. CMS-PAS-EXO-09-002, Jun, 2009.

\bibitem{Rappoccio:1358770}
S.~Rappoccio, {\it A new top jet tagging algorithm for highly boosted top
  jets},  Tech. Rep. CMS-CR-2009-255. CERN-CMS-CR-2009-255, CERN, Geneva, Aug,
  2009.

\bibitem{ATL-PHYS-PUB-2010-008}
{\bf ATLAS} Collaboration, {\it Prospects for top anti-top resonance searches
  using early atlas data.},  Tech. Rep. ATL-PHYS-PUB-2010-008, CERN, Geneva,
  Jul, 2010.

\bibitem{ATL-PHYS-PUB-2009-081}
{\bf ATLAS} Collaboration, {\it Reconstruction of high mass $t\overline{t}$
  resonances in the lepton+jets channel},  Tech. Rep. ATL-PHYS-PUB-2009-081.
  ATL-COM-PHYS-2009-255, CERN, Geneva, May, 2009.

\bibitem{fisher}
R.~Fisher, {\it {The Use of Multiple Measurements in Taxonomic Problems}},
  {\em Annals of Eugenics} {\bf 7-II} (1936) 179--188.

\bibitem{anderson_bahadur}
T.~Anderson and R.~Bahadur, {\it {Classification into Two Multivariate Normal
  Distributions with Different Covariance Matrices}},  {\em Annals of Math.
  Statistics} {\bf 33} (1962) 420--431.

\bibitem{Chekanov:2005cq}
S.~Chekanov, {\it {A New jet algorithm based on the k-means clustering for the
  reconstruction of heavy states from jets}},  {\em Eur.Phys.J.} {\bf C47}
  (2006) 611--616, [\href{http://xxx.lanl.gov/abs/hep-ph/0512027}{{\tt
  hep-ph/0512027}}].

\bibitem{Berger:2002jt}
C.~Berger, E.~L. Berger, P.~Bhat, J.~Butterworth, S.~Ellis, {\em et.~al.}, {\it
  {Snowmass 2001: Jet energy flow project}},
  \href{http://xxx.lanl.gov/abs/hep-ph/0202207}{{\tt hep-ph/0202207}}.

\bibitem{Angelini:2002et}
L.~Angelini, P.~De~Felice, M.~Maggi, G.~Nardulli, L.~Nitti, {\em et.~al.}, {\it
  {Jet analysis by deterministic annealing}},  {\em Phys.Lett.} {\bf B545}
  (2002) 315--322, [\href{http://xxx.lanl.gov/abs/hep-ph/0207032}{{\tt
  hep-ph/0207032}}].

\bibitem{Angelini:2004ac}
L.~Angelini, G.~Nardulli, L.~Nitti, M.~Pellicoro, D.~Perrino, {\em et.~al.},
  {\it {Deterministic annealing as a jet clustering algorithm in hadronic
  collisions}},  {\em Phys.Lett.} {\bf B601} (2004) 56--63,
  [\href{http://xxx.lanl.gov/abs/hep-ph/0407214}{{\tt hep-ph/0407214}}].

\bibitem{Grigoriev:2003yc}
D.~Grigoriev, E.~Jankowski, and F.~Tkachov, {\it {Towards a standard jet
  definition}},  {\em Phys.Rev.Lett.} {\bf 91} (2003) 061801,
  [\href{http://xxx.lanl.gov/abs/hep-ph/0301185}{{\tt hep-ph/0301185}}].

\bibitem{Grigoriev:2003tn}
D.~Grigoriev, E.~Jankowski, and F.~Tkachov, {\it {Optimal jet finder}},  {\em
  Comput.Phys.Commun.} {\bf 155} (2003) 42--64,
  [\href{http://xxx.lanl.gov/abs/hep-ph/0301226}{{\tt hep-ph/0301226}}].

\bibitem{Lai:2008zp}
Y.-S. Lai and B.~A. Cole, {\it {Jet reconstruction in hadronic collisions by
  Gaussian filtering}},  \href{http://xxx.lanl.gov/abs/0806.1499}{{\tt
  arXiv:0806.1499}}.

\bibitem{Volobouev:2009rv}
I.~Volobouev, {\it {FFTJet: A Package for Multiresolution Particle Jet
  Reconstruction in the Fourier Domain}},
  \href{http://xxx.lanl.gov/abs/0907.0270}{{\tt arXiv:0907.0270}}.

\bibitem{Ellis:2001aa}
S.~Ellis, J.~Huston, and M.~Tonnesmann, {\it {On building better cone jet
  algorithms}},  \href{http://xxx.lanl.gov/abs/hep-ph/0111434}{{\tt
  hep-ph/0111434}}.

\bibitem{Ellis:2007ib}
S.~Ellis, J.~Huston, K.~Hatakeyama, P.~Loch, and M.~Tonnesmann, {\it {Jets in
  hadron-hadron collisions}},  {\em Prog.Part.Nucl.Phys.} {\bf 60} (2008)
  484--551, [\href{http://xxx.lanl.gov/abs/0712.2447}{{\tt arXiv:0712.2447}}].

\bibitem{Stewart:2010pd}
I.~W. Stewart, F.~J. Tackmann, and W.~J. Waalewijn, {\it {The Beam Thrust Cross
  Section for Drell-Yan at NNLL Order}},  {\em Phys.Rev.Lett.} {\bf 106} (2011)
  032001, [\href{http://xxx.lanl.gov/abs/1005.4060}{{\tt arXiv:1005.4060}}].

\bibitem{Berger:2010xi}
C.~F. Berger, C.~Marcantonini, I.~W. Stewart, F.~J. Tackmann, and W.~J.
  Waalewijn, {\it {Higgs Production with a Central Jet Veto at NNLL+NNLO}},
  {\em JHEP} {\bf 1104} (2011) 092,
  [\href{http://xxx.lanl.gov/abs/1012.4480}{{\tt arXiv:1012.4480}}].

\bibitem{Jouttenus:2011wh}
T.~T. Jouttenus, I.~W. Stewart, F.~J. Tackmann, and W.~J. Waalewijn, {\it {The
  Soft Function for Exclusive N-Jet Production at Hadron Colliders}},  {\em
  Phys.Rev.} {\bf D83} (2011) 114030,
  [\href{http://xxx.lanl.gov/abs/1102.4344}{{\tt arXiv:1102.4344}}].

\bibitem{Bauer:2011hj}
C.~W. Bauer, N.~D. Dunn, and A.~Hornig, {\it {Subtractions for SCET Soft
  Functions}},  \href{http://xxx.lanl.gov/abs/1102.4899}{{\tt
  arXiv:1102.4899}}.

\bibitem{Bauer:2011uc}
C.~W. Bauer, F.~J. Tackmann, J.~R. Walsh, and S.~Zuberi, {\it {Factorization
  and Resummation for Dijet Invariant Mass Spectra}},
  \href{http://xxx.lanl.gov/abs/1106.6047}{{\tt arXiv:1106.6047}}.

\end{thebibliography}\endgroup
\bibliographystyle{JHEP}

\end{document}